\documentclass[12pt]{article}
\usepackage{graphicx}
\usepackage{amsmath}
\usepackage{amssymb}
\usepackage{rv-macros}
\begin{document}

\title {Factorization of Hard Processes in QCD%
     \footnote{Originally published in A.H. Mueller, ed.,
     ``Perturbative QCD",
     Adv.\ Ser.\ Direct.\ High Energy Phys.\  {\bf 5}, 1 (1988).
     Authors' affiliations updated.
     } 
}

\author{
    John C. Collins
\\
    Penn State University\\
    104 Davey Lab\\
    University Park PA 16802, U.S.A.
\and
    Davison E. Soper 
\\
    Institute of Theoretical Science\\
    University of Oregon\\
    Eugene OR 97403, U.S.A.
\and
    George Sterman
\\
    C.N. Yang Institute for Theoretical Physics\\
    Stony Brook University \\
    Stony Brook NY 11794-3840, U.S.A.
}

\date{27 September 2004}

\maketitle
\begin{abstract}
 We summarize the standard factorization theorems 
for hard processes in QCD, and describe their proofs. 
\end{abstract}



\section{Introduction}
\label{introsec}

In this chapter, we discuss the factorization theorems that enable one
to apply perturbative calculations to many important processes involving
hadrons. In this introductory section we state briefly what the
theorems are, and in Sects.\ \ref{hardsec} to \ref{pdfsec}, we indicate
how they are 
applied in calculations.  In subsequent sections, we present an outline
of how the theorems are established, both in the simple but
instructive case of scalar field theory and in the more complex and
physically interesting case of quantum chromodynamics
(QCD).
 
The basic problem addressed by factorization theorems is
how to calculate high energy cross sections.  Order by order
in a renormalizable perturbation series, any physical
quantity is a function of three classes of variables with
dimensions of mass.  These are the kinematic energy scale(s) of the
scattering, $Q$,
the masses, $m$, and a renormalization scale $\mu$.  We can
make use of the asymptotic freedom of QCD by choosing the
renormalization scale to be large, in which case the
effective 
coupling constant $g(\mu)$ will be correspondingly small,
$g(\mu) \sim 1/ \ln (\mu/\Lambda_{\rm QCD})$. 
The renormalization scale, however, will appear in
ratios $Q/\mu$ and $\mu/m$, and at high
energy at least one of these ratios is 
large.   If we pick $\mu \sim Q$, for instance, then at $n$
loops the coupling will generally appear in the
combination $g^{2n}(Q) \ln^{an}(Q/m)$, with $a=1$ or
$2$. (See Sect.\ \ref{regionsec}.)  As a result,
the perturbation series is
no longer an expansion in a small parameter. 
The presence of logarithms 
involving the masses shows the importance of contributions from 
long distances, where the precise values of masses
(including the vanishing gluon mass!) are relevant.  For such
contributions we do not expect asymptotic freedom to help, since it is a
property of the coupling only at short distances.  In
summary, a general cross section is a combination of short-
and long-distance behavior, and is hence not 
computable directly in perturbation theory for QCD.

There are exceptions to
this rule.  For reasons which will become
clear in Sect.\ \ref{regionsec}, these are inclusive cross sections
without hadrons in the initial state, such as the total
cross section for \epem\ annihilation into hadrons, or
into jets.  

This leaves over, however, the majority
of  experimentally studied lepton-hadron and hadron-hadron
large momentum transfer cross sections, as well as
inclusive cross sections in \epem\ annihilation with detected hadrons.  
Factorization theorems allow us to derive
predictions for these cross sections, by separating
(factorizing) long-distance from short-distance behavior in
a systematic fashion.
Thus almost 
all applications of perturbative QCD use factorization properties of
some kind.  

In this chapter, we will explicitly treat factorization theorems for
inclusive processes in which
(1)  all Lorentz invariants defining the process are large and comparable, 
except for particle masses, and 
(2) one counts all
final states that include the specified outgoing particles or jets.  
The second 
condition means that we consider such processes as $ {\rm
hadron}\ A + {\rm hadron}\ B \to {\rm hadron}\ C + X $,
where the $X$  denotes ``anything else'' in addition to the specified 
hadron $C$.
The first condition means that in this example 
the specified hadron $C$ should have a
transverse momentum comparable to the center-of-mass energy.
For such processes, the theorems show how to factorize long distance
effects, which are not perturbatively calculable, into functions describing
the distribution of partons in a hadron --- or hadrons in a parton in the
case of final-state hadrons.  Not only can these functions be measured
experimentally, but also the same parton
distribution and decay functions will be observed in all such processes.  The
part of the  cross section that remains after the parton distribution and
decay functions have been factored out is the short distance cross section
for the hard scattering of partons.  This hard scattering cross section is
perturbatively calculable, by a method which we describe below.

Some examples of processes for which one expects a factorization theorem 
of this type to
hold include (denoting hadrons by $A$, $B$, $C$ \dots)
\begin{itemize}
\item Deeply inelastic scattering,
   ${\rm lepton} + A \to {\rm lepton}' + X$;
\item $\e^+ + \e^- \to A + X$;
\item The Drell-Yan process,
   \begin{itemize}
      \item $A + B\to \mu^+ + \mu^-  + X$,
      \item $A + B\to \e^+ + \e^-  + X$,
      \item $A + B \to {\rm W}  + X$,
      \item $A + B \to {\rm Z}  + X$;
   \end{itemize}
\item $A + B \to {\rm jet}  + X$;
\item $A + B \to {\rm heavy\ quark}  + X$.
\end{itemize}
In the last example, the heavy quark mass, which must be large
compared to 1 GeV, plays the role of the large momentum transfer.
In the Drell-Yan case, the kinematic invariants are the particle masses,
the square, $s$, of the center-of-mass energy, and 
the invariant mass $Q$ and transverse momentum $\qt$ of the
lepton pair.  The requirement, for the theorems that we discuss, that the
invariants all be large and comparable means that not only should $Q^2$ be
of order $s$, but also that either we integrate over
all $\qt$ or $\qt$ is of order $Q$.  

There are applications of QCD to processes in which there is a large momentum
scale involved but for which the most straightforward sort of factorization
theorem, as discussed in this chapter, must be modified.  However,
the same style of analysis as we will describe applies to these more
general situations.  (The Drell-Yan process when $\qt$ is much less than
$Q$ is an example.)  We will summarize these in Sect.\ \ref{conclsec}.

Some of the factorization properties, 
such as those we describe in this chapter, have been
proved at a reasonable level of rigor within the context of perturbation
theory.  But many of the other results have, so far, been proved less
completely.  

The following three subsections give explicit factorization
theorems for three basic cross sections from the list above, deeply
inelastic scattering, single-particle inclusive annihilation and the
Drell-Yan process.  These three examples illustrate most of the issues
involved in the application and proof of factorization.  We close the
section by relating factorization to the parton model.

\subsection{Deeply Inelastic Scattering}
\label{introsec:DIS}
 
Deeply inelastic lepton scattering plays a central role in any discussion of
factorization, both because this was the first process in
which pointlike partons were ``seen'' inside the hadron, and because much of
the data that determines the parton distribution functions comes from
measurements in this process.  In particular, let us consider the process $\e
+ A \to \e + X$, 
which proceeds via the exchange of a virtual photon with
momentum $q^\mu$.  From the measured cross section, one can extract
the standard hadronic tensor $W^{\mu\nu} (q^\mu, p^\mu)$,
\begin{eqnarray}
  W^{\mu\nu} &=& \frac{1 }{4\pi} \int \d^4y \e^{ i q \cdot y} 
      \sum_X  \langle A|j ^\mu (y) | X \rangle \langle X|j^\nu(0) | A \rangle
\nonumber\\
  &=& F_1(x, Q^2) \left( -g^{ \mu\nu} + \frac{q^\mu q^\nu }{q^2} \right)
\nonumber\\
  && +F_2(x, Q^2) \frac{ \left( p^\mu - q^\mu p \cdot q / q^2 \right)
                         \left( p^\nu - q^\nu p \cdot q / q^2 \right)
                  }{
                      p \cdot q
                 } ,   
\numlabel{eq:str.fn.def}
\end{eqnarray}
where $Q^2 = - q_\mu q^\mu$,
$x = Q^2 / 2 q \cdot p$, $p^\mu$ is the momentum of
the incoming hadron $A$, and $j^\mu(x)$ is the electromagnetic current.  
(More
generally, $j^\mu(y)$  can be any electroweak current, and there will be
more than two scalar structure functions $F_ i$.)  

We consider the process in the Bjorken limit, i.e., large $Q$ at fixed $x$.
The factorization theorem is contained
in the following expression for $W^{\mu\nu}$,
\begin{equation}
W^{\mu \nu}(q^\mu,p^\mu)
= \sum_{a}
\int_{x}^1 \frac{\d \xi }{\xi}  \ f_{a/A}(\xi,\mu)\
H_{a}^{\mu\nu}\!\!\left(
q^\mu,\xi p^\mu,\mu,\al(\mu)  \right)
+ \hbox{remainder}.
\numlabel{eq:2}
\end{equation}
Here $f_{a/A}(\xi,\mu)$ is a parton distribution function, whose 
precise definition is given in Sect.\ \ref{pdfsec}.  There, 
$f_{a/A}(\xi,\mu) \d \xi$ is interpreted as the 
probability to find a parton of
type $a$ (= ${\rm gluon}, {\rm u}, \bar {\rm u}, {\rm d}, \bar {\rm d},
\dots$) in a hadron of type $A$ carrying a fraction $\xi$ to $\xi + \d \xi$
of the hadron's momentum.  In the formula, one sums over all the possible
types of parton, $a$.  We can prove Eq.\ (\ref{eq:2}) in perturbation theory, with a
remainder down by a power of $Q$ (in this case, the power is $Q^{-2}$
modulo logarithms, but the precise value depends on the cross section at
hand, and has not always been determined).

We can project Eq.\ (\ref{eq:2}) onto individual structure functions:
\begin{align}
   F_1(x,Q^2) &= \sum_{a}
      \int_{x}^1 \frac{\d \xi}{\xi}  \ f_{a/A}(\xi,\mu)\
       H_{1a} \left( \frac{x}{\xi}, \frac{Q}{\mu}, \al(\mu) \right) 
       + \hbox{remainder},
\nonumber\\
   \frac{1 }{x} F_2(x,Q^2) &= \sum_{a}
      \int_{x}^1 \frac{\d \xi }{\xi}  \ f_{a/A}(\xi,\mu)\
       \frac{\xi }{x} 
       H_{2a} \left( \frac{x}{\xi}, \frac{Q}{\mu}, \al(\mu) \right) 
       + \hbox{remainder},
\numlabel{eq:3}
\end{align}
The extra factors of $ 1/x$ and $\xi / x$ in the equation for $F_2$ are
needed because of the dependence on target momentum of the tensor
multiplying $F_2$.

Inspired by the terminology of the operator product expansion for the
moments of the structure functions, it is conventional to call the
first term on the right of either of Eqs.\ (\ref{eq:2}) or
(\ref{eq:3}) the leading twist contribution, and to call the remainder
the higher twist contribution.  The same terminology of leading and
higher twist is used for the factorization theorems for other
processes.

It is not so obvious why proving Eq.\ (\ref{eq:2}) in perturbation
theory is useful, given that hadrons are not perturbative objects.
But suppose we do decide on a way of computing the matrix elements in
Eq.\ (\ref{eq:str.fn.def}) perturbatively.  For any such formulation for hadron
$A$, both $W^{\mu\nu}$ and $f_{a/A}$ will depend on phenomena at the
scale of hadronic masses (or some other infrared cutoff), and the
exact nature of these phenomena will depend on our particular choice
of $A$, as well as on the precise values we pick for both hadronic and
partonic masses.  The content of the factorization theorem is that
this dependence of $W^{ \mu \nu }$ on low mass phenomena is entirely
contained in the factor of $f_{a/A}$.

The remaining function, the hard scattering coefficient
$H_{a}^{\mu\nu}$, has two important properties.  First, it depends
only on the parton type $a$, and not directly on our choice of hadron
$A$.  Secondly, it is ultraviolet dominated, that is, it receives
important contributions only from momenta of order $Q$.  The first
property allows us to calculate $H_{a}^{\mu\nu}$ from Eq.\
(\ref{eq:2}) with the simplest choice of external hadron, $A=b$, $b$
being a parton.  (We will see an example of this in our calculations
for the Drell-Yan process in Sect.\ \ref{hardsec}.)  The second
property ensures that when we do this calculation, $H_{a}^{\mu\nu}$
will be a power series in $\alpha_s(Q)$, with finite coefficients.  We
now assume that nonperturbative long-distance effects in the complete
theory factorize in the same way as do perturbative long-distance
effects.  Once this assumption is made, we can interpret our
perturbative calculation of $H_{a}^{\mu\nu}$ as a prediction of the
theory.  Parton model ideas, summarized in
Sect~\ref{introsec:parton.model}, give motivation that the assumption
is valid.  Note that our definition of the parton distributions, which
we will give in Sect.\ \ref{pdfsec}, is an operator definition, which
can be applied beyond perturbation theory.

This ability to calculate the $H_{a}^{\mu\nu}$ results in great predictive
power for factorization theorems.  For
instance, if we measure $F_2(x, Q^2)$ for a particular hadron $A$, 
Eq.\ (\ref{eq:3}) will
enable us to determine $f_{a/A}$.  
We then derive a prediction $F_1(x, Q^2)$ for the same
hadron $A$, in terms of the observed $F_2$ and the calculated functions
$H_{ia}$.  This is the simplest example of the universality of
parton distributions.

The functions $H_{i a}$ may be thought of as
hard-scattering structure functions for parton targets, but this
interpretation should not be taken too literally.  In any case, methods for
putting this procedure into practice, including definitions for
the parton distributions are the subjects of Sects.\ \ref{hardsec} to
\ref{pdfsec}. 
 
Originally, Eq.\ (\ref{eq:2}) was primarily discussed in terms of the moments
of the structure functions, such as
\begin{eqnarray}
   \tilde F_1(n,Q^2)
   &=& \int_0^1 \frac{\d x}{x}\ x^n F_1(x,Q^2) ,
\nonumber\\
   \tilde F_2(n,Q^2)
   &=& \int_0^1 \frac{\d x}{x} \ x^{n-1} F_2(x,Q^2) .
\end{eqnarray}
With this notation, Eq.\ (\ref{eq:3}) becomes
\begin{equation}
\tilde F_i (n,Q^2)
= \sum_{a}
 \tilde f_{a/A}(n,\mu)\
\tilde H_{i a} \!\!\left( n, \frac{Q}{\mu}, \al(\mu) \right).
\numlabel{eq:5}
\end{equation}
In this form of the factorization theorem, when $n$ is an integer, the
$\tilde f_{j/A}(n,\mu)$ are hadron matrix elements of certain local
operators, evaluated at a renormalization scale $\mu$. On the other
hand, the structure function moment $ \tilde F_2(n,Q^2)$ can be
expressed in terms of the hadron matrix element of a product of two
electromagnetic current operators evaluated at two nearby space-time
points.  Equation (\ref{eq:5}) thus appears as an application of the
operator product expansion \cite{Wilson,Z,CHM}. The product of the two
operators is expressed in terms of local operators and some
perturbatively calculable coefficients $\tilde H_{ i a} ( n, Q / \mu,
\al(\mu) )$, called Wilson coefficients.  It was using this scheme
that the $\tilde H_{ i a} ( n, Q / \mu, \al(\mu) )$ were first
calculated \cite{Bardeen}.

\subsection{Single Particle Inclusive Annihilation}
\label{introsec:incl.annih}

In this subsection, we consider the  
process $\gamma^* \to A +X$, where $\gamma^*$ is an off-shell photon.
The relevant tensor for the process, for which structure functions
analogous to those in Eq.\ (\ref{eq:str.fn.def}) may be derived, is
\begin{equation}
 D^{\mu\nu}(x,Q) = \frac{1}{4\pi}
\int \d^4y \ \e^{iq\cdot y} \sum_X \langle 0| j^\mu(y) | A X \rangle
\langle A X| j^\nu(0) | 0\rangle,
\numlabel{eq:6}
\end{equation}
where $q ^\mu $ is now a time-like momentum and $Q^2 = q^2$.  The sum
is over all final-states that contain a particle $A$ of defined momentum
and type.  We define a scaling variable by $z = 2p \cdot q /Q^2$, where $p
^\mu $ is the momentum of $A$, and we will consider the appropriate
generalization of the Bjorken limit, that is, $Q$ large with $z$ fixed.

The factorization theorem here is quite analogous to Eq.\ (\ref{eq:2}),
but incorporates the slightly different kinematics,
\begin{equation}
D^{\mu\nu}(z, Q) = \sum_a \int _z ^1 \frac{\d \zeta }{\zeta} \,
H_a^{\mu\nu} (z/\zeta , Q/\mu, \al(\mu) ) \, d_{A/a} (\zeta) ,
\numlabel{eq:7}
\end{equation}
with corrections down by a power of $Q$, as usual.  We have used the
same notation for the hard functions as in deeply inelastic
scattering, and as in that case they are perturbatively calculable
functions.  Here it is the fragmentation functions $d_{A/a} (\zeta)$
which are observed from experiment, and which occur in any similar
inclusive cross section with a particular observed hadron in the final
state.  For example, single-particle inclusive cross sections in
deeply inelastic scattering cross sections require the factorization
both of parton distributions $f_{a/A}$, with $A$ the initial hadron,
and of distributions $d_{B/a}$, with $B$ the observed hadron in the
final state.  We shall not go into the details of such cross sections
here \cite{EarlyProofs}.

\subsection{Drell-Yan}
 
Our final example to illustrate the important issues of factorization is
the Drell-Yan process:
\begin{equation}
A + B\to \mu^+ +\mu^- + X
\end{equation}
at lowest order in quantum electrodynamics but, in principle, at any order in
quantum chromodynamics.  $q^\mu$ is now the momentum of the muon pair.
We shall be concerned with the cross section $\d\sigma/\d Q^2\d y$, where $Q^2$
is the square of the muon pair mass,
\begin{equation}
Q^2 = q^\mu q_\mu ,
\end{equation}
and $y$ is the rapidity of the muon pair,
\begin{equation}
y = \frac{1}{2}\, \ln\!\left( \frac{ q \cdot P_A }{ q \cdot P_B } \right).
\end{equation}
We imagine letting  $Q^2$ and the center of mass energy $\sqrt s$ become very
large, while $Q^2/s$ remains fixed.
 
The relevant factorization theorem, accurate up to corrections
suppressed by a power of $Q^2$, is
\begin{multline}
   \frac{ \d\sigma }{ \d Q^2 \d y }  \sim
   \sum_{a,b} \int_{x_A}^1 {\d\xi_A} \,
   \int_{x_B}^1{ \d\xi_B}\
   \times
\\
   \times
   f_{a/A}(\xi_A,\mu) \
   H_{ab}\!\left(\frac{x_A }{\xi_A}, \frac{x_B }{\xi_B}, Q;
   \frac{\mu}{Q},\al(\mu)\right)
   \, f_{b/B}(\xi_B,\mu) .
\numlabel{eq:11}
\end{multline}
Here  $a$ and $b$ label parton types and we denote
\begin{equation}
x_A = \e^y \sqrt{\frac{Q^2 }{s}},\quad
x_B = \e^{-y} \sqrt{\frac{Q^2 }{s}}.
\numlabel{eq:12}
\end{equation}
 
The function $H_{ab}$ is the ultraviolet-dominated hard scattering cross
section, computable in perturbation theory.  It plays the role of a parton
level cross section and is often written as
\begin{equation}
H_{ab} = \frac{ \d\sigmahat }{ \d Q^2 \d y }
\end{equation}
when it is not necessary to display the functional dependence of $H_{ab}$ on
the kinematical variables.  The parton distribution
functions, $f$, are the same
as in deeply inelastic scattering.  Thus, for instance, one can measure the 
parton distribution functions in deeply inelastic scattering experiments and
apply them to predict the Drell-Yan cross section. As before, the
parameter $\mu$ is a renormalization scale used in the calculation of
$H_{ab}$.

\subsection{Factorization in the Parton Model}
\label{introsec:parton.model}

Having introduced the basic factorization theorems, we will now try to give
them an intuitive basis.  Here we shall appeal to Feynman's parton
model\cite{Feyn}.  In fact, we shall see that factorization
theorems may be thought of as field theoretic realizations
of the parton model.  
 
In the parton model, we imagine hadrons as extended objects, made up of
constituents (partons) held together by their mutual interactions.  Of
course, these partons will be quarks and gluons in the real world, as
described by QCD, but we do not use this fact yet.  At the level of the
parton model, we assume that the hadrons can be described in terms of
virtual partonic states, but that we are not in a position to calculate the
structure of these states.  On the other hand, we assume that we do know
how to compute the scattering of a free parton by, say, an electron.  By
``free", we simply mean that we neglect parton-parton interactions.  This
dichotomy of ignorance and knowledge corresponds to our inability to
compute perturbatively at long distances in QCD, while having asymptotic
freedom at short distances.

To be
specific, consider inclusive electron-hadron scattering by
virtual photon exchange at high energy and momentum
transfer.  Consider how this scattering looks in the
center-of-mass frame, where two important things happen to
the hadron.  It is Lorentz contracted in the direction
of the collision, and its internal
interactions are time dilated.  So, as the center-of-mass
energy increases the lifetime of any virtual partonic
state is lengthened, while the time it takes the electron
to traverse the hadron is shortened.  When the latter is
much shorter than the former the hadron will be in a
single virtual state characterized by a definite number of
partons during the entire time the electron takes to cross
it.  Since the partons do not interact during this time,
each one may be thought of as carrying a definite
fraction $x$ of the hadron's momentum in the center of mass
frame.  We expect $x$ to satisfy 
$0<x<1$, since otherwise one or more
partons would have to move in the opposite direction to
the hadron, an unlikely configuration.  It now makes sense
to talk about the electron interacting with partons 
of definite momentum, rather
than with the hadron as a whole. In addition, when the
momentum transfer is very high, the virtual photon
which mediates electron-parton scattering cannot travel
far.  Then, if the density of partons is not too high, the
electron will be able to interact with only a single
parton.  Also, interactions which
occur in the final state, after the hard scattering, are
assumed to occur on time scales too long to interfere with
it.

With these assumptions, the high
energy scattering process becomes
essentially classical and incoherent.  That is, the
interactions of the partons among themselves, which occur
at time-dilated time scales before or after the 
hard scattering, cannot interfere
with the interaction of a parton with the electron.  The cross
section for hadron scattering may thus be computed by
combining probabilities, rather than amplitudes.  We 
define a parton distribution $f_{a/H}(\xi)$ as the
probability that the electron will encounter a ``frozen",
noninteracting  parton of species $a$ with fraction $\xi$
of the hadron's momentum.  We take the cross section
for the electron to scatter from such a parton  with
momentum transfer $Q^2$ as the Born cross section $
\sigma_{B} (Q^2,\xi)$.  Straightforward kinematics shows
that  for free partons $\xi>x \equiv 2p \cdot q / Q^2$,
and the total cross section for deeply inelastic scattering
of a hadron by an electron is  
\begin{equation}
\sigma_{e H}(x,Q^2) = \sum_a \int_x^1 \d \xi \, f_{a/H}(\xi)\, 
\sigma_{B} (x/\xi , Q^2).
\numlabel{eq:14}
\end{equation}
This is the parton model cross section for deeply
inelastic scattering.  It
is precisely of the form of Eq.\ (\ref{eq:2}), and is the
model for all the factorization theorems which we discuss in this chapter.


Essentially the same reasoning may be applied to single-particle-inclusive
cross sections and to the
Drell-Yan cross section.
For example, in the parton model the latter process is given by the direct
annihilation of a parton and anti-parton pair, one from
each hadron, in the Born approximation, $\sigma'_B(Q^2,y)$. 
The interactions which produce the distributions of each
such parton occur on a scale which is again much longer than
the time scale of the annihilation and, in addition, final-state 
interactions between the remaining partons take
place too late to affect the annihilation.  We thus
generalize  (\ref{eq:14}) to the parton model Drell-Yan cross
section
\begin{equation}
\frac{ \d \sigma }{ \d Q^2 dy } = \sum_{a} \int_{x_A}^1 \d \xi_A
\int_{x_B}^1 \d \xi_B \ f_{a/A}(\xi_A) \ f_{{\bar a}/B}(\xi_B) 
\ \sigma'_B(Q^2,y) ,
\numlabel{eq:15}
\end{equation}
where $x_{A,B}$ are defined in \ref{eq:12}.  Equation (\ref{eq:15}) is of
the same form as the full factorization formula, 
(\ref{eq:11}), except that there is only a
single sum over parton species, since the hard process here consists of a
simple quark-antiquark annihilation.  In the parton model, the
functions $f_{a/A}(\xi_A)$ in Drell Yan must be the
same as in deeply inelastic scattering, Eq.\ (\ref{eq:14}),
since they describe the internal structure of the hadron,
which has been decoupled kinematically from the
annihilation and from the other hadron.  
It is important to notice
that the Lorentz contraction of the hadrons in  the center
of mass system is indispensable for this universality of
parton distributions.  Without it the partons from
different hadrons would overlap a finite time before the
scattering, and initial-state interactions would then
modify the distributions.

We now turn to the technical
discussion of factorization theorems in QCD, but it is
important not to loose sight of their intuitive basis in
the kinematics of high energy scattering.  In fact, when
we return to proofs of factorization
theorems in gauge theories (Sects.\ \ref{gaugefactsec} and
\ref{gaugeproofsec}) these considerations
will play a central role.



\section{Calculation of the Hard Scattering Cross Section}
\label{hardsec}

In this and the following two sections, we discuss the explicit calculation
of the hard scattering functions for the Drell Yan cross section.  In doing
so, we will cover most of the technical points which are encountered in
applying factorization in other realistic cases as well.
 
\begin{figure}
  \centering
  \includegraphics[scale=0.8]{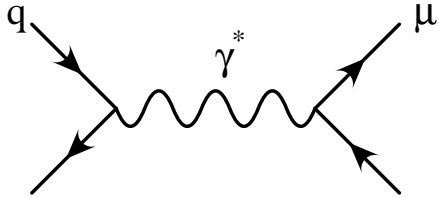}
  \caption{Born amplitude for the Drell-Yan process.  }
  \label{fig:1}
\end{figure}

At order zero in $\al$ for the Drell-Yan cross section,
the hard process described by $H_{ab}$ is
quark-antiquark annihilation, as illustrated in \ref{fig:1}.
One can simply compute this parton level cross section from the Feynman
diagram and insert it into the factorization formula (\ref{eq:11}).
The resulting 
cross section is not itself a prediction of QCD, although it is a
prediction of the parton model.
The factorization
theorem will make the connection
between the two.  At the Born level, it is natural to define
$f_{a/a}(\xi)=\delta(1-\xi)$.  We then find
\begin{eqnarray}
\frac{ \d \sigma^{(0)} }{ \d Q^2 \d y } &=&
H_{ab}^{(0)}\!\left(\frac{x_A }{\xi_A} 
\frac{x_B}{\xi_B}, Q;\frac{\mu}{Q}; \epsilon\right)
\nonumber\\
&=& \delta_{a,\bar b}\ e_a^2\
\frac{4\pi\alpha^2 }{9 Q^4}\
C\! \left(\frac{\mu}{Q}, \epsilon\right)\
\delta\!\left(\frac{x_a }{\xi_a} - 1 \right)\
\delta\!\left(\frac{x_b }{\xi_b} - 1 \right), 
\numlabel{eq:16}
\end{eqnarray}
where the factor $\delta_{a,\bar b}$ indicates that parton a must be the 
antiparticle to
parton b. Here $C(\epsilon)$ is 1 if we work in 4 space-time dimensions.
However, when one wants to calculate higher order contributions, it 
will turn out to be useful
to perform the entire calculation in $4-2\epsilon$ dimensions.  Then
\begin{equation}
C\!\left(\frac{\mu}{Q},\epsilon \right) =
\left(  \frac{\mu^2 }{Q^2}\, e^\gamma\right)^\epsilon
\frac{ (1 - \epsilon)^2 }{(1 - 2 \epsilon / 3) (1 - 2 \epsilon)}
\frac{\Gamma(1 - \epsilon) }{\Gamma (1 - 2 \epsilon)} .
\numlabel{eq:17}
\end{equation}
The $\epsilon$ dependence here arises from three sources.  First, the Dirac
trace algebra gives an angular dependence $1 + \cos^2\theta - 2\epsilon$.
Secondly, one introduces a factor 
$(\mu^2 /(4\pi)\, e^\gamma)^\epsilon$ so as to
keep the cross section at a constant 
overall dimensionality\footnote{
   We use 
   $(\mu^2 /(4\pi)\ e^\gamma)^\epsilon$ rather than $(\mu^2)^\epsilon$ in
   anticipation of our use of \MSbar\ renormalization.
}
of $M^{-4}$.
Finally, the
integration over the lepton angles in $4 - 2\epsilon$ dimensions gives the
remaining $\epsilon$ dependence.  Actually, it is quite permissible to perform
the lepton trace calculation and the integration over lepton angles in 4
dimensions instead of $4 - 2 \epsilon$ dimensions.  This procedure results in
multiplying the Born cross section and the higher order cross section by a
common, $\epsilon$-dependent factor.  As we will see below, such a factor will
drop out in the physical cross section.


\begin{figure}
  \centering
  \includegraphics{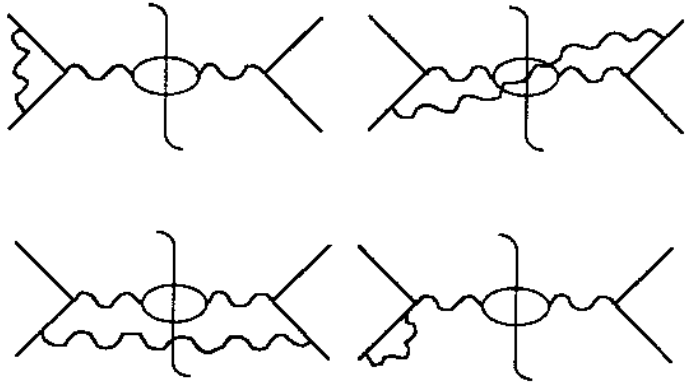}
  \caption{Order $\al$ contributions to the Drell-Yan cross
   section.  }
  \label{fig:2}
\end{figure}

Now let us calculate $H$ at one loop.  
At first order in $\al$, the cross section
gets  contributions from the graphs shown in \ref{fig:2},
along with their mirror diagrams. 
In this figure, we show contributions to both the amplitude and its
complex conjugate, separated by a vertical line which represents the
final state.  We will use this notation frequently below, and refer
to diagrams of this sort as ``cut diagrams".
The situation now is
not so simple, because a straightforward calculation of the cross section
for  $ {\rm quark} + {\rm antiquark} \to \mu^+ +\mu^- + X$ according to the
diagrams shown above yields an infinite result when we use massless,
on-shell quarks as the incoming
particles.

Following Sect.\ \ref{introsec:DIS}, we use the
factorization formula (\ref{eq:11}) applied to incoming partons instead of incoming
hadrons.  
Since the details associated with parton masses are going to factorize,
we can choose to
 calculate the cross section for
${\rm parton}\,a +{\rm parton}\,b \to \mu^+ + \mu^- +X$ 
with the partons having zero mass and transverse momentum.   Let us call
this cross section $G_{ab}$:
\begin{equation}
\frac{ \d\sigma(a\, b\to \mu^+\, \mu^-\, X) }{ \d Q^2 \d y } =
G_{ab}\!\left({x_A}, {x_B}, Q; \frac{\mu}{Q};
\al;\epsilon\right)
\end{equation}
In this calculation there are both ultraviolet and infrared divergences.
Dimensional regularization is used to regulate them both. The
factorization formula is then
\begin{multline}
G_{ab}\!\left({x_A}, {x_B}, Q; \frac{\mu}{Q};
\al; \epsilon\right)
=
\sum_{c,d} \int_{x_A}^1 {\d\xi_A}
            \int_{x_B}^1 {\d\xi_B}\
\\
\times
f_{c/a}(\xi_A;\epsilon) \
H_{cd}\!\left(\frac{x_A }{\xi_A}, \frac{x_B }{\xi_B}, Q;
 \frac{\mu}{Q}, \al (\mu );\epsilon\right)
\, f_{d/b}(\xi_B;\epsilon).
\end{multline}
Both factors in the formula depend on $\mu$, which is the scale factor
introduced in the dimensional regularization and subsequent \MSbar\
renormalization \cite{Bardeen} 
of Green functions of ultraviolet divergent operators.  One
introduces a factor
\begin{equation}
\left( \frac{\mu^2 }{4\pi} e^\gamma \right)^{\epsilon}
\end{equation}
for each integration $\int \d^{4-2\epsilon}k$ in order to keep the
dimensionality of the result independent of  $\epsilon$.  Ultraviolet
divergences then appear as poles in the variable $\epsilon$, which are
subtracted away, as explained in \cite{JCCbook}.  The factor
$e^\gamma/(4\pi)$
that comes along with the $\mu$ is the difference between \MSbar\
renormalization and minimal subtraction (MS) renormalization.
Here $\gamma = 0.577\dots$ is Euler's constant.  

Let us suppose that we have calculated $G_{ab}$ to two orders in perturbation
theory.  We denote the perturbative coefficients by
\begin{equation}
G_{ab}=G_{ab}^{(0)} +\frac{\al }{\pi}\ G_{ab}^{(1)}
+\order (\al^2) .
\end{equation}
Thus $G_{ab}^{(0)}$ is the Born cross section in Eq.\ (\ref{eq:16}).
$G_{ab}^{(1)}$ is the first correction.
 
The first correction $G_{ab}^{(1)}$ will generally have ultraviolet divergences
at $\epsilon = 0$, coming from virtual graphs, and these divergences will
appear as $1/ \epsilon$ poles.  Following the minimal subtraction
prescription, we remove these ultraviolet poles as
necessary.\footnote{
   In the
   particular case of the Drell-Yan cross section (or, more generally, a cross
   section for which the Born graph represents an electroweak interaction), the
   first QCD correction $G_{ab}^{(1)}$  is {\it not} in fact 
   ultraviolet divergent,
   provided that we include the propagator corrections for the incoming quark
   lines  
   This follows from (1) the Ward identity expressing the conservation of the
   electromagnetic current and (2) the fact that the photon propagator does not
   get strong interaction corrections, at lowest order in QED.  
   It can also be verified easily by
   explicit computation.  
}
In general $1/ \epsilon$ poles of infrared origin
will remain in $G_{ab}^{(1)}$, and we shall discuss these infrared poles
presently.
 
Let us similarly denote the perturbative coefficients of the hard
scattering cross section $H_{ab}$ by
\begin{equation}
H_{ab}=H_{ab}^{(0)} +\frac{\al }{\pi}\ H_{ab}^{(1)}
+\order (\al^2) .
\end{equation}
It is these coefficients that we would like to calculate.
 
All we need to know to calculate $H$ from $G$ is the perturbative expansion of
the functions  $f_{a/b}(x,\epsilon)$, which, according to the factorization
theorem, contain all of the sensitivity to small momenta, and are interpreted
as the distribution of parton $a$ in parton $b$.  
These functions can be calculated
in a simple fashion using their definitions (Sect.\ \ref{pdfsec}) 
as matrix elements (here in parton
states) of certain operators.  When the ultraviolet divergences of the
operators are also renormalized using minimal subtraction, one finds 
simply 
\begin{equation}
f_{a/b}(x;\epsilon) =
\delta_{ab}\ \delta(1-x)
- \frac{1 }{2 \epsilon}\ \alpi\ P_{a/b}^{(1)}(x)
+ \order (\al^2) ,
\numlabel{eq:23}
\end{equation}
where $P_{a/b}^{(1)}(x)$ is the lowest order Altarelli-Parisi \cite{AP}
kernel that gives the
evolution with $\mu$ of the parton distribution functions.  We will
discuss the computations that lead to Eq.\ (\ref{eq:23}) in 
Sect.\ \ref{pdfsec}.  For now, let us assume the result.

When we insert these perturbative expansions (\ref{eq:23}) into the
factorization formula, we obtain
\begin{equation}
\begin{split}
   G_{ab}^{(0)}\!\left({x_A}, {x_B}, Q; \frac{\mu}{Q};\epsilon\right)
&
   + \alpi\ G_{ab}^{(1)}\!\left({x_A}, {x_B}, Q; \frac{\mu}{Q};\epsilon\right) 
\\
={}&
    H_{ab}^{(0)}\!\left({x_A}, {x_B}, Q; \frac{\mu}{Q};\epsilon\right)
    + \alpi \ H_{ab}^{(1)}
      \!\left({x_A}, {x_B}, Q; \frac{\mu}{Q};\epsilon\right)
\\
&
    - \frac{1}{2\epsilon}\
    \alpi \
    \sum_{c} \int_{x_A}^1 {\d\xi_A} \,
    P_{c/a}^{(1)}(\xi_A) \
    H_{c b}^{(0)}
    \!\left( \frac{x_A}{\xi_A}, {x_B}, Q; \frac{\mu}{Q};\epsilon\right) 
\\
&
 - \frac{1 }{2 \epsilon}\ \alpi \ \sum_{d} \int_{x_B}^1 {\d\xi_B} \,
   P_{d/b}^{(1)}(\xi_B) \
   H_{ad}^{(0)}
   \!\left({x_A},\frac{x_B }{\xi_B}, Q; \frac{\mu}{Q};\epsilon\right) 
\\
&
    + \order (\al^2) .
\numlabel{eq:24}
\end{split}
\end{equation}
We can now solve for $H_{ab}$.  At the Born level, we find
\begin{equation}
H_{ab}^{(0)}\!\left({x_A}, {x_B}, Q;
\frac{\mu}{Q};\epsilon\right)
=
G_{ab}^{(0)}\!\left({x_A}, {x_B}, Q;
\frac{\mu}{Q};\epsilon\right) .
\end{equation}
Then at the one loop level we obtain
\begin{equation}
\begin{split}
    H_{ab}^{(1)}\!\left({x_A}, {x_B}, Q; \frac{\mu}{Q};\epsilon\right)
={}&
    G_{ab}^{(1)}\!\left({x_A}, {x_B}, Q; \frac{\mu}{Q};\epsilon\right)
\\
& \hspace*{-2pt}
  + \frac{1 }{2 \epsilon}\ \sum_{c} \int_{x_A}^1 {\d\xi_A} \,
    P_{c/a}^{(1)}(\xi_A) \
    G_{c b}^{(0)}
    \!\left( \frac{x_A}{\xi_A}, {x_B}, Q; \frac{\mu}{Q};\epsilon \right) 
\\
& \hspace*{-2pt}
 + \frac{1 }{2 \epsilon}\ \sum_{d} \int_{x_B}^1 {\d\xi_B} \,
   P_{d/b}^{(1)}(\xi_B) \
   G_{ad}^{(0)}
   \!\left({x_A},\frac{x_B }{\xi_B}, Q; \frac{\mu}{Q};\epsilon\right) .
\numlabel{eq:26}
\end{split}
\end{equation}

Thus the prescription is quite simple.  One should calculate the cross
section at the parton level, $G_{ab}^{(1)}$, and subtract from it certain
terms consisting of a divergent factor $1/ \epsilon$, the Altarelli-Parisi
kernel, and the Born cross section (with $\epsilon \ne 0$).  The result is
guaranteed to be finite as $\epsilon \to 0$.
 
Recall that the Born cross section $G^{(0)}$ consists of an $\epsilon$
dependent factor $C(\epsilon)$ times the Born cross section in 4 dimensions,
where $C(\epsilon)$ arises from such sources as the integration over the
lepton angles in the Drell-Yan process.  A convenient way to manage the
calculation is to factor $C(\epsilon)$ out of the first order cross section
$G^{(1)}$ also.  Then the prescription is to remove the $1/\epsilon$ pole in
$G^{(1)}(\epsilon)/ C(\epsilon)$, set $\epsilon = 0$, and multiply by
$C(0) = 1$.  Thus we see that a function of $\epsilon$ that is a common
factor to $G_{ab}^{(0)}$ and  $G_{ab}^{(1)}$ cancels in the physical hard
scattering cross section, as was claimed after Eq.\ (\ref{eq:17}).

When calculating $G^{(1)}$, it should be noted that there are contributions
involving self energy graphs on the external lines, as in \ref{fig:2}.  The
total of all external line corrections gives a factor of $\sqrt {z_2}$ for
each external quark (or antiquark) line and $\sqrt{z_3}$ for each external
gluon line.  Here $z_2$ and $z_3$ are the residues of the poles in the
renormalized quark and gluon propagators.  In the massless theory these
have infrared divergences.  For example the value of $z_2$ in massless QCD
in Feynman gauge
is
\begin{equation}
 z_2 = 1 + \frac{\al }{3 \pi \epsilon } + \order(\al^2).
\end{equation}
Then the contribution of the self energy graphs to $G^{(1)}$ is a factor 
$2 \al /3\pi \epsilon $ times the Born cross section.

%
%

\section{Relation to the renormalization group}
\label{rgsec}
 
The prescription (\ref{eq:26})
for removing infrared poles is intimately related
to  the $\mu$ dependence of $H_{ab}^{(1)}$ --- that is,  to the behavior of
$H_{ab}^{(1)}$ under the renormalization group.  In this section, we display
this connection and show how it leads to the approximate invariance of the
computed cross section under changes of $\mu$.  (Of course, the complete
cross section, to all orders of perturbation theory, is exactly invariant
under changes of $\mu $.  What we are now concerned with is the behavior of
a finite-order approximation.)  
 
We recall  that the Born cross section $G_{ab}^{(0)} = H_{ab}^{(0)}$
contains some $\mu$  dependence from the factor $C({\mu/ Q},\epsilon)$, as
specified in  Eq.\ (\ref{eq:16}).  The one loop  cross section $G_{ab}^{(1)}$
contains this same factor, and we can simply factor  it out of Eq.\
(\ref{eq:26}) and 
set it to 1 when we set $\epsilon = 0$  at the end.  In  addition,
$G_{ab}^{(1)}$ contains a factor $\mu^{2\epsilon}$ from the loop 
integration, 
\begin{equation}
\left( \frac{\mu^2}{4\pi} \e^\gamma \right)^{\epsilon} \int
\d^{4-2\epsilon}k . 
\end{equation}
The $( \e^\gamma \mu^2 / 4\pi )^{\epsilon} $
factor multiplies the $1/\epsilon$ poles in
$G_{ab}^{(1)}$.  Writing
\begin{equation}
\frac{A}{\epsilon}\ \mu^{2\epsilon}
= \frac{A}{\epsilon}
  + 2 A\ \ln(\mu) + \order (\epsilon) ,
\end{equation}
and reading off the value of $A$ from Eq.\ (\ref{eq:26}), we find the
$\mu$ dependence 
of $G_{ab}^{(1)}$ -- and thus of $H_{ab}^{(1)}$:
\begin{equation}
\begin{split}
    H_{ab}^{(1)}\!\left({x_A}, {x_B}, Q; \frac{\mu}{Q}\right)
= &
    H_{ab}^{(1)}\!\left({x_A}, {x_B}, Q;1\right)
\\
&
    - \ln\!\left( \frac{\mu}{Q} \right)\
    \sum_{c} \int_{x_A}^1 {\d\xi_A} \,
    P_{c/a}^{(1)}(\xi_A) \
    H_{c b}^{(0)}\!\left( \frac{x_A}{\xi_A}, {x_B}, Q\right)
\\
& 
    - \ln\!\left( \frac{\mu}{Q} \right)\
   \sum_{d} \int_{x_B}^1 {\d\xi_B} \,
   P_{d/b}^{(1)}(\xi_B) \
   H_{ad}^{(0)}\!\left({x_A},\frac{x_B}{\xi_B}, Q \right).
\end{split}
\numlabel{eq:30}
\end{equation}
Here we have set $\epsilon = 0$ and have suppressed the notation indicating
$\epsilon$ dependence; we have also noted that $H^{(0)}$ does not depend on
$\mu$ when $\epsilon = 0$, so we have suppressed the notation indicating $\mu$
dependence in $H^{(0)}$.
 
We see that $H^{(1)}$ contains logarithms of $\mu / Q$.  If $\mu$ is fixed
while  $Q$ becomes very large, then these logarithms spoil the usefulness of
perturbation theory, since the large logarithms can cancel the small
coupling $\al(\mu)$ that multiplies $H^{(1)}$. For this reason,
one chooses $\mu$ such that $\ln\!\left(\mu / Q \right)$ is not large. 
For example, one chooses $\mu = Q$ or perhaps $\mu = 2Q$ or $\mu = Q/2$.

The freedom to choose $\mu$ results from the renormalization group equations
obeyed by $H$ and $f_{a/A}(\xi)$.  The renormalization group equation for
$H_{ab}$ is
\begin{equation}
\begin{split}
    \mu \frac{\d}{\d\mu}\
&   H_{ab}
    \!\left({x_A}, {x_B}, Q; \frac{\mu}{ Q},\al(\mu)\right)
\\
 =&
  - \sum_{c} \int_{x_A}^1 {\d\zeta_A} \,
    P_{c/a}(\zeta_A,\al(\mu)) \
    H_{c b}
    \!\left( \frac{x_A}{\zeta_A}, {x_B}, Q; \frac{\mu}{Q},\al(\mu)\right)
\\
& 
  - \sum_{d} \int_{x_B}^1 {\d\zeta_B} \,
      P_{d/b}(\zeta_B,\al(\mu)) \
      H_{ad}
      \!\left({x_A},\frac{x_B}{\zeta_B}, Q; \frac{\mu}{Q},\al(\mu)\right).
\end{split}
\numlabel{eq:31}
\end{equation}
Here $P_{c/a}(\xi,\al(\mu))$ is the all orders
Altarelli-Parisi kernel.  It has a perturbative expansion
\begin{equation}
P_{c/a}(\xi,\al(\mu))
= \frac{\al(\mu)}{\pi}\
P_{c/a}^{(1)}(\xi)
+ \dots
\end{equation}
where $P_{c/a}^{(1)}(\xi)$ is the function that appears in Eq.\ (\ref{eq:23}).  Thus
at lowest  order the renormalization group equation (\ref{eq:31})
is a simple consequence
of  differentiating Eq.\ (\ref{eq:23}).
 
Parton distribution functions also have a $\mu$ dependence, which arises
from the  renormalization of the ultraviolet divergences in the products of
quark and  gluon operators in the definitions of these functions, given in
Eqs.~(\ref{eq:43}) and (\ref{eq:44}) below.
The  renormalization  group equation for the distribution functions is
\begin{eqnarray}
\mu \frac{\d}{\d\mu}\
f_{a/A}(\xi,\mu)
&=
\sum_{b} \int_{\xi}^1 \frac{\d\zeta }{\zeta} \,
P_{a/b}\!\left( \zeta,\al(\mu) \right) \
f_{b/A}\!\left(\frac{\xi }{\zeta},\mu \right) .
\numlabel{eq:33}
\end{eqnarray}
 
The physical cross section does not, of course, depend on $\mu$, since
$\mu$ is not one of the parameters of the Lagrangian, but is rather an
artifact of the calculation.  Nevertheless, the cross section
calculated at a finite order of perturbation theory will acquire some
$\mu$ dependence arising from the approximation of throwing away
higher order contributions.  To see how this comes about, we
differentiate Eq.\ (\ref{eq:11}) with respect to $\mu$ and use
Eqs.~(\ref{eq:31}) and (\ref{eq:33}).  This gives
\begin{equation}
\begin{split}
   \mu \frac{ \d }{\d \mu}
&
   ~ \frac{\d\sigma }{\d Q^2 \d y}  
=
\\
={} &
   \sum_{a,b,c}\ \int_{x_A}^1 {\d\xi_A} \,
   \int_{\xi_A}^1 \frac{\d\zeta_A }{\zeta_A} \,
   \int_{x_B}^1{ \d\xi_B} 
\\
&\quad
   P_{a/c}\!\left( \zeta_A,\al(\mu) \right) \
   f_{c/A}\!\left(\frac{\xi_A }{\zeta_A},\mu \right)
   H_{ab}\!\left(\frac{x_A }{\xi_A}, \frac{x_B }{\xi_B}, Q; 
              \frac{\mu}{Q}, \al(\mu) \right)
   \, f_{b/B}(\xi_B,\mu) 
\\
&
-
   \sum_{a,b,c}\ \int_{x_A}^1 {\d \bar \xi_A} \,
   \int_{x_A /\bar\xi_A}^1 {\d\zeta_A } \,
   \int_{x_B}^1{ \d\xi_B}
\\
&
\quad
   f_{a/A}\left({\bar \xi_A },\mu \right)
   P_{c/a}\!\left( \zeta_A,\al(\mu) \right) \
   H_{c b}\!\left(\frac{x_A }{\bar \xi_A\zeta_A}, \frac{x_B}{\xi_B}, Q;
         \frac{\mu}{Q}, \al(\mu) \right) 
   \, f_{b/B}(\xi_B,\mu) 
\\
&
   + \mbox{$B$ terms} .
\end{split}
\numlabel{eq:34}
\end{equation}
Here the two terms shown relate to the evolution of the partons in hadron $A$.
As indicated, 
two similar terms relate to the evolution of the partons in
hadron $B$.  We now change the order of
integration in the second term to put the $\bar \xi_A$ integration inside
the $\zeta_A$ integration, then change the integration variable from
$\bar\xi_A$ to $\xi_A = \bar \xi_A \zeta_A$, and finally reverse the order
of integrations again.  This gives
\begin{equation}
\begin{split}
   \mu \frac{ \d }{\d \mu}\
&
   \frac{\d\sigma }{\d Q^2 \d y}  
=
\\
&
    \sum_{a,b,c}\ \int_{x_A}^1 {\d\xi_A} \,
    \int_{\xi_A}^1 \frac{\d\zeta_A }{\zeta_A} \,
    \int_{x_B}^1{ \d\xi_B}
\\
& ~
    \times
    P_{a/c}\!\left( \zeta_A,\al(\mu) \right) \
    f_{c/A}\!\!\left(\frac{\xi_A }{\zeta_A},\mu \right)
    H_{ab}\!\!
        \left( \frac{x_A}{\xi_A}, \frac{x_B }{\xi_B}, Q; 
               \frac{\mu}{Q}, \al(\mu) 
        \right)
    \, f_{b/B}(\xi_B,\mu)
\\
&  -
    \sum_{a,b,c}\
    \int_{x_A}^1 {\d  \xi_A} \,
    \int_{\xi_A}^1 \frac{\d\zeta_A }{\zeta_A} \,
    \int_{x_B}^1 {\d\xi_B}
\\
& ~
\times
    f_{a/A}\!\!\left({\frac{\xi_A }{\zeta_A} },\mu \right)
    P_{c/a}\!\left( \zeta_A,\al(\mu) \right) \
    H_{c b}
     \!\!\left( \frac{x_A }{\xi_A}, \frac{x_B }{\xi_B}, Q;
                \frac{\mu}{Q}, \al(\mu)
     \right) 
   \, f_{b/B}(\xi_B,\mu) 
\\
& + {B\ \rm terms} .
\end{split}
\end{equation}
We see that the two terms cancel exactly as long as $P_{a/b}$ and $H_{ab}$
obey  the renormalization group equations exactly.  Now, when $H_{ab}$ is
calculated  only to order $\al^N$, it only obeys the
renormalization group equation (\ref{eq:31}) to the
same order. In this case, we will
have
\begin{equation}
\mu \frac{ \d }{\d \mu}\
\frac{\d\sigma }{\d Q^2 \d y}  =
\order (\al^{N+1})\ ,
\end{equation}
when the parton distribution functions obey the renormalization group
equation with the Altarelli-Parisi kernel calculated to order $\al ^N$ 
or better. One thus finds that the result of a Born level calculation
can be strongly $\mu$ dependent, but by including the next order
the $\mu$ dependence is reduced.
 
We have argued that one should choose $\mu$ to be on the order of the large
momentum scale in the problem, which is $Q$ in the case of the Drell-Yan cross
section.  We have the right to choose $\mu$ as we wish because the result would
be independent of $\mu$ if the calculation were done exactly. The choice
$\mu \sim Q$ eliminates the potentially large logarithms in Eq.\ (\ref{eq:30}).
Another choice is often used.  One substitutes for $\mu$ in Eq.\ (\ref{eq:11})
the value $ \sqrt{\hat s} = \sqrt{\xi_A \xi_B s}$.  We now have a value of
$\mu$ that depends on the integration variables in the factorization.
 
Let us examine whether this is valid, assuming that
$P_{a/b}$ and $H_{ab}$ are calculated exactly.  We replace $\mu$ by
\begin{equation}
\mu(\lambda,\xi_A, \xi_B) = \mu_0^{1-\lambda}
\left( \sqrt{\xi_A \xi_B s} \right)^\lambda  ,
\quad\quad 0<\lambda<1 .
\end{equation}
At $\lambda = 0$ we have a valid starting point.  When we get to $\lambda = 1$
we have the desired ending point.  The question is 
whether the derivative of the
cross section with respect to $\lambda$ is zero.  Applying the same calculation
as before, we obtain instead of Eq.\ (\ref{eq:34}) the result
\begin{equation}
\begin{split}
    \frac{\d}{\d\lambda}
   ~\frac{\d\sigma }{\d Q^2 \d y} 
={}&
    \sum_{a,b,c}\ \int_{x_A}^1 {\d\xi_A} \,
    \int_{\xi_A}^1 \frac{\d\zeta_A }{\zeta_A} \,
    \int_{x_B}^1{ \d\xi_B} \,
    \frac{1}{2} \, \ln\!\left(  \frac{ \xi_A \xi_B s }{ \mu_0^2 } \right)
\\
&\hspace*{-10pt}
\times
    P_{a/c}\!\left( \zeta_A,\al(\mu(\lambda,\xi_A, \xi_B)) \right) \
    f_{c/A}\!\left(\frac{\xi_A }{\zeta_A},\mu(\lambda,\xi_A, \xi_B) \right)
\\
&\hspace*{-10pt}
\times
    H_{ab}\!\left(\frac{x_A }{\xi_A}, \frac{x_B }{\xi_B}, Q;
    \frac{\mu(\lambda,\xi_A, \xi_B)}{Q}, \al(\mu) \right)
   \, f_{b/B}(\xi_B,\mu(\lambda,\xi_A, \xi_B)) 
\\
& \hspace*{-20pt}
-
   \sum_{a,b,c}\ \int_{x_A}^1 {\d \bar \xi_A} \,
   \int_{x_A /\bar\xi_A}^1 {\d\zeta_A } \,
   \int_{x_B}^1{ \d\xi_B}\
   \frac{1}{2}\, 
   \ln\!\left( \frac{ \bar\xi_A \xi_B s }{ \mu_0^2 } \right)
\\
&\hspace*{-10pt}
\times
   f_{a/A}\left({\bar \xi_A },\mu(\lambda,\bar\xi_A, \xi_B) \right)
   P_{c/a}\!\left( \zeta_A,\al(\mu(\lambda,\bar\xi_A, \xi_B)) \right)
\\
&\hspace*{-10pt}
\times
   H_{c b}\!\left(\frac{x_A }{\bar \xi_A\zeta_A}, \frac{x_B }{\xi_B}, Q;
    \frac{\mu(\lambda,\bar\xi_A, \xi_B)}{Q}, \al(\mu) \right) \,
   f_{b/B}(\xi_B,\mu(\lambda,\bar\xi_A, \xi_B))
\\
&\hspace*{-20pt}
   + B \ {\rm terms} .
\end{split}
\end{equation}
Now making the same change of variables as before, we obtain
\begin{equation}
\begin{split}
    \frac{ \d }{\d \lambda}
&
   ~\frac{\d\sigma }{\d Q^2 \d y}  
={}
\\
&
    \sum_{a,b,c}\ \int_{x_A}^1 {\d\xi_A} \,
   \int_{\xi_A}^1 \frac{\d\zeta_A }{\zeta_A} \,
   \int_{x_B}^1{ \d\xi_B} \
   \frac{1}{2}\, \ln\!\left( \frac{ \xi_A \xi_B s }{ \mu_0^2 } \right)
\\
&\quad\quad\times
   P_{a/c}\!\left( \zeta_A,\al (\mu(\lambda,\xi_A, \xi_B)) \right) \
   f_{c/A}\!\left(\frac{\xi_A }{\zeta_A},\mu(\lambda,\xi_A, \xi_B)
           \right)
\\
&\quad\quad\times
H_{ab}\!\left(\frac{x_A }{\xi_A}, \frac{x_B }{\xi_B}, Q;
 \frac{\mu(\lambda,\xi_A, \xi_B)}{Q}, \al(\mu) \right)
\, f_{b/B}(\xi_B,\mu(\lambda,\xi_A, \xi_B)) 
\\
&  -
\sum_{a,b,c}\
\int_{x_A}^1 {\d \xi_A} \,
\int_{\xi_A}^1 \frac{\d\zeta_A }{\zeta_A} \,
\int_{x_B}^1{ \d\xi_B}\
\frac{1}{2}\, 
\ln\!\left( \frac{ \xi_A \xi_B s }{ \zeta_A\mu_0^2 } \right)
\\
&\quad\quad\times
f_{a/A}\left(\frac{ \xi_A}{\zeta_A },\mu(\lambda,\xi_A/\zeta_A, \xi_B)
\right)
P_{c/a}\!\left( \zeta_A,\al (\mu(\lambda,\xi_A/\zeta_A, \xi_B))
\right)
\\
&\quad\quad\times
H_{c b}\!\left(\frac{x_A }{\xi_A}, \frac{x_B }{\xi_B}, Q;
 \frac{\mu(\lambda,\xi_A/\zeta_A, \xi_B)}{Q}, \al(\mu) \right) \,
f_{b/B}(\xi_B,\mu(\lambda,\xi_A/\zeta_A, \xi_B)) 
\\
&  + {B\ \rm terms} .
\end{split}
\end{equation}
We see that the cancellation between the two terms has been spoiled, first
by the differences in the values of $\mu(\lambda,\dots)$ in the two terms,
but more importantly by the differences in the arguments of the logarithm
in the two terms.  We conclude that the substitution of $\hat s$ for
$\mu^2$ results in an error of order $ \al$ no matter how accurately the
hard scattering cross section is calculated.  This is not a problem if the
hard scattering cross section is calculated only at the Born level, which
is, in fact, commonly the case.  However, it is wrong to substitute $\hat
s$ for $\mu^2$ when a calculation beyond the Born level is used.

%
%
\section{The parton distribution functions}
\label{pdfsec}
 
The parton distribution functions are indispensable ingredients in the
factorization formula (\ref{eq:11}).  We need to know the distribution
of partons in 
a hadron, based on experimental data, in order to obtain predictions from
the formula. In addition, we need to know the distribution of partons in a
parton in order to calculate the hard scattering cross section $H_{ab}$. 
The hard scattering cross section is
obtained by factoring the parton distribution functions out of the physical
cross section.  Evidently, the result  depends on exactly what it is that
one  factors out.
 
\subsection{Operator Definitions}
\label{pdfsec:operators}

In this section, we describe the definition for the parton distribution
functions that we use elsewhere in this chapter. A more complete
discussion can be found in Ref.\ \cite{pdfs}. In this definition, the
distribution 
functions are matrix elements in a hadron state of certain operators that act
to count the number of quarks or gluons carrying a fraction $\xi$ of the
hadron's momentum.  We state the definition in a  reference frame in which the
hadron carries momentum $P^\mu$ with a plus component $P^+$, a minus
component $P^- = m^2/2P^+$, and transverse components equal to zero. (We
use $P^{\pm}= (P^0 \pm  P^3)/\sqrt 2$).
 
The definition may be motivated by looking at the theory quantized on
the plane $x^+ = 0$ in the light-cone gauge $A^+ = 0$, since it is in
this picture that field theory has its closest connection with the
parton model \cite{null}.  In this gauge, ${\cal G} = 1$, where ${\cal
G}$ is a path-ordered exponential of the gluon field that appears in
the definition of the parton distributions. The light-cone gauge tends
to be rather pathological if one goes beyond low order perturbation
theory, and covariant gauges are preferred for a complete treatment.
However quantization on a null plane in the light-cone gauge provides
a useful motivation for the complete treatment.
 
In this approach the quark field has two components that represent the
independent degrees of freedom;  $\gamma^+ \psi(x)$ contains these
components and not the other two.  One can expand the two independent
components in terms of quark destruction operators $b(k^+,\kt,s)$ and
antiquark creation operators  $d(k^+,\kt,s)^\dagger$ as follows:
\begin{multline}
   \gamma^+ \psi(0,x^-,{\T x}) =
   \frac{1}{(2\pi)^3}
   \sum_s \int_0^\infty \frac{\d k^+ }{2 k^+} \int \d \kt  
\\ \times
   \left[
       \gamma^+ U(k,s) \e^{- i k\cdot x} b(k^+,\kt,s)
       +\gamma^+ V(k,s) \e^{+ i k\cdot x} d(k^+,\kt,s)^\dagger
   \right].
\end{multline}
The quark distribution function is just the hadron matrix element of
the operator that counts the number of quarks.
\begin{equation}
f_{q/A}(\xi)\, \d \xi = 
   \frac{1}{(2\pi)^3}
   \sum_s \frac{\d (\xi P^+) }{2 (\xi P^+)}
\int \d \kt \
\langle P |\,
b(\xi P^+,\kt,s)^\dagger b(\xi P^+,\kt,s)
\, | P \rangle .
\end{equation}
In terms of $\psi(x)$, this is
\begin{displaymath}
f_{q/A}(\xi) = \frac{1}{4\pi} \int \d x^- \e^{-i\xi P^+ x^-}
\langle P | \bar\psi (0,x^-,\zerot)\, \gamma^+\,
\psi (0,0,\zerot) | P \rangle .
\end{displaymath}
We can keep this same definition, while allowing the possibility of computing
in another gauge, by inserting the operator
\begin{equation}
{\cal G} = \P\, \exp
\left\{
i g\int_0^{x^-} \d y^- A_c^+(0,y^-,\zerot)t_c
\right\},
\end{equation}
where $\P$ denotes an instruction to order the gluon field operators
$A_a^+(0,y^-,\zerot)$ along the path. The operator ${\cal G}$ is evidently 1
in the $A^+ = 0$ gauge. With this operator, the definition is gauge
invariant.

We thus arrive at the definition \cite{pdfs,CFP}
\begin{equation}
f_{q/A}(\xi) = \frac{1}{4\pi} \int \d x^- \e^{-i\xi P^+ x^-}
\langle P | \bar\psi (0,x^-,\zerot)\, \gamma^+\, {\cal G}\,
\psi (0,0,\zerot) | P \rangle  .
\numlabel{eq:43}
\end{equation}
 
For gluons, the definition based on the same physical motivation is
\begin{equation}
f_{g/A}(\xi) =
\frac{1 }{2\pi \xi P^+}
 \int \d x^- \e^{-i\xi P^+ x^-}
\langle P | F_a(0,x^-,\zerot)^{+\nu} \, {\cal G}_{ab}\,
F_b(0,0,\zerot)_\nu^{\ +}\, | P \rangle   ,
\numlabel{eq:44}
\end{equation}
where $F_{\mu\nu}^a$ is the gluon field strength operator and where in
${\cal G}$ we now use the octet representation of the SU(3) generating
matrices $t_c$.
 
\subsection{Feynman rules and eikonal lines }
\label{pdfsec:eikonal}

\begin{figure}
   \centering
   \includegraphics{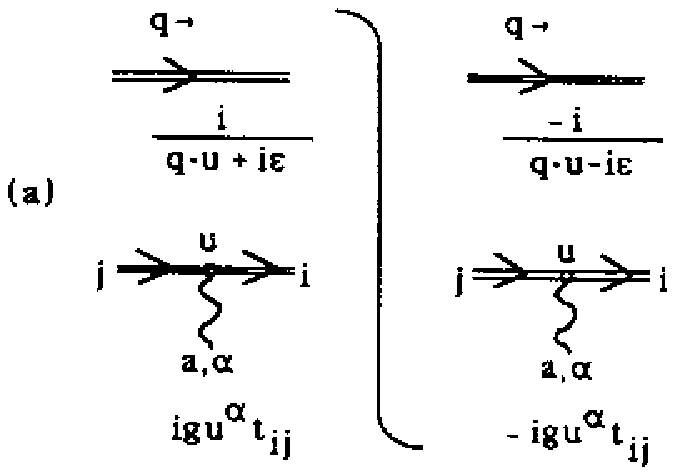}
~~
   \includegraphics{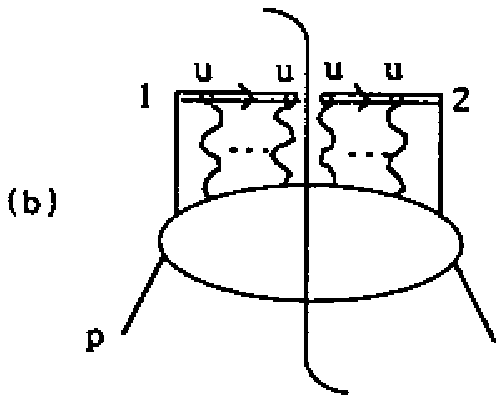}
   \caption{(a) Feynman rules for eikonal lines in the amplitude and
       its complex conjugate.  (b) A general contribution to a parton
       distribution.} 
   \label{fig:3}
\end{figure}

The Feynman rules for parton distributions are derived in
a straightforward manner from the standard Feynman
rules.  Consider, for instance the distribution $f_{q/q}$
of a quark in a quark.  To compute this quantity in
perturbation theory, we use the following identity
satisfied by any ordered exponential,
\begin{multline}
\P\ \exp \Bigl \{  i g \int_0^{\eta} \d\lambda \, n \cdot 
A(\lambda n^\mu) \Bigr \}
= 
\\
\Bigl [ \P\ \exp \Bigl \{ i g \int_0^{\infty} \d\lambda  \, n
\cdot  A((\lambda+\eta) n^\mu) \Bigr \} \Bigr ]^\dagger 
\ P\ \exp \Bigl \{ i g \int_0^{\infty} \d\lambda \, n \cdot 
A(\lambda n^\mu) \Bigr \}\ . 
\numlabel{eq:45}
\end{multline}
Using (\ref{eq:45}) in Eq.\ (\ref{eq:43}), for instance, enables us to
insert a complete set of states and write
\begin{equation}
f_{q/q}(\xi)= \frac{1 }{4\pi} \int \d x^- e^{-\xi P^+ x^-}
\sum_n \langle P|{\bar \Psi}(0,x^-,\zerot) | n \rangle  \, \gamma^+ \,
\langle n|\Psi(0,0,\zerot) | P \rangle \ ,
\numlabel{eq:46}
\end{equation}
where we define $\Psi$ as the quark field times an
associated ordered exponential,
\begin{equation}
\Psi(x) \equiv \psi(x) 
\ \P\, \exp \left[ i g \int_0^{\infty} \d\lambda \, v \cdot 
A(x+\lambda v^\mu) \right] \ ,
\numlabel{eq:47}
\end{equation}
where $v^\mu \equiv g^\mu_-$, and $A^\mu(x) \equiv A^\mu_c(x)t_c$.
To express the matrix elements in Eq.\ (\ref{eq:46}) in terms
of diagrams, we note that by (\ref{eq:47}) the gluon fields in
the expansion of $\Psi$ are time ordered by
construction.  Expanding the ordered exponentials, and
expressing them in momentum space we find
\begin{multline}
\P\ \exp \left[ i g \int_0^{\infty} \d\lambda \, n \cdot 
A(\lambda n^\mu) \right]
=
\\
1 +
\P \sum_{n=1}^\infty \prod_{i=1}^n
\int \frac{\d^4 q_i }{(2\pi)^4 }
\, g \, n \cdot {\tilde A}(q_i^\mu) 
\frac{ 1 }{n \cdot \sum_{j=1}^i q_j +i\epsilon }\ ,
\numlabel{eq:48}
\end{multline}
where we define the operator P on the right-hand side
of the equation to order the fields with the lowest value
of $i$ to the left.  From Eq.\ (\ref{eq:48}) we can read off
the Feynman rules for the expansion of the ordered
exponential \cite{pdfs,CFP}.  They are illustrated in Fig.\ \ref{fig:3}.

The denominators $n \cdot \sum_j q_j +i\epsilon$ are
represented by double lines, which we shall refer to as
``eikonal" lines.  These lines attach to gluon
propagators via a vertex proportional to $-i g n^\mu$.  Fig.\ \ref{fig:3}(a)
shows the formal Feynman rules for eikonal lines and
vertices.   In Fig.\ \ref{fig:3}(b), we show a general
contribution to $f_{q/q}$, as defined by Eq.\ (\ref{eq:46}).

\begin{figure}
   \centering
   \includegraphics[scale=0.9]{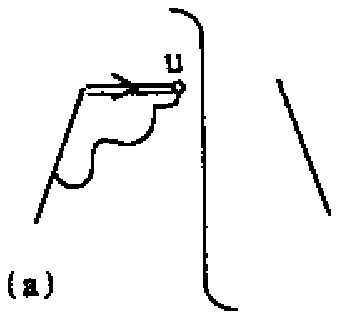}
~~
   \includegraphics[scale=0.9]{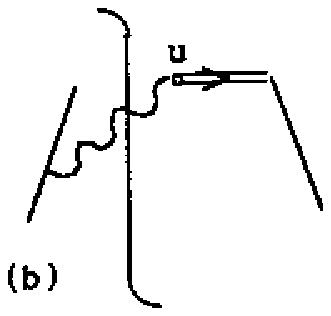}
~~
   \includegraphics[scale=0.9]{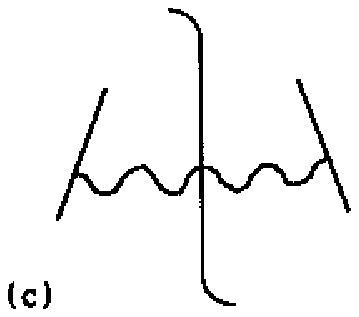}
~~
   \includegraphics[scale=0.9]{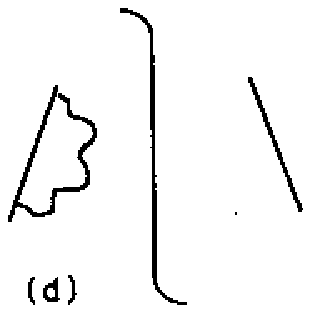}
   \caption{One loop corrections to quark distribution, Eq.\ (\ref{eq:43}).  }
   \label{fig:4}
\end{figure}

The positions of all the explicit fields in
Eq.\ (\ref{eq:46}) differ only in their plus
components.  As a result, minus and transverse momenta are
integrated over.  (They may thought of as flowing freely
through the eikonal line.)  The plus momentum flowing out
of vertex $1$ and into vertex $2$, however, is fixed to be
$\xi P^+$.  No plus momentum flows across the cut
eikonal line in the figure.  Fig.\ \ref{fig:4} shows the one
loop corrections to $f_{q/q}(\xi)$.  

To be explicit,
Fig.\ \ref{fig:4}(b) is given in $n$ dimensions by
\begin{multline}
   \frac{1 }{4\pi} \frac{4}{3}
   \int \frac{\d^n q }{(2\pi )^n } \,
    N_{\alpha\beta}(q)\,
   2\pi \delta(q^2) \,
   \frac{i }{(p-q)^2+i\epsilon}\ 
\times 
\\ \times
   \tr [(\rlap /p-\rlap/q)(-i g\gamma^\alpha) \rlap /p \gamma^+)]\,
   (i g\mu^\epsilon n^\beta)\ 
   \frac{ -i }{ u \cdot q - i\epsilon } ,
\numlabel{eq:49}
\end{multline}
where $N_{\alpha\beta}$ is the polarization tensor of
the gluon. By applying minimal subtraction to Eq.\ (\ref{eq:49})
and the similar forms for the other diagrams in Fig.\ \ref{fig:4},
we easily verify Eq.\ (\ref{eq:23}) for $f_{q/q}$. Gluon
distributions are calculated perturbatively in a similar
manner.  We will need the concept of eikonal lines again,
when we discuss the proof of factorization in gauge
theories.
 
\subsection{Renormalization}

The operator products in the definitions 
(\ref{eq:43}) and (\ref{eq:44}) require renormalization, as
discussed in Ref.\ \cite{pdfs}.  We choose to renormalize using the \MSbar\ scheme. Of
course, renormalization introduces a dependence on the renormalization scale
$\mu$.  The renormalization group equation for the $f_{a/A}$ is the
Altarelli-Parisi equation (\ref{eq:33}).  A complete derivation of this result
may be found in Ref.\ \cite{pdfs}.
 
The one-loop result, Eq.\ (\ref{eq:23}),
can actually be understood without looking at the details of the calculation.
At order $\al $, one has simple one loop diagrams that contain an
ultraviolet divergence that arises from the operator product, but also contain
an infrared divergence that arises because we have massless, on-shell partons
as incoming particles.  The transverse momentum integral is zero, due to a
cancellation of infrared and ultraviolet poles, which we may exhibit
separately: 
\begin{equation}
 \left( \frac{\mu^2}{4\pi} e^\gamma \right)^{\epsilon}
\int \frac{ \d^{2-2\epsilon}\kt }{ (2\pi)^{2-2\epsilon} }
 \ \frac{1}{\kt^2}
= \frac{1 }{4\pi}
\left\{
\frac{ 1 }{ \epsilon_{\rm\scriptscriptstyle UV} } -
\frac{ 1 }{ \epsilon_{\rm\scriptscriptstyle IR} }
 \right\} .
\end{equation}
In this way, we obtain
\begin{equation}
f_{a/b}(\xi;\epsilon) =
\delta_{ab}\ \delta(1-\xi)
+ \left\{\frac{1 }{\epsilon_{\rm\scriptscriptstyle UV}}
 -\frac{1 }{\epsilon_{\rm\scriptscriptstyle IR}}\right\}
 \alpi \ P_{a/b}^{(1)}(\xi)
 - {\rm counterterm}
+ \order (\al^2) .
\end{equation}
The coefficient of $1 / \epsilon_{\rm\scriptscriptstyle UV}$ is the
`anomalous dimension' that appears in the renormalization group equation,
that is, the Altarelli-Parisi kernel.  Following the \MSbar\ renormalization
scheme, we use the counter term to cancel $1 / \epsilon_{\rm
\scriptscriptstyle UV}$ term.   This leaves the infrared $1 / \epsilon$,
which is {\it not} removed by renormalization,
\begin{equation}
f_{a/b}(\xi;\epsilon) =
\delta_{ab}\ \delta(1-\xi)
-\frac{1 }{\epsilon}
 \alpi \ P_{a/b}^{(1)}(\xi)
+ \order (\al ^2) .
\numlabel{eq:52}
\end{equation}
 
\subsection{ Relation to Structure Functions}

Let us now consider the relation of the parton distribution functions to the
structure functions measured in deeply inelastic lepton scattering. If we use
the definition of parton distribution functions given above, then the
structure
function $F_2$ is given by the factorization equation (\ref{eq:2}).
At the Born
level, the hard scattering function is simply zero for gluons and the quark
charge squared, $e_j^2$, times a delta function for quarks.  Thus the formula
for $F_2$ takes the form
\begin{equation}
\begin{split}
   x^{-1} F_2(x,Q) 
={}&
   \sum_j e_j^2\ f_{j/A}(x,\mu) 
  + \sum_{j,b} e_j^2
      \int_{x}^1 \frac{\d \xi }{\xi} \ f_{b/A}(\xi,\mu)\,
      \frac{\al  }{\pi}\,
      C_{j b}\left( \frac{x }{\xi}, \frac{Q }{\mu}  \right)
\\&
 + \order (\al ^2).
\numlabel{eq:53}
\end{split}
\end{equation}
The sums over $j$ run over all flavors of quarks and antiquarks. Gluons do not
contribute at the Born level, but 
they do at order $\al $, through virtual quark-antiquark pairs.
The hard scattering  coefficients $C_{j b}$ can be obtained by calculating (at
order $\al $)   deeply inelastic scattering from on-shell massless partons,
then removing  the infrared divergences according to the scheme discussed in
Sect.\ \ref{hardsec}.

The explicit form of the perturbative coefficients $C_{j b}$ is
\cite{Bardeen}
\begin{align}
   C_{j k}(z,1)
={}&
   \delta_{j k}\, \frac{4}{3} 
   \left[
      - \frac{1}{2} \frac{1+z^2 }{1-z}\, \ln\!\left( \frac{ z }{ 1-z } \right)
      + \frac{3 }{4} \frac{1 }{1-z} - \frac{3 }{2} - z
   \right]_+
\nonumber\\
   C_{j g}(z,1)
={}&
   -\frac{1}{2} 
   \left\{
       \frac{1}{2}
         \left[ z^2 + (1-z)^2 \right]\,
         \left[ \ln\!\left( \frac{ z }{ 1-z } \right) + 1 \right]\,
       - 3 z(1-z)
   \right\} ,
\numlabel{eq:54}
\end{align}
where the plus subscript to the bracket in the first equation denotes a
subtraction that regulates the $z \to 1$ singularity,
\begin{eqnarray}
\int_x^1 \d z\ [C(z)]_+\ h(z)
&=& \int_0^1 \d z\ [C(z)]_+\ h(z)\, \Theta(z>x) \nonumber\\
&=& \int_0^1 \d z\ C(z)\ \Bigl\{ h(z)\, \Theta(z>x) - h(1) \Bigr\}.
\end{eqnarray}
 
\subsection{Other Parton Distributions}

The definitions (\ref{eq:43}) and (\ref{eq:44}) are the most natural for many
purposes.  They are not, however, unique.  Indeed, any function
$g_{b/A}(y)$, which can be related to $f_{a/A}(x)$ by
convolution with ultraviolet functions $D_{ab}(x/y,Q/\mu)$ in a form like
\begin{equation}
g_{a/A}(x)= \sum_b \int_x^1 (\d y/y) \, D_{ab}(x/y,Q/\mu, \al(\mu)) \,
f_{b/A}(y)\ ,
\end{equation}
is an acceptable parton distribution \cite{St87}.  The hard scattering
functions calculated with the distributions $g_{b/A}$ will differ
from those calculated with $f_{a/A}(x)$, but this difference will
itself be calculable from the functions $D_{ab}$ as a power series in
$\al (Q)$.

The most widely used parton distribution of this type is based on deeply
inelastic scattering, and may be called the DIS definition.  The
definition is
\begin{equation}
f_{j/A}^{\rm DIS}(x,\mu) =
f_{j/A}(x,\mu)
+\sum_{b}
\int_{x}^1 \frac{\d \xi }{\xi} f_{b/A}(\xi,\mu)\,
\frac{\al  }{\pi} \,
C_{j b}\left( \frac{x }{\xi}, 1  \right)
 + \order (\al ^2).
\end{equation}
for quarks or antiquarks of flavor $j$.  Comparing this definition with
Eq.\ (\ref{eq:53}), we see that
\begin{equation}
x^{-1}F_2(x,Q) =
\sum_j e_j^2\ f_{j/A}^{\rm DIS}(x,Q)
+ \order (\al ^2).
\end{equation}
That is, we adjust the definition so that the order $\al $ correction
to deeply inelastic scattering vanishes when $\mu =Q$.  It is not so clear
what one should do with the gluon distribution in the DIS scheme.  One
choice \cite{DFLM} is
\begin{equation}
f_{g/A}^{\rm DIS}(x,\mu) =
f_{g/A}(x,\mu)
 - \sum_j \sum_b
\int_x^1 \frac{\d \xi }{\xi} f_{b/A}(\xi,\mu)
\frac{\al  }{\pi}
C_{j b}\left( \frac{x }{\xi}, 1  \right)
 + \order (\al ^2).
\numlabel{eq:59}
\end{equation}
This has the virtue that it preserves the momentum sum rule that is obeyed
by the \MSbar\ parton distributions \cite{pdfs},
\begin{equation}
\sum_a \int_0^1 \d \xi\ f_{a / A}(x,\mu) = 1 .
\end{equation}
 
If one wishes to use parton distribution functions with the DIS
definition, then one must modify the hard scattering function for the
process under consideration.  One should combine Eqs.\ (\ref{eq:52})
and (\ref{eq:59}) to get the DIS distributions of a parton in a
parton, then use these distributions in the derivation in
Sect.\ \ref{hardsec}.
 
It should be noted that there is some confusion in the literature
concerning the term $+1$ that follows the logarithm in $C_{j g}$ in
Eq.\ (\ref{eq:54}).  The form quoted is the original result of Ref.\
\cite{Bardeen}, translated from moment-space to $z$-space.  In the
calculation with incoming gluons, one normally averages over
polarizations of the incoming gluons instead of using a fixed
polarization.  This means that one sums over polarizations and divides
by the number of spin states of a gluon in $4 - 2 \epsilon$
dimensions, namely $2 - 2 \epsilon$.  If, instead, one divides by 2
only, one obtains the result (\ref{eq:54}) without the $+1$, which may
be found in Ref.\ \cite{Altarelli}.  This does no harm if, as in the
case of Ref.\ \cite{Altarelli}, one wants to express the cross section
for a second hard process in terms of DIS parton distribution
functions and if one consistently divides by 2 instead of $2 - 2
\epsilon$ in {\it both} processes.  However, it is not correct if one
wants to relate the DIS structure functions to \MSbar\ parton
distribution functions, defined as hadron matrix elements of the
appropriate operators, renormalized by \MSbar\ subtraction.


\section{Factorization for $\phi^3$ Theory}
\label{phidissec}

In this and the next section, we study the factorization
theorem in a $\phi^3$ theory for $n \le 6$ space-time dimensions. 
First we show how the factorization theorem comes about for one-loop
corrections in deeply inelastic scattering, and
compare the field theory to the parton model. In the next section, we
will present a reasonably complete but compact derivation of the
factorization theorem in deeply inelastic scattering to all orders of
perturbation theory.

The scalar theory allows us to study these issues in a simplified but highly
nontrivial context. As emphasized above, the purpose of the
factorization  theorems is to separate long-distance behavior in
perturbation theory.  In  the scalar theory, as we shall see, this
behavior is associated with  partons that are collinear to the observed
hadrons.  The organization of such ``collinear divergences" is central to
factorization in all field theories, but in gauge theories they
are joined by ``soft" partons, associated with infrared divergences. Indeed,
the basic problem in gauge theories is to show how that infrared or ``soft"
divergences cancel (see Sect.\ \ref{gaugeproofsec}). 
In $\phi^3$ theory the infrared problem is absent, so that studying this
theory allows us to study the  
basic physics of factorization 
in the simplest possible setting.

The Lagrangian is
\begin{equation}
\calL
= \half \left(\partial\phi\right)^2- \half m^2\phi^2 - 
\frac16 g (\mu^2 \e^\gamma/4\pi)^{\epsilon/2} \phi^3 + \mbox{counterterms}\ .
\numlabel{eq:61}
\end{equation}
We will use, where necessary, dimensional regularization, with space-time
dimension $n= 6 - 2 \epsilon $.
It is worth recalling that at $n=6$ the theory is renormalizable, while for
$n<6$ it is superrenormalizable.  We shall not concern ourselves with the
theory for $n>6$ where it is nonrenormalizable by power-counting.  
$\mu$ is a mass
which enables us to keep $g$ dimensionless as we vary $n$.
We will renormalize the theory with the \MSbar\ prescription. 
We use the factor $(\mu^2 \e^\gamma/4\pi)^{\epsilon/2} $ rather than the
more conventional $\mu^\epsilon $, so that we can implement \MSbar\
renormalization as pure pole counterterms.
(For convenience, we will define the $h\phi$ counterterm that
renormalizes the tadpole graphs by requiring the sum of the tadpoles and
their counterterm to vanish.)
We define
\begin{equation}
\bar\mu = \mu \sqrt{\e^\gamma/4\pi} .
\end{equation}

\subsection{Deeply inelastic scattering}

\begin{figure}
   \centering
   \includegraphics{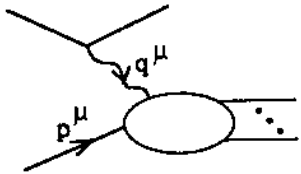}
   \caption{Deeply inelastic scattering.}
   \label{fig:5}
\end{figure}

Our model for deeply inelastic scattering
consists of the exchange of a
weakly interacting boson, $A$, not included in the Lagrangian (\ref{eq:61}).
This is illustrated diagrammatically in the same way as for QCD, in
Fig.\ \ref{fig:5}.  
The weak boson couples to the $\phi$ field through an interaction
proportional to $\half A \phi^2$.  There is then a single structure function
which we define by
\begin{equation}
 F(x,Q) = \frac{Q^2}{2\pi}
\int \d^6y \, \e^{iq\cdot y} \langle p| j(y)\, j(0) | p \rangle ,
\numlabel{eq:63}
\end{equation}
where $ j = \half \phi^2$.  The momentum transfer is $q^\mu$, and the usual
scalar variables are defined by $Q^2 = - q^2$ and $x = Q^2 / 2p \cdot q$,
with $p^\mu$ the momentum of the target.
We will investigate the structure function in the Bjorken limit of large
$Q$ with $x$ fixed, and our calculations will be for the case that the
state $| p \rangle $ is a single $\phi$ particle (with non-zero mass, as given
in Eq.\ (\ref{eq:61})).  

When $Q$ is large, each graph for the structure
function behaves like a polynomial of $\ln(Q/m)$ plus corrections that
are nonleading by a power of $Q$.  Factorization is possible because only a
limited set of momentum regions of the space of loop and final state phase
space momenta contribute to the leading power.  First we will explain the
power counting arguments that determine these ``leading regions", and how
they are related to the physical arguments of the parton model.

\begin{figure}
   \centering
   \includegraphics{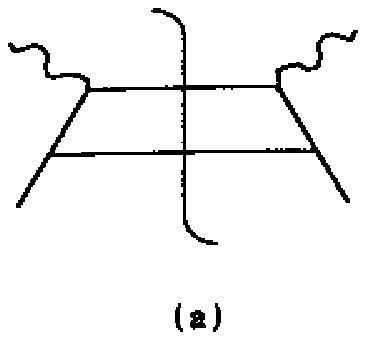}
\quad
   \includegraphics{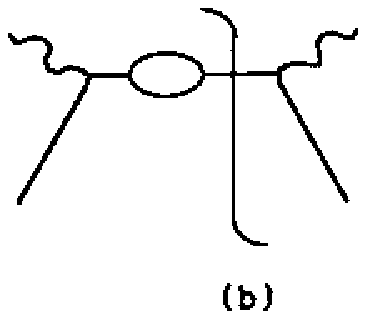}
\quad
   \includegraphics{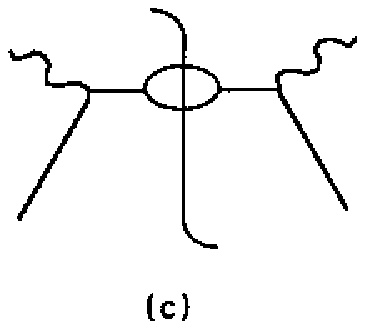}
\quad
   \includegraphics{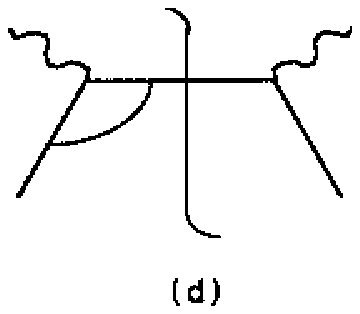}
\quad
   \includegraphics{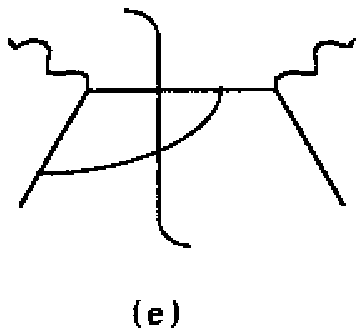}
   \caption{One-loop corrections to deeply inelastic scattering.  For graphs (b), (d) and (e), we also have the hermitian conjugate graphs.  }
   \label{fig:6}
\end{figure}

The tree graph for the structure function is easy to calculate.  It is
\begin{equation}
  F_0 = Q^2 \delta(2p\cdot q +q^2) = \delta(x-1).
\end{equation}
The one-loop ``cut diagrams" (as defined in Sect.\ \ref{hardsec} 
above) which contribute
to $F$ are given in Fig.\ \ref{fig:6}.  

Each of these
diagrams illustrates a different bit of the physics, so we shall treat them in
turn, starting with the ``ladder" correction, Fig.\ \ref{fig:6}(a).

\subsection{Ladder Graph and its Leading Regions}

The Feynman integral for the cut diagram Fig.\ \ref{fig:6}(a) is
\begin{equation}
   F_{2(a)} = 
    \frac{ g^2Q^2 {\bar\mu}^{2\epsilon} }{ (2\pi)^{ 5-2\epsilon} }
    \int \d^{6-2\epsilon }k 
     \frac{ \delta ((p-k)^2-m^2)\, \delta ((q+k)^2-m^2) }
          { (m^2-k^2)^2 } .
\numlabel{eq:65}
\end{equation}
Although for nonzero $m$ this integral is finite, it will prove convenient to
retain the dimensional
regularization, in order to display some 
very important dimension-dependent features of the
$Q \to \infty $ limit.

Equation (\ref{eq:65}) is calculated conveniently in terms of light-cone coordinates.
Without loss of generality, we may choose the
external momenta, $q^\mu$ and $p^\mu$ to be $p^\mu  = (p^+, 
m^2/2p^+,\zerot)$ and
$q^\mu  = (-x p^+, Q^2/2x p^+, \zerot)$.  Notice that this formula for
$q^\mu$ corresponds to a slight change in the definition of $x$, which we
now define by $Q^2 /2 p \cdot q \equiv x / ( 1 - x m^2 /Q^2)$.  At leading
power in $Q$, there is no difference, but at finite energy our formulas
will be simplified by this choice.  

The $\delta $-functions in
(\ref{eq:65}) can be used to perform the $\kt$ and $k^-$ integrals.  Then if we
set $\xi = k^+ / p^+$, we find
\begin{eqnarray}
   F_{2(a)} 
&=&
   \frac{g^2 }{64 \pi^3 }
     \left( \frac{ Q^2 }{ \e^\gamma \mu^2 x (1-x) } \right)^{-\epsilon}
   \frac{1}{ \Gamma (2-\epsilon) } 
     \, \left( 1 - \frac{m^2 x }{ Q^2} \right)^{1-\epsilon}
\nonumber\\
  &&\times
     \int_{\xi_{\rm min}}^{\xi_{\rm max}} \d\xi \,
     \frac{ x [(\xi_{\rm max} - \xi ) (\xi - \xi_{\rm min}) ]^{1-\epsilon}
     }{ [\xi - x - x^2(1-\xi ) m^2 / Q^2]^2 }\ ,
\numlabel{eq:66}
\end{eqnarray}
where the limits $\xi_{\rm min}$ and $\xi_{\rm max}$ are given by
\begin{equation}
  \frac{1+x }{ 2} \pm \frac{1-x }{ 2}
  \sqrt{1 - \frac{4m^2x }{ (1-x) (Q^2 + m^2x)}}\ .
\end{equation}
In this form, we can look for the leading regions of the
ladder corrections.  To do this, it is simplest
to set the mass to zero, find the leading regions, and then check back as to
whether we must reincorporate the mass in the actual calculation.
So, to lowest order in $\Delta \equiv m^2/Q^2$, (\ref{eq:66}) becomes
\begin{eqnarray}
F_{2(a)}
&=&
   \frac{g^2 }{ 64 \pi^3 }
  \left( \frac{ Q^2 }{ \e^\gamma \mu^2 x (1-x) } \right)^{-\epsilon} \!
   \frac{1 }{ \Gamma (2-\epsilon) } 
\nonumber\\
&&\times
   \int_{x(1+\Delta)}^1 \! \d\xi \, 
      \frac{ x  \left\{(1 - \xi) [\xi-x(1+\Delta)] \right\}^{1-\epsilon } }
           { \left\{ \xi - x [1+x(1-\xi)\Delta] \right\}^2 } .
\numlabel{eq:68}
\end{eqnarray}
To interpret this expression, we must distinguish between the renormalizable
($n=6, \epsilon=0)$ and superrenormalizable $(n<6, \epsilon>0)$ cases.

 In the
super-renormalizable case, ($\epsilon > 0$), the leading-power contribution
$(Q/\mu)^0$ comes from near the endpoint $\xi = x(1+\Delta)$.  The bulk of
the integration region, where $\xi -x=O(1)$ is suppressed by a power of $Q$. 
The integral is power divergent when $m=0$, and clearly
 we cannot neglect the mass.   

Now consider the renormalizable case, $n=6$.  When we set $\epsilon =0$,
Eq.\ (\ref{eq:68}) has leading power ($Q^0$) contributions from both the region $\xi-x$
near zero, where, as above, the mass may not be neglected, and  the region
$\xi - x =O(1)$, where it may.  In the former region, the integral is
logarithmically divergent for zero mass, but since the nonzero mass acts as
a cutoff, the two regions $\xi \sim x$ and $\xi - x = O(1)$ should be thought
of as giving contributions of essentially equal importance.  We now
interpret these dimension-dependent leading regions.

\subsection{ Collinear and Ultraviolet Leading Regions; the Parton Model}

To see the physical content of the leading regions identified above, it is
useful to relate the variable $\xi$ in (\ref{eq:66}) to 
the momentum $k^\mu$, by the
relations
\begin{eqnarray}
   k^+  &=& \xi p^+ ,
\nonumber\\
   k^-  &=& \frac{ -1 }{ 2 p^+ (1-x)}
          \left[ \left( \frac{ \xi }{ x} -1 \right) Q^2 - m^2 (1-\xi ) 
\right] ,
\nonumber\\
   \kt^2 &=& \frac{Q^2 (1-\xi ) (\xi - x) }{ x (1-x) }
          - \frac{m^2 [(1-\xi )^2 + \xi (1-x)] }{ 1 - x }.
\end{eqnarray}
Changing variables to $\kt^2$, we
now rewrite the integral Eq.\ (\ref{eq:68}) in a form which is
accurate to leading power in $\Delta$ for $n<6$,
\begin{equation}
F_{2(a)}=
   \frac{g^2 }{ 64 \pi^3 }
   \frac{1 }{ \Gamma (2-\epsilon) } 
   \int_0 \d \kt^2 \, 
            \frac{ [\kt^2]^{1-\epsilon }\, (\mu^2 \e^\gamma )^\epsilon 
                }{ (\kt^2 + m^2(1-x+x^2))^2 }\ . 
\numlabel{eq:70}
\end{equation}
We emphasize that this expression is accurate to leading power
in the region $\kt^2/Q^2 = O(\Delta)$, which is sufficient to give the
full leading power for $n<6$, although not for $n=6$, where larger $\kt $
also contribute.
 
\begin{figure}
   \centering
   \includegraphics{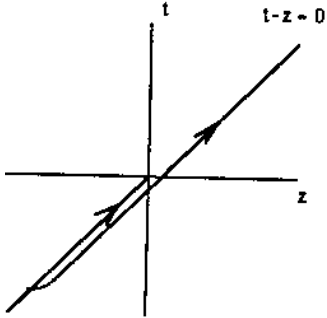}
   \caption{Space-time structure of collinear interaction.}
   \label{fig:7}
\end{figure}

Now let us choose a frame in which $p^+$ is
of order $Q$. When $\xi \to x$, the components of $k^\mu$ are of order
$(Q, (\xi -x)Q,  Q\sqrt{\xi -x})$, and 
at its lower limit, $\xi -x$ is of order $m^2/Q^2$.  Hence, in the region
that gives the sensitivity to $m$, $\T k$ is small, and $k^\mu$ is
ultrarelativistic and represents a particle moving nearly collinear to the
incoming momentum, $p^\mu$.   In addition,
the on-shell line, of momentum $p^\mu-k^\mu$, is nearly collinear to the
incoming line as well.  In fact, when $m$ and $\T k$ are both zero, 
$k^\mu$ is also on the mass shell. The energy deficit necessary to put
both the momenta $k^\mu$ and $p^\mu-k^\mu$ on shell is of order $\kt^2/Q$
in this frame.  Thus, in this frame, the intermediate state represented by
the Feynman diagram lives a time of order $Q/\kt^2$, which diverges 
in the collinear limit.   The space-time picture for such a process is
illustrated in Fig.\ \ref{fig:7}, and we see a close relation to the
parton model,  
as discussed in Sect.\ \ref{introsec:parton.model}, which
depends on the time dilation of partonic states.  Partonic
states whose energy deficit is much greater than $m$ in the chosen frame
correspond to $\xi - x$ of order unity, and do not contribute at
leading twist. 
Thus here, as in the parton model, there is a 
clear separation between long-lived,
time-dilated states which contribute to the distribution of partons from which
the scattering occurs, and the hard scattering itself, which occurs on a short
time scale.   

From this discussion, the collinear region, which is the only leading region 
when $n$ is less than 6, is naturally described in parton model
language.  

When $n=6$, the collinear region remains leading.  In addition,
however, all scales between $\kt =m$ and $\kt =Q$ contribute at
leading power, and there is no natural gap between long- and
short-distance interactions. When $\xi-x$ is order unity, $k^\mu$ is
separated from $p^\mu$ by a finite angle, and corresponds to a
short-lived intermediate state, where $(p-k)^2 \gg m^2$.  This leading
region, which is best described as ``ultraviolet", is not naturally
described by the parton model.  But, in an asymptotically free theory
(as ${(\phi^3)}_6$ is), such short-lived states may still be treated
perturbatively.  We shall see how to do this below.

In summary, the ladder diagram shows two important features: a strong
correspondence with the parton model from leading collinear regions for both
superrenormalizable and renormalizable theories and, for the renormalizable
theory only, leading ultraviolet contributions, not present in the parton
model.  

\subsection{Parton distribution functions and parton model}

\begin{figure}
   \centering
   \includegraphics{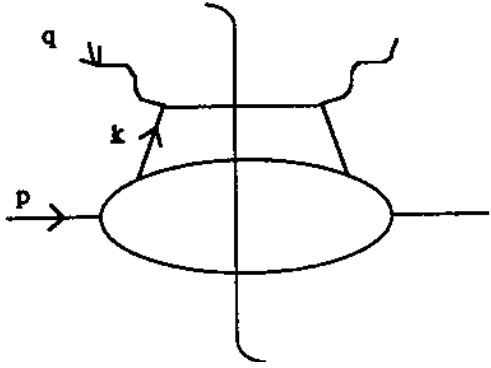}
   \caption{Dominant graphs for deeply inelastic scattering in parton model.}
   \label{fig:8}
\end{figure}

We shall now freely generalize the results for the
one loop ladder diagram.  Indeed, as 
we shall see in Sect.\ \ref{regionsec}, some of the 
dominant contributions to the structure
function arise from 
(two-particle-reducible) graphs of the form of Fig.\ \ref{fig:8}.  A single parton of
momentum $k^\mu$ comes out of the hadron and undergoes a collision in the
Born approximation.  If we temporarily neglect all other contributions, we
find that
\begin{equation}
   F (x, Q) = \int \frac{\d^6k }{ (2 \pi )^6 }
       \Phi(k,p) H(k,q) + \order (1/Q^a),
\end{equation}
where $\Phi$ represents the hadronic factor in the diagram and $H$ the hard
scattering (multiplied by the factor of $Q^2/2\pi$ in the definition of the
structure function):
\begin{equation}
  H (k, q) = Q^2 \delta ( (k+Q)^2 - m^2).
\end{equation}

\begin{figure}
   \centering
   \includegraphics{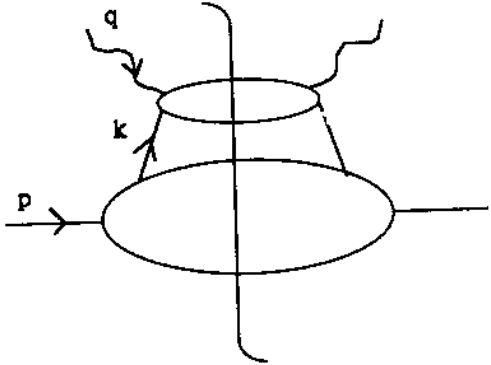}
   \caption{Dominant regions for deeply inelastic scattering in
            ${(\phi^3)}_6$ theory.} 
   \label{fig:9}
\end{figure}

For $\phi^3$ with $n<6$, as in the parton model, the parton momentum
$k^\mu$ is nearly collinear to the hadron momentum $p^\mu$. 
This implies that we can neglect $m$ and the minus and transverse 
components of $k^\mu$ in
the hard scattering, so that we can write
\begin{equation}
   H = H (x / \xi, Q) = \delta (\xi /x - 1) ,
\end{equation}
and hence
\begin{equation}
   F (x, Q) = \int_0^1 \d \xi 
    \left[ \int p^+ \frac{ \d k^- \d^4\kt }{ (2 \pi )^6} \Phi(k,p) \right]
    \delta (\xi /x - 1) + \order (1/Q^a).
\numlabel{eq:74}
\end{equation}
Here we define $\xi = k^+ / p^+$.  The limits on the $\xi$ integral are
0 to 1, since the final state must have positive energy.
We therefore define the parton distribution function (or number density):
\begin{eqnarray}
  f (\xi) &=& \xi p^+ \int \frac{\d k^- \d^4\kt }{ (2 \pi )^6} \, \Phi (k, p)
\nonumber\\
        &=& \int \frac{\d^6k }{ (2 \pi )^6}  \, 
             \Phi (k, p) \, \delta (\xi p^+ / k^+ - 1).
\numlabel{eq:75}
\end{eqnarray}
With this definition (\ref{eq:74}) becomes
\begin{eqnarray}
   F (x, Q) &=& \int \frac{\d \xi }{ \xi} \, f( \xi) \, H (x / \xi, Q) 
      + \order (1/Q^a)
\nonumber\\
            &=& f (x) + \order (1/Q^a) \qquad (n<6).
\end{eqnarray}
As we shall see in the next section, the factorization theorem is also
true in the renormalizable theory, 
\begin{equation}
F (x, Q) = \int \frac{\d \xi }{ \xi} \, f( \xi) \, H (x / \xi, Q) 
      + \order (1/Q^a) \qquad (n=6)\ ,
\numlabel{eq:77}
\end{equation}
where now $H$ is nontrivial.  The dominant processes that contribute
are illustrated by Fig.\ \ref{fig:9}, which generalizes the parton
model only to the extent of having more than just the Born graph for
the hard scattering.  These processes first involve interactions
within the hadron that take place over a long time scale before the
interaction with the virtual photon.  Then one parton out of the
hadron interacts over a relatively short time scale.

\begin{figure}
   \centering
   \includegraphics{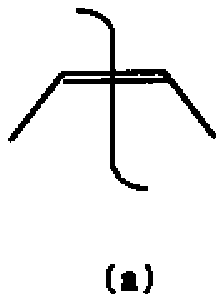}
   \includegraphics{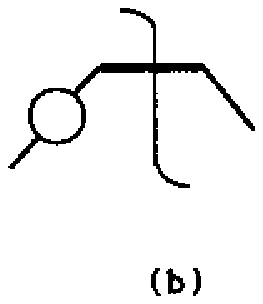}
   \includegraphics{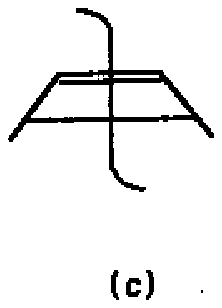}
   \caption{Low-order graphs for parton distribution in $\phi^3$ theory.}
   \label{fig:10}
\end{figure}

We now note that Eq.\ (\ref{eq:75}) can be expressed in
operator form as
\begin{equation}
f(\xi) = \frac{ \xi p^+ }{ 2 \pi} \int_{-\infty}^\infty  
  \d y^- \e^{-i \xi p^+ y^-}
  \langle p| \, \phi (0, y^-, \zerot) \, \phi(0) \, | p \rangle .
\numlabel{eq:78}
\end{equation}
This is the definition which we use for all $n \le 6$.  Of course in
the renormalizable theory $n=6$ renormalization will be necessary
\cite{pdfs}.  The definition (\ref{eq:78}) is precisely the analog for
$\phi^3$ theory of those we gave in Sect.~\ref{pdfsec} for QCD. It
involves an integral over a bilocal operator along a light-like
direction.  The graphs for $f (\xi)$ up to one-loop order are shown in
Fig.\ \ref{fig:10}.  Feynman rules are the same as for the gauge
theory, but without eikonal lines.

It is natural to interpret $f (x) \d x$ as the number of partons with
fractional momenta between $x$ and $x + \d x$.  This interpretation is
justified by the use of light front quantization \cite{null}, as we
saw in Sect.~\ref{pdfsec}.  Note that although the definition picks
out a particular direction as special to the problem, it is invariant
under boosts parallel to this direction.
 
The ladder graph, Fig.\ \ref{fig:10}(c) gives
\begin{equation}
   f_{c} = \frac{ g^2 {\bar \mu}^{ 2 \epsilon} }{ (2 \pi)^{6 - 2 \epsilon} }
         \int \d^{6 - 2 \epsilon} k \, \delta (\xi p^+ / k^+ - 1)
         \, \frac{ 2 \pi \, \delta \left( (p-k)^2 - m^2) \right) }
                 { (m^2 - k^2)^2 }.
\end{equation}
The $\delta$-functions may be used to perform the $k^+$ and $k^-$
integrals, after which we obtain
\begin{equation}
f_{c} = \frac{g^2  }{ 64 \pi^3} 
         \frac{ (\e^\gamma \mu^2)^\epsilon \xi (1 - \xi) }
              { \Gamma (2- \epsilon) }
         \int_0^\infty \d \kt^2 
         \, \frac{ (\kt^2)^{1 - \epsilon} }
                 { \left[ \kt^2 + m^2 (1 - \xi + \xi^2) \right]^2 } ,
\numlabel{eq:80}
\end{equation}
which matches Eq.\ (\ref{eq:70}) in the $\kt  \rightarrow 0$ limit.  That is, we
have constructed the parton distribution to look like the structure function
at low transverse momentum.  The significance of this fact will become clear
below.

For $n<6$, Eq.\ (\ref{eq:80}) is the same as the full leading structure function (\ref{eq:70}),
and it exemplifies  the validity of the parton model in a 
super-renormalizable theory.  
When $n=6$, however, there is a logarithmic ultraviolet
divergence from large $\kt$ in (\ref{eq:80}).  So, in the renormalizable theory 
we must renormalize $f( \xi)$.  (Since $f$ is a theoretical construct
defined to make treatments of high-energy behavior simple and convenient,
we are entitled to change its definition if that is useful; in particular,
we are allowed to include renormalization in its definition.)
If we use
the \MSbar\ scheme, then the renormalized value of $f_c$ for nonzero mass
is: 
\begin{equation}
R[f_c] = - \frac{g^2  }{ 64 \pi^3} \xi (1 - \xi) 
         \, \ln \left[ \frac{ m^2 (1 - \xi + \xi^2) }{ \mu^2} \right],
\numlabel{eq:81}
\end{equation}
while for zero mass it is (compare Eq.\ (\ref{eq:23}))
\begin{equation}
R[f_c] = - \frac{g^2  }{ 64 \pi^3} \xi (1 - \xi) 
         \, \frac{ 1 }{ \epsilon }.
\end{equation}
Now let us see what this means in the calculation of the hard part, as in
Sect.\ \ref{hardsec}.
To calculate the hard part, we expand Eq.\ (\ref{eq:77}) in powers of $g^2$, as in
Eq.\ (\ref{eq:24}), and solve for $H^{(1)}(x/\xi,Q)$.  There is some question about what
to do with the higher-twist terms, proportional to powers of $m/Q$.  The
simplest method is to simply define
\begin{equation}
H^{(1)}(x/\xi,Q) = \left[F^{(1)}(x/\xi,Q) - f^{(1)} \right]_{m=0}\ ,
\numlabel{eq:83}
\end{equation}
Comparison of Eqs.\ (\ref{eq:70}) and (\ref{eq:80}) shows that the low
$\T k$ region, which is the only leading region which is sensitive to
the mass, cancels between $F^{(1)}$ and $f^{(1)}$, at the level of
integrands.  Thus, for the combination on the right hand side of Eq.\
(\ref{eq:83}), it is permissible to set the mass to zero.  It is thus
practical to set the mass to zero at the very beginning.  It should be
kept in mind, however, that this is a matter of calculational
convenience, rather than principle.  The factorization theorem allows
us to calculate mass-insensitive quantities whatever the masses we
choose, since all sensitivity to these masses will be factored into
the parton distributions.

Now let us return to the remaining diagrams in Fig.\ \ref{fig:6},
treating first the ``final state" interactions, Fig.\ \ref{fig:6}(b)
and (c).

\subsection{Final state interactions}

The graphs of Fig.\ \ref{fig:6}(b) and (c) have a self-energy
correction on the outgoing line, the final state cut either passing through
the self energy or not.  As we will show, these graphs have contributions
that are sensitive to low virtualities and long distances.  However, they are
not of the parton model form, and do not naturally group themselves into
the parton distribution for the incoming hadron.   We will see, however, that
there is a cancellation between the two graphs such that they are either
higher twist ($n<6$), or may be absorbed into the one-loop hard part
($n=6$).  

The self energy graphs give simply the lowest order graph, $\delta (x-
1)$, times the one-loop contribution to the residue of the propagator pole:
\begin{equation}
\begin{split}
    F_{2(b)} 
= \delta(x-1) 
    &\left[
         \frac{-g^2 }{ 128 \pi^3}
        \int_0^1 \d z \int_0^\infty \d \kt^2 \, \kt^2 
        \frac{z (1-z) }{ [\kt^2 + m^2 (1-z+z^2)]^2} \right.
\\
       & ~ + \mbox{counterterm} ~ \Big].
\numlabel{eq:84}
\end{split}
\end{equation}
We may derive this expression in either of two ways.  One way is to
combine the denominators of the two propagators in the loop by a
Feynman parameter before performing the $k^+$ and $k^-$ integrals.
Then $z$ is the Feynman parameter.  Alternatively, we may first use
contour integration to perform the $k^-$ integral.  Then we get
(\ref{eq:84}) by writing $k^+ = z \, (p^++q^+)$.  The integral is the
same by either derivation.  But the second method shows that we may
interpret $z$ as a fractional momentum carried by one of the internal
lines.  Since we will be concerned with the low $\kt^2$ region, while
the counterterm, if computed with \MSbar\ renormalization, is governed
by the $\kt \to \infty$ behavior of the integrand, we do not write the
counterterm explicitly.

There is clearly a significant contribution in (\ref{eq:84}) from
small $\kt$, where the mass $m$ is not negligible.  The cut
self-energy graph, in Fig.\ \ref{fig:6}(c), will also contribute in
this region.  Now the region of low $\kt$ represents the effect of
interactions that happen long after the scattering off the virtual
photon, and it is reasonable to expect that interactions happening at
late times cancel, since the scattering off the virtual photon
involves a large momentum transfer $Q$ and therefore should take place
over a short time-scale.  However, the uncut self-energy graph only
contributes when $x$ is exactly equal to 1, while the cut self energy
graph has no $\delta$-function and thus contributes at all values of
$x$.

This mismatch is resolved when we recognize that we should treat the values of
the graphs as distributions rather than as ordinary functions of $x$.  That
is, we consider them always to be integrated with a smooth test function.
Mathematically, this is necessary to define the $\delta$-function.
Physically, the test function corresponds to an averaging with the
resolution of the apparatus that measures the momentum of the lepton that
is implicitly at the other end of the virtual photon.  After this
averaging, a measurement of the lepton momentum does not distinguish the
situation where a single quark goes into the final state from the situation
where the quark splits into two.

We therefore consider an average of the structure function $F (x)$ with a
smooth function $t (x)$:
\begin{equation}
 \langle t, F \rangle \, \equiv \, \int \d x \, t(x) \, F(x).
\end{equation}
Then the contribution of the self-energy graph is
\begin{equation}
  \langle t, F_{2(b)} \rangle \, = \, t(1) \, F_{2(b)}\ .
\end{equation}

Next we compute the cut graph, Fig.\ \ref{fig:6}(c).  Its value is
\begin{multline}
  \langle t, F_{2(c)} \rangle 
=
  \frac{ g^2 Q^2 }{ 64 \pi^3}
  \int \d k^+ \int \d \kt^2 \,  \kt^2 \int \d x \int \d k^- \, 
\\ \times
    \frac{ t(x) \, \delta (k^2 - m^2) \, \delta ((p+q-k)^2 -m^2)
    }{ [ (p+q)^2 - m^2]^2}.
\end{multline}
To make this correspond with the form of (\ref{eq:84}), we define 
$z = k^+ / (p^+ + q^+)$, and then use the $\delta$-functions to do the 
$k^-$ and $x$ integrals.  After some algebra, and after the neglect of
terms suppressed by a power of $Q$, we find
\begin{equation}
\langle t, F_{2(c)} \rangle \, = \, \frac{g^2 }{ 128 \pi^3}
    \int_0^1 \d z \int_0^\infty \d \kt^2 \, \kt^2 \, 
    \frac{z (1-z) t(x) x^2 }{ [\kt^2 + m^2 (1-z+z^2)]^2} ,
\end{equation}
where the Bjorken variable $x$ satisfies
\begin{equation}
  x = \left[ 1 + \frac{\kt^2 + m^2 }{ Q^2 z (1-z)} \right]^{-1}.
\end{equation}
We now add the two diagrams to obtain:
\begin{equation}
\begin{split}
    \langle t,  F_{2(b)} & {}+{} F_{2(c)} \rangle 
\\
 {}={}&
    \frac{g^2 }{ 128 \pi^3}
    \int_0^1 \d z \int_0^\infty \d \kt^2 \, \kt^2 \, 
    [t(x) x^2 - t(1)] \,
    \frac{z (1-z) }{ [\kt^2 + m^2 (1-z+z^2)]^2} 
\\
     & + \mbox{counterterm} .
\numlabel{eq:90}
\end{split}
\end{equation}
In the region $\kt \ll Q$, $x$ is close to one, and there is a cancellation
in the integrand of Eq.\ (\ref{eq:90}).  The cancellation fails
when $z$ is close to zero or one, but the contribution of that region is
suppressed by a power of $Q$.  We are therefore permitted to set $m=0$ in
the calculations of the graphs, 
after which a calculation (with dimensional regularization to regulate the
infrared divergences that now appear in each individual graph
at $\kt=0$) is much easier.  

\subsection{Vertex correction}

Finally, we consider the vertex correction Fig.\ \ref{fig:6}(d).  It
has the value
\begin{eqnarray}
   F_{2(d)} &=& \delta (x-1) 
    \frac{ -i g^2 {\bar \mu}^{2\epsilon } }{ (2\pi )^{6-2\epsilon }}
    \int \d^{6-2\epsilon }k 
\nonumber\\
&& \hspace*{1cm}
\times
  \frac{1}{ [m^2-k^2]\, [m^2-(p+k)^2]\, [m^2-(p+k+q)^2] }
\nonumber\\
   &&  {} + {}\mbox{counterterm}
\nonumber\\
  &=& - \delta (x-1) \frac{g^2 }{ 64 \pi^3} 
        \int_0^1 \d \alpha_1  \int_0^{1-\alpha_1 } \d \alpha_2 
\nonumber\\
&&\qquad \times
   \ln \left[ \frac{m^2(1-\alpha_1 -\alpha_2 -(\alpha_1 +\alpha_2 )^2) + 
                 Q^2\alpha_2 \alpha_1 
        }{ {\bar \mu}^2} \right],
\end{eqnarray}
where we work in $ d = 6 - 2\epsilon$ space-time dimensions to regulate the
ultraviolet divergence.
When $Q \to \infty $, we can clearly neglect the mass, so that we have (at
$\epsilon=0$)
\begin{equation}
 F_{2(d)} = -\delta (x-1) \frac{g^2 }{ 128\pi^3} 
             \left[ \ln \frac{Q^2 }{{\bar \mu}^2} - 3 \right]
        + \order \left( \frac{1 }{ Q^2} \right) .
\end{equation}
$F_{2(d)}$ is higher twist for $\epsilon > 0$.  

The graph Fig.\ \ref{fig:6}(e) is related to Fig.\ \ref{fig:6}(d) by
moving the final state cut so that it cuts the inner lines of the
loop.  We will not calculate it explicitly.  But when that is done,
the quark mass can be neglected, just as for the uncut vertex.

In summary, the only diagram from Fig.\ \ref{fig:6} which corresponds
to the parton distribution is the ladder diagram, Fig.\ \ref{fig:6}(a).
Non-ladder diagrams are either higher twist, or contribute only to the
hard part (renormalizable case).  These results are consistent with
the structure of Fig.\ \ref{fig:8} and Fig.\ \ref{fig:9}, which show the
structure of regions which give leading regions for $n<6$ and $n=6$,
respectively.  As we shall show in the next section, it is this
structure which enables us to prove that the parton distributions Eq.\
(\ref{eq:78}) absorb the complete long-distance dependence of the
structure function.

\section{Subtraction Method}
\label{subtrsec}

To establish a factorization theorem one must first find the leading
regions for a general graph.  We will see how to do this in
Sect.~\ref{regionsec}.  The result, for deeply inelastic scattering in
a nongauge theory, has been summarized by the graphical picture in
Fig.\ \ref{fig:9}, and it corresponds closely to our detailed
examination of the order $g^2$ graphs.  It can be converted to a
factorization formula if one takes sufficient care to see that
overlaps between different leading regions of momentum space do not
matter.

An approach that makes this process clear is due to
Zimmermann \cite{Z}.  To treat the operator product expansion (OPE), he
generalized the methods of Bogoliubov, Hepp, Parasiuk, and Zimmermann
(BPHZ) \cite{Z,BPH} that were used to renormalize Feynman graphs.
(Although the original formulation was for completely massive theories
with zero momentum subtractions, it can be generalized to use
dimensional continuation with minimal subtraction \cite{MSOPE}.  This
allows gauge theories to be treated simply.)  In the case of deeply
inelastic scattering a very transparent reformulation can be made in a
kind of Bethe-Salpeter formalism \cite{BS}, although it is not clear
that in the case of a gauge theory the treatments in the literature
are complete.  In this section, we will explain these ideas in their
simplest form.

There are two parts to a complete discussion: the first to obtain the
factorization, and the second to interface this with the renormalization.
We will treat only the first part completely.  In ${(\phi^3)}_6$ theory, 
renormalization is a relatively trivial affair.  Moreover, if we regulate
dimensionally, with $\epsilon$ just slightly positive, one can choose to
treat as the leading terms not only contributions that are of order $Q^0$
(times logarithms) as $Q \to \infty$, but also those terms that are of
order $Q$ to a negative power that is of order $\epsilon$.  The remainder
terms are down by a full power of $Q^2$, and can be identified as ``higher
twist''.  In this way one has the same structure for the factorization,
without the added complications of renormalization.

Zimmermann's approach is to subtract out from graphs their leading behavior
as $Q \to \infty$.  This is a simple generalization of the renormalization
procedure that subtracts out the divergences of graphs.  From the structure
function $F (x, Q)$ one thereby obtains the remainder $\frem (x, Q)$,
which forms the higher twist contributions.  The leading twist terms are $F
- \frem $.  It is a simple algebraic proof to show that $F - \frem$
has the factorized
form $ f * H$, with $f$ being the parton distribution we have defined
earlier, and with `$*$' denoting the convolution in Eq.\ (\ref{eq:77}).

\subsection{Bethe-Salpeter decomposition }

\begin{figure}
   \centering
   \includegraphics{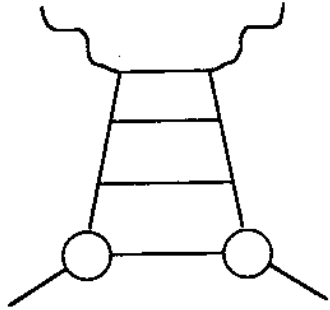}
   \caption{Example of ladder graph with several rungs.  }
   \label{fig:11}
\end{figure}

In the graphical depiction of a leading region, Fig.\ \ref{fig:9},
exactly one line on each side of the final state cut connects the
collinear part and the ultraviolet part.  So it is useful to decompose
amplitudes into two-particle-irreducible components.  This will lead
to a Bethe-Salpeter formalism.  Consider, for example, the two-rung
ladder graph, Fig.\ \ref{fig:11}, for deeply inelastic scattering off a
{\it composite} particle.  We can symbolize it as
\begin{equation}
\mbox{Fig.\ \ref{fig:11}} 
= \gamma_s \times \gamma \times \gamma \times \gamma_h .
\end{equation}
Here $\gamma_s$ represents the graph that is two-particle-irreducible in the
vertical channel and is attached to the initial state particle, 
$\gamma$ represents a rung, and $\gamma_h$ represents the 
two-particle-irreducible graph where the virtual photon attaches.  It is
necessary to specify where the propagators on the sides of the ladder
belong.  We include them in the component just below.  Thus 
$\gamma_s$ and $\gamma $ have two propagators on their upper external
lines.  The purpose of having a composite particle for the initial
state is to give an example with a non-trivial $\gamma_s$, as in QCD with
a hadronic initial state.  The vertex joining the initial particle is 
a bound-state wave function.

We now decompose the complete structure function as
\begin{eqnarray}
   F &=& \sum_{N=0 }^\infty
          G_s  \, G_r^N \, G_h
\nonumber\\   
     &=& G_s \, \frac{ 1 }{ 1 - G_r} \, G_h.
\numlabel{eq:94}
\end{eqnarray}
Here $G_s$ is the sum of all two-particle-irreducible graphs attached to
the initial state particle, $G_h$ is the sum of all 
two-particle-irreducible graphs coupling to the virtual photon, and $G_r$
is the sum of all graphs for a rung of the ladder.  Thus $G_r$ 
is the sum of all two-particle-irreducible graphs with two upper lines and
two lower lines, multiplied by full propagators for the upper lines.  

The second line of Eq.\ (\ref{eq:94}) has the inverse of $ 1 - G_r$, and it
clearly suggests a kind of operator or matrix formalism.  Indeed, if we
make explicit the external momenta of two ladder graphs, $\gamma_1 (k, l)$
and $\gamma_2 (k, l)$, then their product is
\begin{equation}
(\gamma_1 \gamma_2) (k, l) =
 \int \frac {\d^{ 6 - \epsilon } k' }{ (2 \pi )^{ 6 - \epsilon }}
    \gamma_1 (k, k') \gamma_2 (k', l).
\end{equation}
The rung graphs can thus be treated as matrices whose indices have a
continuous instead of a discrete range of values, while $G_s$ and $G_h$
can be treated as row- and column-vectors.  

In the case that the initial hadron is a single parton, as in the low order
examples in Sect.~\ref{phidissec}, the soft part $G_s$ is trivial: $G_s = 1$,
where `1' represents the unit matrix.

\subsection{Extraction of higher twist remainder}

We can now symbolize the operations used to extract the contribution of a
graph to the hard scattering coefficient.  Consider the example that lead
to Eq.\ (\ref{eq:83}).  We took the original graph and subtracted the
contribution of the graph to $f^{(1)} H^{(0)}$, 
where $H^{(0)}$ is the lowest order hard part.
Then we took the large $Q$ asymptote of the result, by
setting all the masses to zero.

\begin{figure}
   \centering
   \includegraphics{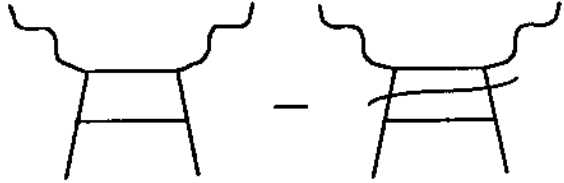}
   \caption{Hard scattering coefficient from Fig.\ \ref{fig:6}(a).  }
   \label{fig:12}
\end{figure}

We represent this in a graphical form in Fig.\ \ref{fig:12}.  There, the wavy line
represents the operation of short circuiting the minus and transverse
components of the loop momentum coming up from below, and of setting
all masses above the line to zero.  Symbolically, we write this as:
\begin{eqnarray} 
   \mbox{Contribution of Fig.\ \ref{fig:6}(a) to $H^{(1)}$ } 
&=& 
   P \gamma \gamma_h - P \gamma P \gamma _h
\nonumber\\
&=&    P \gamma \, (1-P) \, \gamma _h ,
\numlabel{eq:96}
\end{eqnarray}
where the operator $P$ is defined by
\begin{equation}
\begin{split}
   P (k, l) 
{}={}&
   (2 \pi )^{ 6-\epsilon } \,  \delta (k^+ - l^+)
          \, \delta (l^-) \, \delta^{6-\epsilon } (\T l)
\\
& {}\times{}
     (\mbox {Set masses to zero in the part of the graph above $P$}).
\end{split}
\end{equation}
In Eq.\ (\ref{eq:96}) we have ignored the need for renormalization
that occurs if $\epsilon =0$.  Either we can assume that we are only
making the argument when $\epsilon $ is slightly positive, or assume
that all necessary renormalization is implicitly performed by minimal
subtraction.

\begin{figure}
   \centering
   \includegraphics{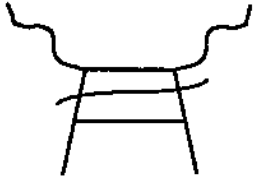}
   \caption{Contribution of Fig.\ \ref{fig:6}(a) to $f^{(1)} H^{(0)}$.  }
   \label{fig:13}
\end{figure}

Fig.\ \ref{fig:6}(a) gives two contributions to the factorization: a
contribution to the one-loop hard part $H^{(1)}$ given in Eq.\
(\ref{eq:96}) or (\ref{eq:83}), and a contribution to $f^{(1)}
H^{(0)}$.  The second of these we picture in Fig.\ \ref{fig:13} and
symbolize as
\begin{equation}
\gamma P \gamma_h.
\end{equation}
Thus we can write the remainder for Fig.\ \ref{fig:6}(a), after subtracting its
leading twist contribution, as
\begin{eqnarray}
   \rem (\mbox {Fig.\ \ref{fig:6}(a)} ) 
&=& \gamma \gamma_h \, - \, 
     \gamma P \gamma_h \, - \, 
     P \gamma (1-P) \gamma_h
\nonumber\\
&=&
   (1-P) \gamma (1-P) \gamma_h.
\end{eqnarray}
Clearly the operator $1-P$ subtracts out the leading behavior.  

In general, we can write the remainder for the complete structure function
as
\begin{eqnarray}
   \frem  &=& \sum_{N=0 }^\infty 
          G_s  \, [(1-P) G_r ]^N \, (1-P) \, G_h
\nonumber\\   
     &=& G_s \, (1-P) \, \frac {1 }{1 - G_r (1-P) } \, G_h.
\numlabel{eq:100}
\end{eqnarray}
This formula is valid without renormalization, even at $\epsilon =0$.  In
the first place, renormalization of the interactions can be done inside the
$\gamma $'s.  This is because there is nesting but no overlap between,  on the
one hand, the graphs to which the operation $1-P$ is applied and, on the other
hand, the vertex and self-energy graphs for which counterterms are needed in the
Lagrangian of the theory.  Further divergences occur because of the
extraction of the asymptotic behavior, and these give rise to the need to
renormalize the parton distribution.  But the regions that give rise to
such divergences are of the form where lines in some lower part of a graph
are collinear relative to lines in the upper part.  All such regions are
canceled in Eq.\ (\ref{eq:100}) since to the operator $1-P$ they behave just like the
regions that give the leading twist behavior of the structure function.  

\subsection{Factorization}

It is now almost trivial to prove factorization for the leading twist part
of the structure function, which is
\begin{equation}
F - \frem =
G_s \, \frac {1 }{1 - G_r } \, G_h -
G_s \, (1-P) \, \frac {1 }{1 - G_r (1-P) } \, G_h.
\end{equation}
Simple manipulations give
\begin{eqnarray}
    F - \frem &=&
    G_s \, \frac {1 }{1 - G_r } 
    P  \frac {1 }{1 - G_r (1-P) } \, G_h
\nonumber\\
   &=& f * H.
\end{eqnarray}
We now have an explicit formula for the hard scattering coefficient:
\begin{equation}
    H =
    P \, \frac {1 }{1 - G_r (1-P) } \, G_h ,
\end{equation}
while the parton distribution $f$ satisfies
\begin{equation}
    f \times P =
    G_s \, \frac {1 }{1 - G_r } \, P.
\numlabel{eq:104}
\end{equation}

One somewhat unconventional feature of our procedure is that not only do we
define $P$ to set to zero the minus and transverse components of the
momenta going into the subgraph above it, but we also define it to set
masses to zero.  Setting the minus and transverse momenta to zero while 
preserving the plus component is exactly the appropriate generalization of
BPH(Z) zero-momentum subtractions to the present situation.  
Setting the masses to zero as well is a convenient way of extracting the
asymptotic large-$Q$ behavior of a graph, as we saw in our explicit
calculations.  Moreover, particularly in QCD, it greatly simplifies
calculations if one works with a purely zero-mass theory.  Of course,
setting masses to zero gives infrared divergences in all but purely
ultraviolet quantities.  The momentum-space regions that give the
divergences associated with the structure function all have the same form
as the leading regions for large $Q$, Fig.\ \ref{fig:9}, so that the $1-P$ factors in
Eq.\ (\ref{eq:104}) kill all these divergences.  Note that, just as with Zimmermann's
methods, the $P$ operator can be applied at the level of integrands.  In
practical calculations, dimensional continuation serves as both an 
infrared and an ultraviolet regulator.  

In the one-loop example of Sect.\ \ref{phidissec}, the external hadron
is a parton, so that in $G_s=\delta(x-1)$ in (\ref{eq:104}).  At one
loop, $G_r$ corresponds exactly to $f_c$, Eq.\ (\ref{eq:80}).  This
expression, and the distribution $f$ as a whole in (\ref{eq:104}) is
still unrenormalized, and contains ultraviolet divergences.  These may
be removed by minimal subtraction, as in Eq.\ (\ref{eq:81}) at one
loop, or as discussed more generally in Ref.\ \cite{pdfs}.  We should
mention, however, that it would be advantageous to have a subtraction
procedure which combined factorization and renormalization into a
single operation.  The particular procedure outlined by Zimmermann
\cite{Z} does this, but is not immediately applicable when all
particles are massless.  Duncan and Furmanski \cite{DF} have discussed
some of these issues at length.

\subsection{Factorization for Inclusive Annihilation in ${(\phi^3)}_6$}

It is easy to generalize the general arguments of this and the previous
section to other
processes, such as those listed in the introduction.  An important example, is
the cross section in $\phi^3$ theory that is analogous to one-particle 
inclusive
annihilation in \epem\ annihilation, that was discussed in 
Sect.\ \ref{introsec:incl.annih}.  
In the scalar theory, the structure function for this process is 
\begin{equation}
 D(x,Q) = \frac{Q^2 }{ 2\pi}
\int \d^6y \, \e^{iq\cdot y} \sum_X \langle 0| j(y) | H X \rangle  
\langle H X| j(0) | 0 \rangle  ,
\end{equation}
which is exactly analogous to the QCD version, Eq.\ (\ref{eq:6}).  

\begin{figure}
   \centering
   \includegraphics{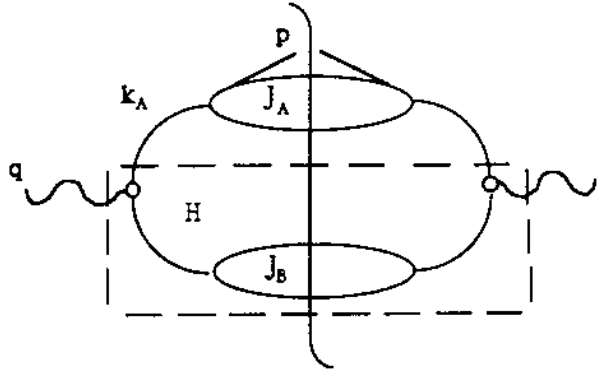}
   \caption{Example of leading region for inclusive
      annihilation. Regions with more than one jet inside the hard
      subdiagram $H$ are also leading.} 
   \label{fig:14}
\end{figure}

It is relatively easy to check that the leading regions for this
process have a form that generalizes Fig.\ \ref{fig:9} for deeply
inelastic scattering, that is, they have the form of Fig.\
\ref{fig:14}.  This was shown in Ref.\ \cite{Lead1PI} (for the case of
a non-gauge theory).

\begin{figure}
   \centering
   \includegraphics{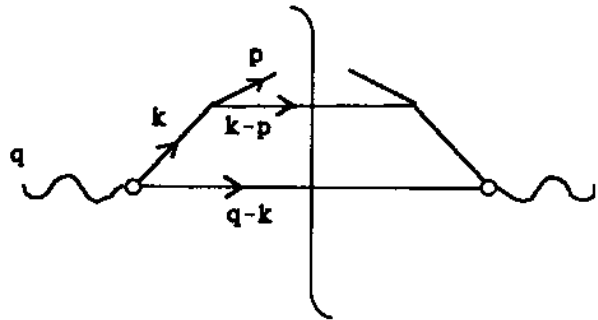}
   \caption{Ladder graph.  }
   \label{fig:15}
\end{figure}

An example is given by the ladder graph of Fig.\ \ref{fig:15}.  We
must integrate over all values of the momentum $(k-p)^\mu$.  When
$(k-p)^\mu$ is collinear to $p^\mu$, the line $k$ has low virtuality.
Then in the overall center-of-mass, the remaining particle $q - k$ has
large energy, approximately $Q/2$, and is moving in the opposite
direction to the first two particles.  We therefore consider the lines
$p$, $k-p$ and $k$ as forming the jet $J_A$ in Fig.\ \ref{fig:14} and
$q - k$ together with the vertex where the `virtual photon' attaches
as forming the hard part $H$.  When $(k-p)^\mu$ has transverse
momentum of order $Q$, we put both $k$ and $k-p$ into the hard part.

In a non-gauge theory, these two regions are the only significant ones,
together with a region that interpolates between them.  As we shall see in
Sect.\ \ref{regionsec}, this statement generalizes to all orders of
perturbation 
theory.  In a gauge theory, like QCD, all kinds of complication arise because
there are also `leading twist' regions involving soft gluons.

\subsection{Factorization, fragmentation function}

Simple generalizations of the arguments for deeply inelastic scattering
give the scalar factorization theorem:
\begin{equation}
D (z, Q) = \int_z^1 \frac{\d \zeta }{\zeta} \,
H (z/\zeta, Q) \, d (\zeta)
+ \order (1/Q^a).
\end{equation}
analogous to Eq.\ (\ref{eq:7}).
Here the fragmentation function is defined in exact analogy to the parton
distribution.  We choose axes so that the momentum $p^\mu$ of the
detected particle is in the positive $z$-direction.  Then we define:
\begin{eqnarray}
   d(z) &=& \frac{ p^+ }{2 \pi z} 
      \int \d y^-  \e^{i p^+ y^- / z} \sum_X
      \langle 0|  \phi (0, y^-, 0_T)  | H X \rangle
      \langle H X|  \phi (0)  | 0 \rangle  
\nonumber\\
   &=& \frac{p^+}{z} \int \frac{\d k^- \d^4 k_T }{ (2 \pi )^6}  \Phi_D (k, p).
\end{eqnarray}
This is interpreted as the number density of hadrons $H$ in a parton.
The formulae are exactly analogous to Eqs.\ (\ref{eq:75}) and
(\ref{eq:78}) for the parton distribution.  Renormalization is needed
here also.

%
%
%
%

\section{Leading Regions}
\label{regionsec}

As we saw in Sects.~\ref{phidissec} and \ref{subtrsec}, 
the first step in constructing a complete
proof of a factorization theorem is to derive the leading regions of
momentum space for a graph of arbitrary order. 
This section begins with
a brief description of a general 
approach to the long- and
short-distance behavior of
Feynman diagrams that results in a derivation of the leading regions.
We apply this method to 
describe the origin of 
high-energy logarithms in scalar 
theories, and go on to discuss the cancellation of final
state interactions, and the infrared finiteness of jet cross
sections.

\subsection{Mass dependence and leading regions}

Consider, then, an arbitrary
 Feynman integral $G({p_i^\nu}/\mu,m/\mu)$,
corresponding to a graph $G$,
 which is a function
of external momenta $p_i^\nu$, mass $m$ (possibly
zero), and renormalization scale $\mu$.   Without loss
of generality, we may take $G$ to be  
dimensionless.  We also assume that 
the invariants formed from different $p_i^\nu$ are all
large, while the $p_i^\nu$ themselves have
invariant mass of order $m$.  Thus:  
\begin{equation}
p_i \cdot p_j = Q^2 \eta_{i j} ,\ \ p_i^2= \zeta_i m^2 ,
\end{equation}
where $Q$ is a high-energy scale, $Q \gg m$, and
the $\eta_{i j}$ and $\zeta_i$ are numbers of order unity.  In the 
following, it will not be necessary to consider 
the $\eta_{i j}$  and $\zeta_i$ dependence, and we will
write $G$ as $G(Q^2/\mu^2,m^2/\mu^2)$.   We will be
interested in the leading term in an expansion in powers of
$1/Q^2$.   (Always we will allow the possibility of a polynomial in $\ln Q$
multiplying the power of $Q$, in each order of perturbation theory.)  

Suppose $G$ is the result of $L$ loop momentum
integrations acting on a product of $I$ Feynman 
propagators, times a function $N$, which is
a polynomial in the internal and external 
momenta.  For simplicity, we absorb into $N$ 
the numerator factors associated with the 
internal propagators, as well as overall kinematic
factors, etc.
$G$ may then  be represented schematically as
\begin{equation}
G(Q^2/\mu^2,m^2/\mu^2)=
\prod_{i=1}^L \int \d^n \ell_i\ N(\{k_j\},\{p_i\})
\prod_{j=1}^I \frac{1}{ (k_j^2-m^2+i\epsilon) } .
\numlabel{eq:109}
\end{equation}
The line momenta $\{k_j^\mu\}$, of course, are functions of the
$\ell_i^\mu$ and the $p_i^\mu$.  Any region in $\ell^\mu$ space which
contributes to $G$ at leading power in $Q^2$ will be called a ``leading
region".  In addition, by a ``short-distance" contribution to (\ref{eq:109}) we
will mean that we have a region of loop momenta in which some subset of the
line momenta, $\{k_j^\mu\}$, are off-shell by at least $\order (Q^2)$; the
short-distance contribution is the factor in (\ref{eq:109}) given by these far
off-shell lines.  Short-distance contributions are independent of masses to
the leading power in $Q^2$, since the integrand can usefully be expanded in
powers of $m^2$ when propagators are far off-shell.  A general leading
region has both short-and long-distance contributions, the latter
associated with lines which are nearer the mass shell.  Roughly speaking,
factorization is the statement that the cross section is a product of
parton distributions, in which all the long-distance contributions are
found, and a hard-scattering coefficient, which has purely short-distance
contributions.  To study factorization, we must characterize all ``long-
distance" contributions.

Our analysis depends on two observations.  The first concerns the close
relation between the high-energy and zero-mass limits.  That is, if the
renormalization scale $\mu$ is chosen to be of $\order (Q)$, then 
the two limits
are equivalent in the function $G(1,m^2/Q^2)$.  Short-distance 
contributions to the $ Q \to \infty$ limit 
are those involving lines for which $k_j^2$ is of
order $Q^2$.  Long-distance contributions
are parts of the $\ell_ i^\mu $ integrations
for which $k_j^2$ is much less than $Q^2$.  
If we scale all momenta down by a factor proportional to $Q$, then
we are considering the $m \to 0$ limit instead.  The short-distance
contributions now have fixed $k_j^2$ and the long-distance contributions
have Feynman denominators
$k_j^2+i\epsilon$ in Eq.\ (\ref{eq:109}) that vanish in the $m \to 0$ limit.  

Note that if $G$ is such that it only has short distance contributions,
then the $Q \to \infty$ limit is $G(1, 0)$, i.e., we can just set
$m=0$.  The QCD coupling, $\al (Q)$, is an implicit argument for $G$, and
we have already chosen to set the renormalization scale $\mu $ equal to
$Q$.  Thus in this case the detailed large $Q$ behavior is 
renormalization-group controlled in a simple way.  

When there are long-distance contributions to $G$, an expansion in powers
of $m$ will often fail.  
So to find the long-distance contributions to $G$, one
must look for singularities in the $m \to 0$ limit.
There are apparent exceptions to this rule, exemplified by 
the integral 
\begin{equation}
\int_0^\infty \d k^2 \frac{ m^2 }{(k^2 + m^2)^2 }.
\end{equation}
However, if we factor out the numerator factor $m^2$, we are left with an
integral that is singular like $1/m^2$.  This singularity is governed by
the denominator.  So what we are looking for is singularities in the $m$
dependence in the integral over the denominators of $G$.

Our second observation is that the integrals in (\ref{eq:109}) are
defined in complex $\ell_i^\mu$-space.  As a result, it is not enough
for a set of denominators to vanish in the integrand of (\ref{eq:109})
for the integral to produce a singularity at $m=0$ in $G$.  We must
have, in addition, a pinch of one or more of the $\ell_i^\mu$
integrals at the position of the singularity, between coalescing
poles.  This fact enables us to apply the simple but powerful analysis
due originally to Landau \cite{Landau,ELOP} on the relation of
singularities in Feynman integra{\it nds} to the singularities of
Feynman integra{\it ls}.  In the next subsection we explain the
application of this argument.

\subsection{Pinch surfaces}

We begin by using Feynman
parameterization to
combine the denominators of Eq.\ (\ref{eq:109}) by
\begin{eqnarray}
    G(Q^2/\mu^2,m^2/\mu^2) 
&=& 
    (I-1)! \, \prod_{j=1}^I \int_0^1 \d \alpha_j
    \, \delta\!\!\left( 1-\sum_{j=1}^I \alpha_j \right)
    \prod_{i=1}^L \int \d^n \ell_i\ 
\nonumber\\
&& \times
    \frac{ N(\{k_j\},\{p_i\}) }
         { \left[ \sum_{j=1}^I \alpha_j 
                  (k_j^2(\ell_i^\nu)-m^2) +i\epsilon 
           \right]^I
         } ,
\end{eqnarray}
where we have exhibited the 
loop-momentum dependence of the line momenta.  
There is now a single
denominator $D(\ell_i,\alpha_j)$, which is quadratic in
loop momenta and linear in Feynman parameters.  Suppose
$D(\ell_i,\alpha_j)$ vanishes for some value of loop
momenta and Feynman parameters.  We will now derive necessary conditions
for this zero to produce a singularity in $G$.  Then we will apply these
conditions to the case $m=0$.

A pole from $D=0$ will {\it not} give a singularity in $G$
if $D$ can be changed from zero by a deformation that does
not cross a pole in {\it any} {\it one} of the 
momentum or parameter contours.  Consider first the
parameter integrals.
Because $D$ is linear in the $\{\alpha_j\}$, a deformation
of the $\alpha_j$ integral will change $D$ away from
zero, unless 
\begin{equation}
k_j^2 = m^2 \ \ {\rm or}\ \ \alpha_j=0 
\numlabel{eq:112}
\end{equation}
for each line.  
In the first case, $D$ is independent of $\alpha_j$, while
in the second we note that $\alpha_j=0$ is an
endpoint  of the $\alpha_j$ integral, away from which it
cannot be deformed.  

Now suppose (\ref{eq:112}) is satisfied, and consider the momentum
integrals.  $D$ will be independent of those loop momenta which flow
only through lines whose Feynman parameters are zero.  The contours of
the remaining loop momenta must be pinched between singularities
associated with the vanishing of $D$.  Since $D$ is a quadratic
function of the remaining momenta, each momentum component sees only
two poles in its complex plane due to the vanishing of $D$.  The
condition for a pinch is thus the same as the condition that the two
zeros of the quadratic form be equal.  That is, in addition to $D=0$
we must have $\partial D/ \partial \ell_i^\mu=0$ for all $\ell_i^\mu$
which flow through one or more on-shell lines.  For each such loop
momentum, the extra condition is \cite{Landau,ELOP}
\begin{equation}
\sum_j\alpha_j k_j^\mu = 0 ,
\numlabel{eq:113} 
\end{equation}
where the sum goes over all 
lines through which the loop momentum $ \ell_i^\mu$ 
flows.  (Note that any line which is not on shell has $\alpha_j = 0$, by
Eq.\ (\ref{eq:112}), so the condition (\ref{eq:113}) can be applied to every loop.)  
Together, (\ref{eq:112}) and (\ref{eq:113}) are known as
the ``Landau equations".   We shall 
refer to any surface in momentum space on
which the Landau equations are satisfied 
as a ``pinch  surface" of the diagram
$G$.  With each pinch surface 
we associate a ``reduced diagram", in
which all off-shell lines are shrunk 
to points.  By construction, the reduced diagram contains
only those loop momenta of the original 
diagram which satisfy (\ref{eq:113}) with nonzero $\alpha$'s.

\subsection{Physical propagation}

The Landau equations are surprisingly 
restrictive, especially in the zero-mass 
limit.  To see why, let us rederive the
observation of Coleman and Norton \cite{CN,St78} 
that Eqs.\ (\ref{eq:112}) and (\ref{eq:113}) have an
appealing physical interpretation. Consider a given
pinch surface.   
We rewrite (\ref{eq:113}) on this surface as 
\begin{equation}
\sum_j (\alpha_j\
\omega_j) v_j^\mu = 0 , 
\end{equation}
where $v_j^\mu$ and $\omega_j$ are the four-velocity and energy 
associated with the 
momentum $k_j^\mu$. The units of the Feynman parameters 
are arbitrary, so  suppose we may, if we wish, 
interpret $\alpha_j$ 
as the frame-independent ratio 
of a time to the energy of line $j$.  
Then each of the components of the
vector
$(\alpha_j\omega_j) v_j^\mu$ has 
the units of
a distance in space-time.  It is the distance traversed in
time $(\alpha_j\omega_j)$ by a free particle moving
classically with velocity $v^\mu_j$.

Now suppose 
we associate
a definite position $x_1^\mu$ to one of 
the vertices in the reduced diagram 
associated with the pinch
surface.  Then, if line $j$ attaches to the vertex at
$x_1^\mu$, $x_1^\mu+(\alpha_j\ \omega_j) v_j^\mu$
may be interpreted as the position of the vertex at the other end of line
$j$.  
Continuing in this manner,
we can associate with the reduced diagram a position in
space-time for every one of its vertices, and
a physical process in which free particles move 
between these points. 
Equation (\ref{eq:113})  ensures that this program can be carried
out consistently, by requiring that in going around any
closed loop we come back to the same position.  We can use
this construction as a necessary condition for a pinch
surface.  

Finally, note that, because Eqs.\ (\ref{eq:112}) and (\ref{eq:113})
are homogeneous in the $\alpha$'s, a rescaling of the $\alpha$'s
leaves the Landau equations satisfied. Hence the vertices in the
physical picture are an indefinite distance apart, and, in particular,
this distance may be arbitrarily large.

\subsection{Collinear and infrared pinches; power counting}

For a general diagram with arbitrary masses and external
momenta the criterion of physical propagation allows a
very rich analytic structure.  In the massless limit,
however, this structure actually simplifies since
multiparticle thresholds become degenerate.  The physical processes of
which an isolated massless (but not massive) particle is
kinematically capable are as follows.
First, a
massless particle of momentum $p^\mu$ may split into two
(or more) massless particles of momenta $\alpha p^\mu$ and
$(1-\alpha) p^\mu$, and vice-versa.  This is the source of
collinear divergences.  Second, a particle may emit or
absorb one (or more) zero energy particles.  This is the source of
infrared divergences.  We easily check that arbitrary loops
involving only collinear and zero-momentum particles can 
satisfy the Landau equations.  Generally, we will describe a subdiagram
consisting of mutually collinear particles as
a ``jet" subdiagram.  Lines that have zero momentum in the massless limit
we will call ``infrared''. 
A jet subdiagram describes the
evolution of a set of collinear lines, as they absorb
and emit other collinear and infrared lines.  

\begin{figure}
   \centering
   \includegraphics{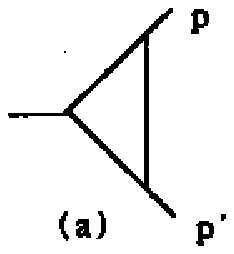}
\quad
   \includegraphics{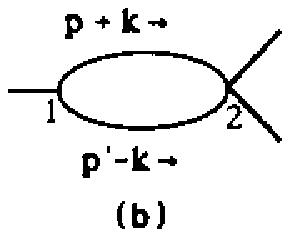}
\quad
   \includegraphics{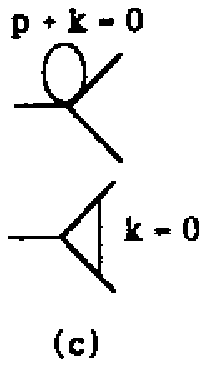}
\quad
   \includegraphics{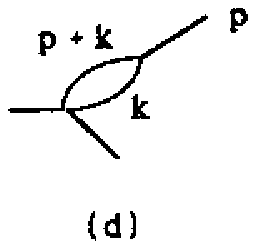}
   \caption{(a) Vertex correction,  (b) Reduced diagram corresponding
      to Eq.\ (\ref{eq:117}), (c) Infrared reduced diagram, (d) collinear
      reduced diagram.  } 
   \label{fig:16}
\end{figure}

As an example, let us consider the vertex
correction Fig.\ \ref{fig:16}a, in dimensionally regulated $\phi^3$
theory,  
\begin{equation}
V(p,p') = 
\int \frac{ \d^n k }{ (2\pi)^n }
\ \frac{1}{ [(p'-k)^2-m^2+i\epsilon] \, 
[(p+k)^2-m^2+i\epsilon] \,  (k^2-m^2+i\epsilon) } ,
\numlabel{eq:115}
\end{equation}
where we assume a production process, $Q^2\equiv
(p'+p)^2>0$. This example is used for illustrative purposes only. 
The term ``leading" will refer here only to this diagram, and not
to the behavior of the Born diagram.  In fact, Fig.\ \ref{fig:16}(a) is
nonleading compared to the Born process for all $n<6$.

The Landau equation for Fig.\ \ref{fig:16}(a) is
\begin{equation}
\alpha_1(p'-k)^\mu + \alpha_2(p+k)^\mu 
+ \alpha_3k^\mu = 0 .
\numlabel{eq:116}
\end{equation}
For on-shell ($p^2={p'}^2=m^2$) scattering with $m \ne 0$,
Eq.\ (\ref{eq:116}) has no solutions at all.  Note that this is
the case even though there is a singular surface
illustrated in Fig.\ \ref{fig:16}(b), with 
\begin{equation}
(p+k)^2=(p'-k)^2=m^2 ,\ k^2<0 .
\numlabel{eq:117}
\end{equation}
This singular surface corresponds to the production of
two particles, followed by a subsequent spacelike
scattering.  Although such a process is kinematically
possible, it clearly cannot correspond to physical
propagation, because the two 
particles produced at vertex $1$ propagate in
different directions, and would therefore not be able to
meet at vertex $2$ to scatter again.

Now let us consider the case that $m=0$.  By the same reasoning,
(\ref{eq:117}) does not give a pinch surface, if $k \ne 0$.  There are
nevertheless two sets of solutions.  First, there are infrared
solutions where one line has zero momentum,
\begin{align}
    k^\mu     = 0, & ~~ \alpha_1 = \alpha_2 = 0 , \nonumber\\
    (p+k)^\mu = 0, & ~~ \alpha_1 = \alpha_3 = 0 , \nonumber\\
    (p-k)^\mu = 0, & ~~ \alpha_2 = \alpha_3 = 0 . 
\numlabel{eq:118}
\end{align}
Second, there are collinear solutions, where two of
the lines are parallel to one of the outgoing external
particles,
\begin{align}
    \alpha_1(p-k)^\mu+\alpha_3k^\mu &= 0,~ \alpha_2=0,\  k^2=p \cdot k=0 ,
\nonumber\\
    \alpha_2(p'+k)^\mu+\alpha_3k^\mu &= 0,~ \alpha_1=0,\  k^2=p' \cdot k=0 . 
\numlabel{eq:119}
\end{align}
The physical pictures associated with typical
infrared and collinear pinch surfaces are
shown in Figs.\ \ref{fig:16}(c) and (d), respectively.  In each case,
there is physically realizable propagation between
vertices.  

Now we observe that 
even though solutions to the Landau equations like
Eqs.\ (\ref{eq:118}) and (\ref{eq:119}) give pinch 
surfaces, they still do not necessarily produce mass
dependence that is relevant to the leading power of $Q$, and 
hence are not necessarily leading regions.  

The Born graph for the vertex behaves like $Q^0$.  The contribution to the
one-loop graph from the pure short-distance region, from the region,
$|k^\mu|=\order (Q)$, is $Q^{n-6}$.  Thus this region is leading when the
theory is renormalizable, at $n=6$, but is non-leading relative to the Born
graph when the theory is super-renormalizable, $n<6$.  

Next we consider the one-loop graph near its singular surfaces.
For example,
consider the integral (\ref{eq:115}) near the surface defined by
the first of Eqs.\ (\ref{eq:118}).
To be
specific, let $|k^\mu|<k_{\rm max}$, $\mu=0 \cdots 3$, with
$k_{\rm max}$ being some fixed scale (which must $\gtrsim m$). 
In this region the integral
behaves as   
\begin{equation}
\frac{1}{Q^2}
\int_{k^\mu<k_{\rm max}} \frac{ \d^n k }{k^4 } \sim 
\frac{ k_{\rm max}^{n-4} }{ Q^2 } .
\end{equation}
Compared to the short-distance region, 
this infrared region is leading only for
$n \le 4$. (The other two infrared regions 
in (\ref{eq:118}) require 
$n \le 2$.)   (We remind the reader that, compared to the Born graph,
none of these regions contribute to the leading power of $Q$.)  

Similarly, near the collinear pinch surfaces
of Eq.\ (\ref{eq:118}), the integral behaves as 
\begin{equation}
\frac{1}{Q^2}
 \int_{k_\perp<k_{\rm max}} \frac{\d k^2}{k^2}
\frac{ \d^{n-2} k_\perp }{ k_\perp^2 }\ \sim 
\frac{ k_{\rm max}^{n-4} }{ Q^2} ,
\end{equation}
so that again only for $n \le 4$ do we find 
collinear
contributions from this diagram that are leading compared with the 
short-distance contribution. 

In summary, 
for the scalar theory in six dimensions, only short 
distance regions are (relatively) leading for
Fig.\ \ref{fig:16}.  This
result generalizes to all orders in the vertex correction
for this theory \cite{VertexPhi}.

The process of estimating the strength of a singularity
is known as ``power counting".  We will give more low
order examples below, while more general arguments can be
found in Ref.\ \cite{St78}.  We can, however, summarize the
basic result of these arguments briefly.  Let $D$ be a
reduced diagram with $S$ ``infrared loops" and $I_S$
infrared lines whose momenta vanish at the corresponding
pinch surface, and with $C$ ``collinear loops" and $I_C$
collinear lines whose momenta become proportional to an
external momentum at the pinch surface.  Finally, let
$N_2$ denote the number of two point subdiagrams in $R$.
In $\phi^3$ theory in $n$ space-time dimensions, let us define \cite{St78} 
the ``degree of divergence" by
\begin{equation}
\omega(R) = n S + (n/2) C - 2I_S - I_C + N_2. 
\numlabel{eq:122}
\end{equation}
This corresponds to a power law
\begin{equation}
Q^{(n-6) V} \lambda^{\omega (R)}
\numlabel{eq:123}
\end{equation}
as $Q \to \infty$.  Here $V$ is the number of loops and $\lambda $ is a small
parameter that parameterizes the approach to the singularities in the
massless theory.  The power law is measured relative to the power for the
Born graph.  

In the renormalizable case, $n=6$, we get leading behavior
only if $\omega (R) = 0$ and there are no graphs that give negative $\omega
(R)$. 
The term $N_2$, which
tends to suppress the behavior of the integral at the
singular point, is due to the fact that the renormalized
two-point function must vanish on-shell if the particle is
to have zero mass.  In this sense, the infrared behavior
of the theory is dependent on renormalization \cite{St78}.

In the superrenormalizable case, $n<6$, the first factor in (\ref{eq:123})
gives a negative power of $Q$ that just corresponds to normal ultraviolet
power counting; this is the power that comes from a purely short-distance
contribution to the graph.  A sufficiently strong power law singularity in
the massless theory is needed to overcome this if one is to get a leading
contribution.  

\subsection{Leading regions for deeply inelastic scattering in $\phitheory$}

As an application, we now discuss the general
leading regions for the basic inclusive cross sections in
$\phitheory $.  The criterion of
physical propagation shows why the considerations of
Sects.~\ref{phidissec} and \ref{subtrsec} take into account all relevant
leading regions. To see this, we must generalize our
concept of leading regions to include cut diagrams, of the
type discussed in Sects.~\ref{phidissec} and \ref{subtrsec}.  
There is no problem in
doing this, and power counting may be estimated for
collinear and infrared cut lines 
with the same degree of divergence Eq.\ (\ref{eq:122}) as
for virtual lines.  

It is useful to apply the optical theorem
to reexpress the deeply inelastic scattering structure
function Eq.\ (\ref{eq:63}) as the discontinuity of the forward
Compton scattering amplitude $T(q,p)$, 
\begin{eqnarray}
    F(x,Q^2) &=& \disc T(q,p) ,
\nonumber\\
    T(q,p) &=& \frac{Q^2}{2\pi} \int \d^6 y \, \e^{iq \cdot y}
\langle p| T(j(y) j(0) | p \rangle  . 
\end{eqnarray}
This relation holds diagram-by-diagram, once cuts are
summed over, so that a necessary condition for a 
region $L$ to be leading in $F$ is that it be leading in
$T$.

\begin{figure}
   \centering
   \includegraphics{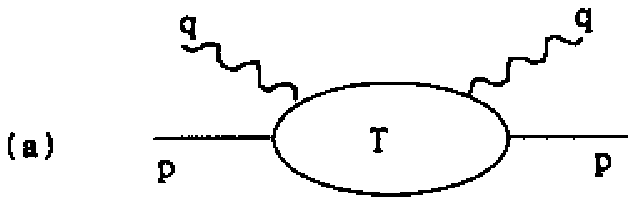}
\\*[5mm]
   \includegraphics{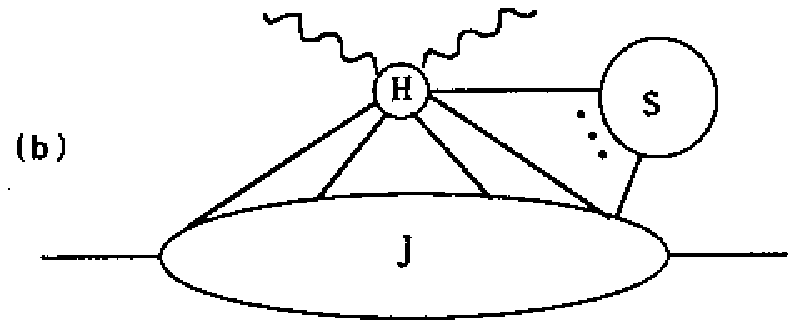}
\\*[5mm]
   \includegraphics{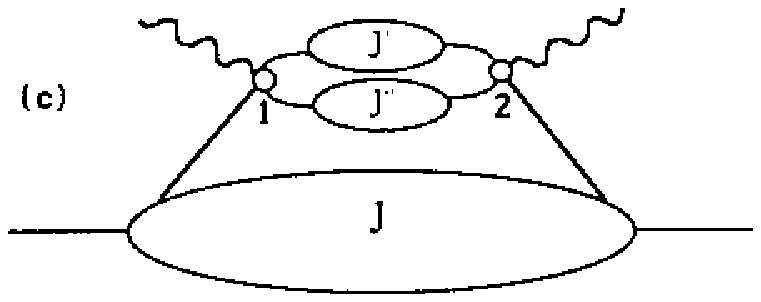}
   \caption{(a) Diagrams for forward Compton scattering.  (b) Reduced
   diagrams for pinch surfaces.  (c) Reduced diagram for a singular
   surface which is not pinched.  } 
   \label{fig:17}
\end{figure} 

  We thus need to consider the
leading regions of the diagrams illustrated by Fig.\ \ref{fig:17}(a),
which represents
 forward Compton scattering.  The pinch surfaces are symbolized
in Fig.\ \ref{fig:17}(b).  The incoming hadron can form a jet of
lines, and in addition, may
interact with any number of soft lines, connected to a
subdiagram $S$ consisting of only lines with zero
momentum.  To see why this is the general form, consider
a singular surface not of this kind, as in Fig.\ \ref{fig:17}(c).   
Here, one or more lines of the jet may
scattering with the incoming photon to form a set of
on-shell outgoing jets, which then rescatter to emit the
outgoing photon and reform the outgoing jet, which
eventually evolves into the outgoing hadron.  Such
a process is certainly consistent with momentum
conservation.  Surfaces of this type are not pinch surfaces for the
amplitude $T (q,p)$, 
however, because the outgoing jets can
never collide again once they have gone a finite distance
from the point at which they are produced.  As a
result, in every pinch surface, the incoming and outgoing
photons attach at the $same$ point  in space-time, and we
derive the picture of Fig.\ \ref{fig:17}(b).  This result shows that
all divergences associated with final state interactions
cancel in the sum over final states.

As indicated above, not every
pinch surface will correspond to a leading region.  In
particular, power counting using Eq.\ (\ref{eq:122}) shows that
there are no infrared divergences in $\phitheory $, that is,
no zero-momentum lines for any leading
region \cite{St78}. In addition, we can show that only
the minimum number of jet lines (two) can attach the jet
to the hard part in Fig.\ \ref{fig:17}(b).  It is a straightforward
exercise in counting to prove these results, using
(\ref{eq:122}), the Euler identity (loops = lines - vertices + 1),
and the observation that every internal line of a graph begins and 
ends at a vertex.  In summary, we can show that Fig.\ \ref{fig:9} is 
indeed the reduced diagram of the most general leading
region for deeply inelastic scattering in the scalar theory.

\subsection{Unitarity and jets: the cancellation of final state
  interactions}
\label{regionsec:unitarity}

The cancellation of final state interactions in deeply
inelastic scattering plays an important role in the
analysis for deeply inelastic scattering just described. 
This cancellation is a general feature of inclusive hard
scattering cross sections, and is used repeatedly in
factorization proofs.  The physics behind this cancellation
has already been pointed out in Sect.~\ref{phidissec}: a hard scattering
is well localized in space-time, and, as a result, it cannot
interfere with long-distance effects which describe the
further evolution of the system.  Thus, when we sum over
final states in an inclusive cross section, we lose
information on the details of evolution in the final state,
and are left with the constraint that, by unitarity, the sum of
probabilities of $all$ final states is unity.  As a result,
at each order in perturbation theory, long-distance
contributions to final states must cancel.

It is worth noting that it is not always necessary to
sum over all final states to cancel long-distance
interactions.  There are three kinds of cross sections, among those
mentioned in Sect.\ \ref{introsec}, for which the cancellation of final state
interactions is important.  In deeply inelastic scattering and 
Drell-Yan, for instance, we sum over all hadronic final states.  In
single-particle inclusive cross sections, on the other hand, we shall
find in Sect.\ \ref{gaugeproofsec} that cancellation requires the use of
Ward identities.  Finally, in jet cross sections, the cancellation
comes about in a sum over all final states which satisfy certain
criteria in phase space.  Let us hint at how this happens.

If all non-forward particles in the
final state emerge from a single hard scattering, the
criterion of physical propagation requires that the
long-distance contributions will come entirely from soft
and collinear interactions.  This is because, as in the
low order example of Fig.\ \ref{fig:16}, jets emerging from a
single point and propagating freely cannot meet
again to produce a new hard scattering.  In this case,
once the energy and direction of a set of jets is
specified, the sum over only those final states consistent
with these jets will also give unity, and their collinear
and infrared divergences cancel in the sum.  The technical
proof of these statements may be given in a number of
ways.  The simplest is based on a truncation of the
hamiltonian to describe only collinear and infrared
interactions.  Then, since the truncated hamiltonian is
hermitian, it generates a unitary evolution operator whose
divergences cancel by the ``KLN" theorem \cite{Kinoshita,LN}. 
It is also useful to see that this cancellation is
manifested on a diagram-by-diagram basis within each
leading region in perturbation theory \cite{St78}.  Proofs of
this type are most easily given in terms of time-ordered
or light-cone ordered perturbation theory \cite{null,CM}.  

\begin{figure}
   \centering
   \includegraphics{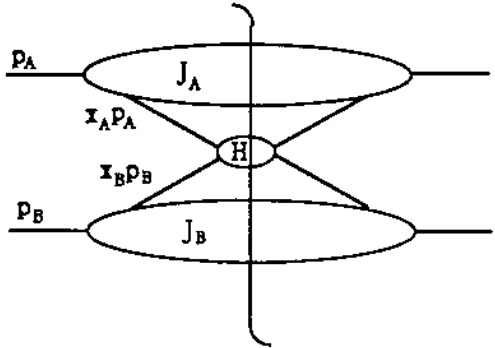}
   \caption{Leading regions for the Drell-Yan cross section in $\phitheory $.}
   \label{fig:18}
\end{figure}

Technicalities aside, the cancellation of final state
interactions at the level of jets has a number of important
consequences.  The simplest of these is the finiteness of
jet cross sections in \epem-annihilation cross sections
\cite{SW}.  We have already seen its importance for the
analysis of deeply inelastic scattering in $\phitheory $. 
It has a similar simplifying effect for single-particle
inclusive cross sections, as well as for the Drell-Yan and
related cross sections.  To illustrate this, we show,
in Fig.\ \ref{fig:18} the reduced diagrams for leading regions
in the scalar
``Drell-Yan" cross section, defined for the scalar theory by
analogy to Eq.\ (\ref{eq:63}) in $\phitheory $, after the sum
over final states. 
We see that all information about the final state has been
absorbed into a single hard part $H$.  Note that this
result holds not only for the fully inclusive Drell-Yan
cross section, but also for semiinclusive cross sections
such as hadron-hadron $\rightarrow$ Drell-Yan pair + jets.

%


\section{Factorization and Gauge Invariance}
\label{gaugefactsec}

In this and the following section, we discuss the
extension of factorization proofs to gauge theories.  We begin with a
discussion of the {\it classical} Coulomb field of a fast moving charge, an 
example that anticipates what happens in the full quantum theory.

We next summarize the Ward identities we will need.  Then we discuss the
leading regions of Feynman graphs in a gauge theory.  There are great
differences from the case of the $\phi ^3$ theory discussed in the previous
section.  With the  aid of the example of the vertex graph, we show how, 
after an appropriate eikonal approximation, Ward identities are applicable
that will combine graphs into a factorized form.  

In the next section, we will show how factorization may
be proved for a variety of experimentally important cross
sections, which can be measured in deeply inelastic
scattering, \epem\ annihilation and hadron-hadron
scattering.  We should emphasize at the outset that
although we regard existing proofs in all these cases as
reasonably satisfying, there is still room for improvement,
especially for hadron-hadron cross sections.  We will
point out the shortcomings of existing arguments in Sect.\ \ref{conclsec}.

\subsection{Classical considerations}
\label{gaugefactsec:classical}

Before getting into a detailed discussion of Feynman
diagrams, it is worth noting that insight can be gained
into the physical content of factorization theorems from 
purely classical considerations.  This discussion will at
once highlight an important difference between gauge and
scalar theories, and at the same time show why this
difference, important though it is, 
respects factorization.

As we observed in Sect.~\ref{introsec:parton.model}, the parton model
picture of 
hadron-hadron scattering rests in part on the
Lorentz contraction of colliding hadrons.  Now a
simplified  classical analog of a hadron is a
collection of point charges, each acting as a source of a
classical scalar field.  We would expect that
if the parton model, or factorization, is to make
sense, these fields ought to be Lorentz contracted
themselves, and this is just what happens.  Let us see
how.

Consider first a static classical scalar field
$\phi_{\rm cl}({\bf x})$, associated with a point particle of
charge $q$ at the origin.  If we 
assume that the field obeys
Laplace's equation, it is given in the rest frame of the 
particle by     
\begin{equation}
\phi_{\rm cl}({\bf x}) = \frac{ q }{ |{\bf x}| } .
\numlabel{eq:125}
\end{equation}
Now consider the same field in a frame
where the particle is moving at velocity $c\beta$ 
along the $z$-axis.  
Then the field at ${x'}^\mu$ in this frame is
\begin{equation}
\phi_{\rm cl}' (x' )
 = \frac{q}{ \bigl [ x_T^2 + \gamma^2 (\beta ct'-x'_3)^2
            \bigr ]^{1/2} },
\end{equation}
where, as usual, $\gamma=(1-\beta^2)^{-1/2}$.  For an
observer at $t' = 0$ in the primed system, the $\phi$
field decreases as $\gamma^{-1}$ as $\beta$ approaches
unity, except near $x'_3 =0$.  Thus, the $\phi$ field is indeed
Lorentz contracted, and any force proportional to the
$\phi$ field is also Lorentz contracted into a small longitudinal distance 
about $x'_3 =0$.  This means that in the rest frame
of a scalar ``hadron", the forces due to another such
hadron approaching at nearly the speed of light are
experienced in a Lorentz-contracted fashion, just as
supposed in the parton model.

Now let us apply this reasoning to a classical gauge
theory, in this case classical electrodynamics.  Here,
the field in the rest frame of a point particle of charge
$q$ is precisely analogous to Eq.\ (\ref{eq:125}),
\begin{equation}
A_{\rm cl}^{\mu}({\bf x}) =  \frac{ q \delta_{\mu 0} }{ |{\bf x}| }. 
\end{equation}
Because this is a vector field, however, there is a big
difference from the scalar case in a frame in which the
particle moves with velocity $c\beta{\hat {\bf x}}_3$.  In
this frame, we find
\begin{eqnarray}
  {A'_{\rm cl}}^0(x')
&=&
   \frac{q \gamma }
   { \left[ \T x^2 + \gamma^2 (\beta ct'-x'_3)^2 \right]^{1/2}} ,
\nonumber\\
   {A'_{\rm cl}}^3(x') 
&=& 
    \frac{ - q \beta \gamma }
         { \left[ \T x^2 + \gamma^2 (\beta ct'-x'_3)^2 \right]^{1/2}}
    ,
\nonumber\\
   {A'_{\rm cl}}_T(x') 
&=&
   0.  
\end{eqnarray}
For large $\gamma$, the field in the zero and
three directions are actually independent of $\gamma$ at
fixed times before the collision.  It might therefore
seem that a vector field is not Lorentz contracted, and
would not respect the assumptions of the parton model.  If
we look, however, at the field strengths rather than the
vector potential, we find a different story.  The electric
field in the three direction, for instance, is given in
the primed frame by 
\begin{equation}
E_3'(x') = \frac{ -q \gamma (\beta c t' - x_3')
}{\bigl [ \T x^2 + \gamma^2 (\beta c t' - x_3')^2
\bigr ]^{3/2}  } ,
\end{equation}
which shows a $\gamma^{-2}$ falloff.  Since
the force experienced by a test charge (or parton) in the
primed frame is proportional to the field strength rather
than the vector potential itself, the physical effects of
the moving charge are much smaller than its vector
potential at any fixed time before the collision.  This
in turn may be understood as the fact that, as $\gamma
\rightarrow \infty$, the vector potential approaches the
total derivative 
\begin{equation}
q \partial^\mu \ln (\beta c t' - x_3') .
\end{equation}
That is, for any fixed time the vector potential becomes
gauge equivalent to a zero potential.  

We can conclude from this excursion into special
relativity that factorization will be
 a more complicated issue in gauge theories than in scalar
theories.  Only for gauge invariant quantities will the
gauge-dependent, large vector potentials of
moving charges, which naively break
factorization, cancel.  So, in particular, we cannot
expect factorization to be a property of individual
Feynman diagrams, as it was in scalar theories.  On the
other hand, we should look for the solution to these
problems in the same techniques which are used to show
the gauge independence of physical quantities.  

\subsection{Ward identities}

The Ward-Takahashi identities of QED and the
Taylor-Slavnov identities of nonabelian gauge theories
ensure the perturbative unitarity of these theories.  We
shall refer to them collectively as ``Ward identities"
below.  

\begin{figure}
   \centering
   \includegraphics{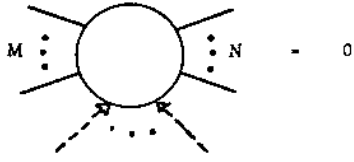}
   \caption{Ward identity.}
   \label{fig:19}
\end{figure}

Ward identities may be expressed in various forms, for
instance, as identities between renormalization
constants (the familiar $Z_1=Z_2$ of QED).  For our
purposes, however, the basic Ward identity is given
graphically by the equation
\begin{equation}
\langle N| \ {\rm T} \ \p_{\mu_1} A^{\mu_1} (x_1)
\times \cdots \times \p_{\mu_n} A^{\mu_n} (x_n)\ 
| M \rangle =0 , 
\numlabel{eq:131}
\end{equation}
where $A^\mu (x)$ is an abelian or nonabelian gauge field,
and where $M$ and $N$ are physical states, that is, states
involving on-shell fermions and gauge particles, all with
physical polarizations.  In particular, physical states do
not include ghosts.  Equation (\ref{eq:131}) will be represented
graphically by Fig.\ \ref{fig:19}, in which the scalar operator
$\p_\mu A^\mu (x)$ is represented by a dashed line ending
in an arrow.  In momentum space, this operator is
associated with a standard perturbation theory vertex in
which one gluon field is contracted into its own
momentum.  Here and below, we refer to such a gluon as
``longitudinally polarized".  Note that this is to be taken
as referring to the four-momentum.

Proofs of Eq.\ (\ref{eq:131}) are most easily given in a path
integral formulation using BRST invariance, as in, for
instance Ref.\ \cite{IZ}.  They can also be proved in a purely
graphical form, as in the original proofs of Refs.\
\cite{Hooft} and \cite{HooftVeltman}.  
Here we need not concern ourselves with the
details of these proofs, although it may be worthwhile
to exhibit the very simplest example of Eq.\ (\ref{eq:131}).  This
is the lowest order contribution to the electron
scattering amplitude with a single longitudinally
polarized photon.  At this order, we have
\begin{equation}
q^\mu {\bar u}(p+q) \gamma_\mu u(p)
= {\bar u}(p+q) \bigl [ (\st p +\st q + m) 
- (\st p  +m) \bigr ] u(p) = 0.
\numlabel{eq:132}
\end{equation}
The first equality is sometimes referred to as the
``Feynman identity", and the overall result is current
conservation at lowest order.  This is not surprising,
since classical current conservation is a consequence
of gauge invariance.  In the quantum theory, it appears
as a matrix element relation, whose validity
is ensured by the Ward identity.  A helpful
exercise is to construct the analog of Eq.\ (\ref{eq:132}) for the
scattering of a physically polarized gluon. The
graphical proof consists essentially of repeated
applications of identities like Eq.\ (\ref{eq:132}). 

Even without going into the details of the proof of Eq.\ (\ref{eq:131}), 
we can elucidate its interpretation. 
First, it is true order-by-order in
perturbation theory, although not graph-by-graph in
perturbation theory.  In addition, we may imagine
constructing a path integral in which only certain
momenta are included, for instance ultraviolet momenta
and/or momenta parallel to a given direction.  Then at a
given order, the Ward identities hold for both
internal and external lines in this restricted portion of
momentum space.  This heuristic argument may be
verified by a close look at the graphical proof of Ward
identities in Refs.~\cite{Hooft} and \cite{HooftVeltman}.  

So far, we have discussed Ward identities for the
S-matrix.  As we saw in Sect.~\ref{pdfsec}, however, we will
sometimes be interested in matrix elements involving a
gauge invariant but nonlocal operator which includes the
ordered exponential of the gauge field.  Such matrix
elements also obey Ward identities, which may be proved
by either of the methods mentioned in connection with
Eq.\ (\ref{eq:131}).  The simplest generalization of (\ref{eq:131}) to
this case is  
\begin{equation}
\langle N| {\rm T} \ \bigl (\Pi_i \p_{\mu_i} A^{\mu_i} (x_i)  
\ \P \exp \bigl \{ i g \int_0^{\infty} \! \d y^- 
A^+(0,y^-, \zerot) \bigr \} \Phi(0)
\bigr ) \  | M \rangle =0, 
\numlabel{eq:133}
\end{equation}
where in the ordered exponential $A$ refers to the gauge
field in the representation of field $\Phi$, which may be 
a fermion or gauge particle.  Equation (\ref{eq:133}), in various
guises, will be useful in our proofs of factorization.

\subsection{Singularities in Gauge Theories}

Discussions of factorization start with a catalog
of the pinch surfaces of the relevant Feynman diagrams, as described in 
Sect.~\ref{regionsec}.  
They then proceed, by power counting, to estimate the
strength of singularities encountered in each such
surface.  The same procedure may be carried out for gauge
theories, but, as we will now see, many of the regions that are nonleading
in $\phi^3$ are now leading.  Thus the results 
are much richer than in $\phi^3$ theory.

The one-loop vertex graph illustrates the origin of the 
infrared and collinear singularities.  
Ignoring overall factors, including group structure, we find that the graph
is given by
\begin{equation}
V_\mu(p,p') = 
\int \frac{\d^4 k }{ (2\pi )^4}
\ \frac{ {\bar v} (p') \gamma^\alpha ( - \st p'+ \st k + m) 
      \gamma_\mu (\st p+\st k+m) \gamma_\alpha u(p) 
  }{ 
      [(p'-k)^2-m^2+i\epsilon] \, 
      [(p+k)^2-m^2+i\epsilon] \, (k^2+i\epsilon)  
   } 
 .  
\numlabel{eq:134}
\end{equation}
The singularity structure (with one minor exception) 
is the same as in $\phi ^3$ theory; what changes
is the strength of the singularities.  To discuss the large $Q$ region, we
will consider, as before, the massless limit.  Of the solutions (\ref{eq:118})
and (\ref{eq:119}) to the Landau equations, the first of the infrared solutions 
($k ^\mu =0$) and both of the collinear solutions (\ref{eq:119}) give leading
power behavior at large $Q$, as we will see.  
These are singularities in the fully massless
theory, and, by our discussion in Sect.~\ref{regionsec}, they correspond to 
long-distance contributions when $Q$ is large.  

In addition to these singularities, there is a genuine singularity, at
$k ^\mu =0$, even when the fermion mass is nonzero.  This is an
example of the usual infrared divergences of QED and is caused by the
masslessness of the gluon.  This singularity survives the $Q \to
\infty$ limit, of course, and becomes the first of the solutions
(\ref{eq:118}).  The methods that we use to treat both the collinear
and especially the infrared singularities in the fully massless
nonabelian theory are explicitly motivated by the elegant methods
given by Grammer and Yennie \cite{GY} to treat the ordinary infrared
problem in QED.  In QED, the infrared divergences correspond directly
to the real physics of the long range of the Coulomb field and the
genuinely massless photon.  But in QCD the infrared divergences are
cut off by confinement.  Since this is a nonperturbative phenomenon,
the resulting cutoff is not easily accessible (if at all) in
perturbation theory.  Perturbative calculations must be restricted to
sufficiently short-distance phenomena so that asymptotic freedom is
useful.  The singularity structure of the massless theory is just a
convenient tool to aid in the factorization of long-distance
phenomena.

\subsection{Infrared divergences in gauge theories: the eikonal
  approximation} 
\label{gaugefactsec:IR}

We now consider the infrared singularity at $k ^\mu =0$.  Although our
ultimate aim is to treat the large $Q$ limit, our discussion will not need
to assume this limit initially.
Let us see how the integral (\ref{eq:134}) behaves
near this point.  As $k^\mu \rightarrow 0$, it is
valid to make the following two approximations in $V_\mu(p,p')$,
\begin{align}
    &\mbox{(1) Neglect $k^\mu$ compared to $m$ and $p^\mu$ in
          numerator factors, }
\nonumber\\ 
   &\mbox{(2) Neglect $k^2$ compared to $p' \cdot k$ and 
              $p' \cdot k$ in denominator factors.  }
\numlabel{eq:135}
\end{align}
Together, these two prescriptions define the ``eikonal
approximation" for the graph.  Simple manipulations show
that in the eikonal approximation $V_\mu$ is given
by
\begin{equation}
   V_\mu(p,p') = 
   -4 p \cdot p' {\bar v} (p')\gamma_\mu u(p) 
   \int \frac{\d^4 k }{(2\pi )^4}
   \ \frac{1}
          { (-2p' \cdot k+i\epsilon) \, 
            (2p \cdot k+i\epsilon) \, (k^2+i\epsilon) 
          } . 
\end{equation}
In this form it is apparent that
the $k$ integral is logarithmically divergent from the
region near $k^\mu=0$.  Notice also that, because the
numerator in proportional to $p' \cdot
p$, this divergent integral behaves as a constant at as
$p' \cdot p \rightarrow \infty$, that is, with the same
power as the elementary vertex.  This is to be contrasted with the
situation in $\phi ^3$ theory that was explained in Sect.~\ref{regionsec}.
Infrared behavior with the same power law
behavior in $Q$ as the elementary vertex
is a characteristic of theories with 
vector particles.

The eikonal approximation is, not surprisingly, closely
related to the ordered exponentials of Sect.~\ref{pdfsec:eikonal}, with
their eikonal Feynman rules.  In fact in making the
eikonal approximation (\ref{eq:135}), we are precisely replacing
fermion propagators by eikonal propagators of the type
shown in Eq.\ (\ref{eq:48}) for the parton distribution functions.
We can anticipate the importance of the eikonal
approximation by relating it to the classical discussion
given in Sect.~\ref{gaugefactsec:classical}.  Consider a gluon of momentum
$k^\mu$ interacting with an eikonal line in the $v^\mu$
direction.  The only component of the gluon momentum
$k^\mu$ on which the eikonal propagator depends is $v \cdot
k$, and the only component of the gluon polarization
$\epsilon^\mu(k)$ on which the eikonal vertices depends is
 $v \cdot \epsilon$.  So, as far as the eikonal line is
concerned, the gluon acts in the same way as a
fictitious gluon of momentum $(v \cdot k) u^\mu$ and
polarization $(v \cdot \epsilon) u^\mu$, where $u^\mu$ is
any vector for which $u \cdot v=1$.  But this fictitious 
gluon is longitudinally polarized.  That is, any gluon
interacts with an eikonal line in the same way as a
longitudinally polarized, and therefore unphysical,
gluon.  But we have argued above that such gluons,
although they may be expected to break factorization on a
graph-by-graph basis, should be consistent with it in
gauge invariant quantities.

When $Q$ is large, we consider not just the actual infrared singularity at
$k ^\mu =0$, but the whole infrared region $k ^\mu  \ll Q$. That is, 
as $Q \to
\infty$, we consider the region $k ^\mu /Q \to 0$.  It is possible for the
different components of $k ^\mu /Q$ to go to zero at such different rates
that the eikonal approximation fails.  Since we will rely on this
approximation in proving factorization we will need to evade this failure.

To get an idea of what is involved, let us return to Eq.\
(\ref{eq:134}), and justify the eikonal approximation Eq.\
(\ref{eq:135}) in this simplest of cases.  Failure of Eq.\
(\ref{eq:135}) is caused by failure of the second of the
approximations of which it is comprised: dropping factors of $k^\mu$
in the numerator is a safe bet, because, as we have seen, the factors
$p^\mu$ combine to form large invariants.  So, the issue is whether or
not we may neglect $k^2$ compared to $p \cdot k$ and $p' \cdot k$.
This is nontrivial, because it is easy to find vectors $k^\mu$ for
which $p \cdot k$ and $p' \cdot k$ are small, while $k^2$ remains
relatively large.  This will be the case whenever its spatial momentum
transverse to the $\bf p$ and $\bf p'$ directions is large,
\begin{equation}
    \kt ^2 \sim -k^2 \sim p \cdot k,\ p' \cdot k . 
\numlabel{eq:137}
\end{equation}
This region was called the ``Glauber" region in Ref.\ \cite{BBL}. It
is easy to check that in this region the $k^\mu$ integral of Eq.\
(\ref{eq:134}) is logarithmically divergent.  If we were to put in a
gluon mass (as is consistent for an abelian theory), the divergence
would disappear, but we would still have a contribution from the
region (\ref{eq:137}) to the leading-power behavior at large $Q$, that
is, a contribution of order $Q^0$ times logarithms.

Does this mean that the
eikonal approximation is wrong?  In fact it does not.  To see
this, we appeal to our freedom to deform
momentum space contours.  Suppose,
for definiteness that $\bf p$ and $\bf p'$ are in the $\pm
\bf z$ directions, respectively, and that $|{\bf p}|=|{\bf
p}'|$.  We then change variables from the set $\{k_0,{\bf
k}_i \}$ to the set $\{\kappa^{\pm} =2^{-1/2}(k_0 \pm ({\bf
p} \cdot {\bf k})/\omega_p), \ \kt \}$. (These
become light cone variables in the high energy limit.) 
Then in the region defined by Eq.\ (\ref{eq:137}), the $k ^\mu \to 0$ singularity of
(\ref{eq:134}) is given by   
\begin{equation}
\int \d^2\kt \frac{1 }{(-\kt ^2 + i \epsilon ) }
\d \kappa^+ \d \kappa^- 
\frac{1 }{( 2^{3/2} \omega_p \kappa^- -\kt ^2 + i\epsilon) \, 
  \ ( -2^{3/2} \omega_p' \kappa^+ -\kt ^2 + i\epsilon)
} ,
\end{equation}
where the variables $\kappa^{\pm}$ appear in only
one denominator each.  In this form we see explicitly that
the $\kappa^{\pm}$ integrals are not trapped in the region 
$\kappa^{\pm} \ll \kt$, since they
each encounter only a single pole in this region.  As
a result, these contours may be deformed away from the
origin into the region $|\kappa^{\pm}| \sim |\kt|$.  
But in this region the eikonal
approximation is valid, provided only that $|k ^\mu | \ll Q$.
So, we may relax our
criteria for the eikonal approximation to include the
possibility that, even if it is not valid everywhere
along the undeformed contours, these integrals can be
deformed in such a way that it holds along the deformed
contours.

\subsection{Collinear divergences and choice of gauge}

In addition to infrared divergences, we have to consider collinear
divergences in the massless limit.  The nature of the collinear
contributions 
to leading regions depends
on the gauge, as we will now show.

Consider the gluon propagator in a axial gauge $n \cdot
A=0$.  It has the form 
\begin{equation}
D_{\mu\nu}(k)=\frac{-i }{k^2+i\epsilon} 
\left( g_{\mu\nu} - 
\frac{k_\mu n_\nu + n_\mu k_\nu }{n \cdot k}
+ \frac{n^2 k_\mu k_\nu }{(n \cdot k)^2 } \right)  ,
\end{equation}
which satisfies
\begin{equation}
k^\mu D_{\mu\nu}(k) = i \left( \frac{n_\nu }{n \cdot k}
- \frac{n^2 k_\nu }{(n \cdot k)^2 } \right)  .
\numlabel{eq:140}
\end{equation}
Such a gauge is ``physical" because its
propagator has no particle pole when contracted into any vector
proportional to its momentum.  Another way of putting
this is that in such a gauge longitudinal degrees of
freedom do not propagate.  This is to be contrasted with
a covariant gauge like the Feynman gauge, for which
$k^\mu D_{\mu\nu}(k)=-i k_\nu/k^2$.  As a result, leading
regions in which longitudinal degrees of freedom
propagate are present in covariant gauges but absent in
physical gauges.  Let us see what this means in
practice. To do so, we turn again to the vertex
correction, (\ref{eq:134}).

\begin{figure}
   \centering
   \includegraphics{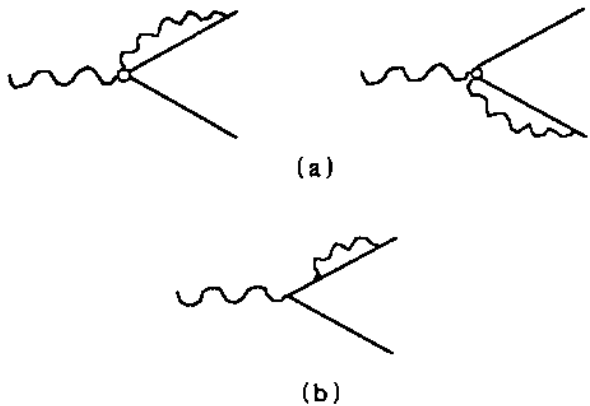}
   \caption{Leading collinear reduced diagrams at one loop: 
            (a) covariant gauge,  (b)  physical gauge.  }
   \label{fig:20}
\end{figure}

We have already stated the locations of the collinear singularities, in
Eq.\ (\ref{eq:119}).  The two possibilities are that $k ^\mu $ is proportional to $p
^\mu $ and that it is proportional to ${p'}^\mu $.  
The corresponding reduced diagrams are
shown in Fig.\ \ref{fig:20}(a).   

By doing 
the $k^-$ integral by contour integration, we
easily find that in Feynman gauge the contribution of momenta close to the
singularity where $k$ is proportional to $p$ 
is given by 
\begin{equation}
V_\mu(p,p') \approx \frac{i }{(2 \pi )^3}
\int_{-p^+}^0 \frac{ \d k^+ }{k^+ }
{\bar v}(p') 
\gamma_\mu u(p) \ g_{-+} \ (1 + k^+ / p^+)
\frac{\d^2 \kt  }{ \kt ^2} ,
\numlabel{eq:141}
\end{equation}
and similarly in the second region.  We have exhibited
explicitly the numerator of the gluon propagator.  For fixed $k^+ =
-x p^+$, the $\kt $ integral diverges, and is leading
power, that is, independent of $Q$.  This is the
collinear divergence.  (There is an additional infrared
divergence as $k^+$ vanishes; this region we have already discussed.
This result for the $Q$ dependence 
is known as a ``Sudakov" double logarithm; it is 
associated with the overlap of collinear and infrared
divergences.)  

These
regions, summarized by the reduced graphs
of Fig.\ \ref{fig:20}(a) in which two collinear lines attach
to a hard subdiagram, would not be leading in $\phitheory $,  because of
the lack of the numerator factor.  
Note, however, that in the numerator of the gluon propagator, the term
which gives the leading behavior in the collinear region is
$g_{-+}$.  Since the gluon is moving, by assumption,
parallel to $p^\mu$, which is in the plus direction, this
corresponds to an unphysical polarization at the vertex
adjacent to the antiquark line.  Thus the collinear
divergence is associated with a longitudinally polarized
gluon, and we might expect it to be absent in a physical
gauge --- at least in this particular diagram.

To verify this, we can compute $V_\mu$ in an
axial gauge.  The leading term in Eq.\ (\ref{eq:141}) is
then replaced by 
\begin{eqnarray}
   V_\mu(p,p') 
&\approx&
   \frac{i }{(2 \pi )^3}
   \int_{-p^+}^0 \frac{ \d k^+ }{ k^+ }
   {\bar v}(p') 
   \gamma_\mu u(p) \ (1 + k^+ / p^+)
\nonumber\\
   && \times
   \int \frac{\d^2 \kt  }{ \kt ^2} 
   \left( g_{-+} - \frac{ n_-k_++k_-n_+ }{ n \cdot k } 
           + \frac{k_-k_+ }{(n \cdot k)^2 }
   \right)
 , 
\numlabel{eq:142}
\end{eqnarray}
Using Eq.\ (\ref{eq:140}) (and remembering that $k_\pm=k^\mp$), we
easily check that the collinear divergence in (\ref{eq:141}) is
absent in (\ref{eq:142}), and that the vertex diagram therefore lacks the
Sudakov double logarithm in axial gauge.  Of course,
since the theory is gauge invariant, the corresponding physics, and in
particular 
the double logarithm, has to show up
somewhere, and in axial gauge it occurs in the one loop
fermion self energy.  We leave it as a simple but
instructive exercise to check that this is indeed the
case.  Thus, in axial gauge the reduced diagrams of Fig.\ \ref{fig:20}(a) 
do not correspond to a collinear divergence, while
that of Fig.\ \ref{fig:20}(b) does.  We emphasize here the fact that in 
the Feynman gauge the jets are one-particle irreducible,
while in the axial gauge they are reducible.  As we shall see
below, this result generalizes to all orders.  In this
sense, the gauge theory in a physical gauge behaves, from
the point of view of reduced diagrams for collinear lines, 
like $\phitheory $.  

This suggests that an axial gauge is the most appropriate one for proving
factorization.  However, the singularities at $n \cdot k =0$ cause a lot of
trouble.  In the first place they obstruct \cite{Bo,CSS2,CSS4}
the contour deformations that we
have already seen are essential to demonstrating factorization; this is
equivalent to saying that the 
singularities violate relativistic causality on a graph-by-graph basis.  
Furthermore, the singularities have to be defined by some kind of principal
value prescription, and it is difficult to ensure that the products of
these singularities that occur in higher order graphs can be defined
properly \cite{difficulties}.

\subsection{Power counting for gauge theories}

As in the scalar theory, we must use power counting to
identify those pinch surfaces which actually give leading
regions.  Again, this approach is discussed in detail in
Ref.\ \cite{St78}.  Here, we once again quote the general
result.
Assuming the eikonal approximation, for any leading
region with reduced diagram $R$, we compute the infrared degree of
divergence, $\omega (R)$, analogous to Eq.\ (\ref{eq:122}),
\begin{equation}
\omega(R) = 4S + 2C - 2I_S - I_C + N_2 + \half N_3 + F.  
\end{equation}
We have assumed a space-time dimension 4.
As in (\ref{eq:122}), $S,\ I_S,\ C$, and $I_C$ are, respectively the 
numbers of soft loops and lines, and collinear loops and
lines in $R$ at the associated pinch surface.  $N_2$ is
the number of two-point functions in $R$,
while $N_3$ the number
of three-point functions all of whose external lines are
in the same jet.  $F$ is derived from the numerator factors where soft lines
connect to collinear lines.  
It is positive except when all soft lines connecting to collinear
lines are gluons.  
The suppression terms, $\half N_3$ and $F$, are the only
differences from Eq.\ (\ref{eq:122}).  

We note that the $N_3$ term 
is present diagram-by-diagram in physical gauges \cite{EarlyProofs}, 
but that in Feynman
gauge the computation holds in general only when gauge invariant
sets of diagrams are combined for the hard scattering
subdiagrams.  In fact, in covariant gauges,
individual diagrams may be much more divergent in the
presence of infrared and collinear interactions than is
the cross section, and may even grow with energy
\cite{LabasSt}.  This is a consequence of the well-known fact
that unitarity bounds on energy
growth are only a property of gauge invariant sets of
diagrams.
 
\subsection{General leading regions}

The general leading regions for \epem-annihilation 
processes, for deeply inelastic
scattering and for hard inclusive hadron-hadron
scattering are quite analogous to those for the
$\phitheory $ theory.  The basic difference is that lines
which participate in infrared logarithms must be added
to the corresponding reduced diagrams. 
 
\begin{figure}
   \centering
   \includegraphics{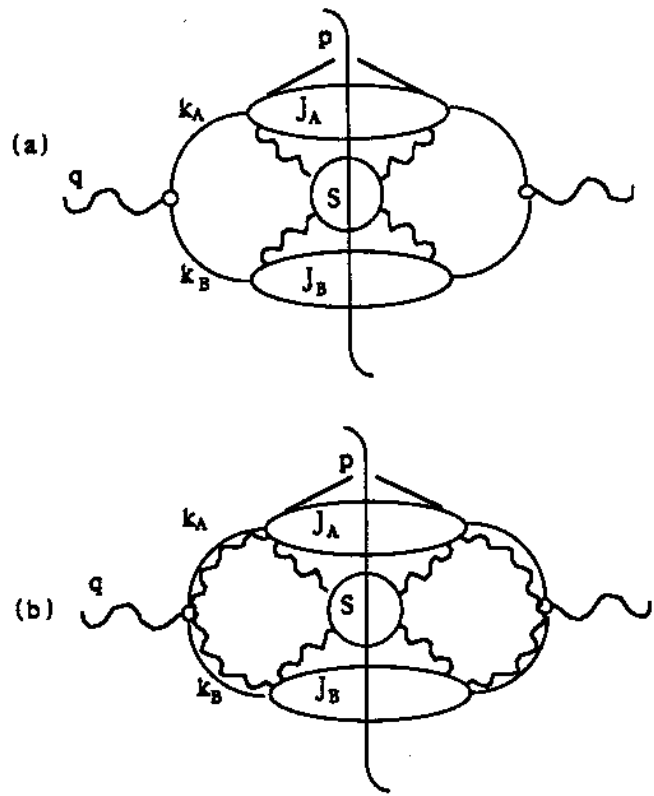}
   \caption{ Typical leading regions for annihilation processes.  (a)
     physical gauge.  (b) covariant gauge.  The most general leading
     region has the possibility of extra jets beyond the two shown here.
   } 
   \label{fig:21}
\end{figure}

Fig.\ \ref{fig:21}(a) shows a general
leading region for a single particle inclusive  cross
section in \epem\ annihilation for a physical gauge, and
Fig.\ \ref{fig:21}(b) for a covariant gauge.

Compared to the leading regions for $\phitheory $, summarized in
Fig.\ \ref{fig:14}, which include only jets of collinear lines and hard
subdiagrams, to get Fig.\ \ref{fig:21}(a) we simply add a ``soft"
subdiagram, consisting of lines whose momenta vanish at
the pinch singular surface in question.  The soft subdiagram
contains in general both soft gluon lines and soft quark
loops (as well as ghost loops in covariant gauges); its
external lines, however, are always gluons.  These
external gluons always attach to (energetic) lines and
not to the hard subdiagrams.  The lines attaching jet
subdiagrams to the hard subdiagrams may be either gluons
or fermions, but at leading power only a single line from
each jet enters a given hard subdiagram, just as in
$\phitheory $.  The physical picture is also the same as in
$\phitheory $; several hard particles recede from a hard
scattering at the speed of light, forming jets by their
self-interactions.  These particles can never interact
with each other except by transfer of soft momenta $|k^\mu | \ll Q$.  
The
presence of vector particles in the gauge theory,
however, does give leading power contributions from 
the exchange of soft particles.  Since
$k^\mu /Q \approx 0$ for each of these particles, they do not affect
those of the Landau equations, (\ref{eq:112}) and (\ref{eq:113}), 
which involve only the jet
subdiagrams.  Of course, it should be kept in mind that
the soft lines have zero momentum only at the exact pinch
singular surface. Feynman integrals get contributions from
an entire region near this surface where the soft momenta
are much smaller than a typical energy of a jet, but may
approach a nonzero fraction of that energy.  

For a covariant gauge, the leading regions are
essentially the same, except that, just as in the
one-loop case, arbitrary numbers of longitudinally
polarized gluons may attach the jet to the hard part, as
shown in Fig.\ \ref{fig:21}(b).  

\begin{figure}
   \centering
   \includegraphics{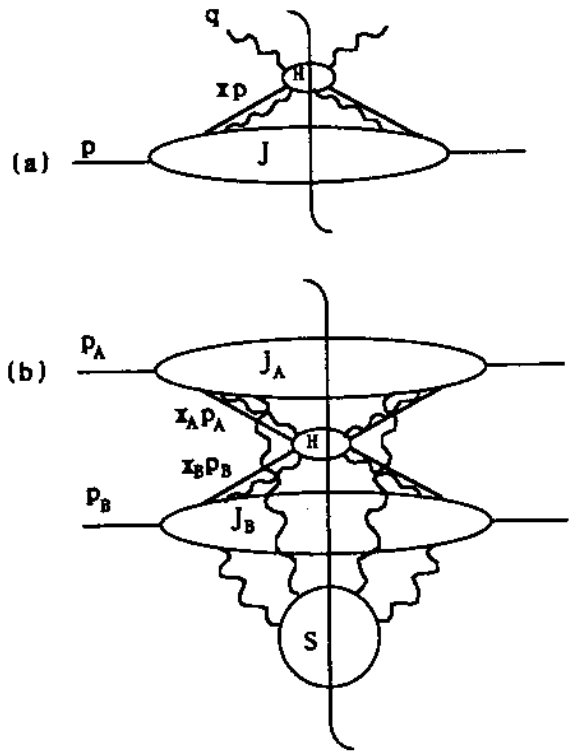}
   \caption{Leading regions in Feynman gauge for (a) inclusive deeply
      inelastic scattering and (b) Drell-Yan  cross sections.  } 
   \label{fig:22}
\end{figure}

Figures \ref{fig:22}(a) and \ref{fig:22}(b) show general leading regions for
inclusive deeply inelastic scattering, and
the Drell-Yan cross sections in Feynman gauge.  
As with $\phitheory $, the sum over final states eliminates
pinch singularities involving final state jets.  The
remaining on-shell lines make up the jets associated with
the incoming particles 
including soft exchanges within and between the jets. 
In covariant gauges, longitudinally polarized gluons may
connect the jets to the hard part.

\section{Factorization Proofs in Gauge Theories}
\label{gaugeproofsec}

We are now ready to discuss the extension of 
factorization theorems to gauge theories, for the basic
cross sections discussed above: deeply inelastic
scattering, single particle inclusive annihilation and
Drell-Yan production.  Each of these will require new
reasoning relative to the scalar case.  

Compared to the proof in Sect.~\ref{subtrsec} for 
$\phitheory $, our treatment of factorization in gauge
theories will be much more modest.
  Rather than derive closed
expressions for the factorized forms in terms of explicit
subtraction operators, we will deal with the cross
sections on a region-by-region basis.  We
will show that an arbitrary leading region either
contributes to the factorized form of the cross section,
or cancels to leading power when gauge invariant sets of
diagrams are combined.  

\subsection{ Deeply inelastic scattering and collinear factorization}
\label{gaugeproofsec:DIS.collinear}

We start with the deeply inelastic scattering cross section,
$h(p)+\gamma^*(q) \rightarrow X$, with $p^\mu (q^\mu)$ being the
momentum of the incoming hadron $h$ (virtual photon
$\gamma^*$).  Here, as we shall see, the question of
factorization reduces to a treatment of collinear singularities
associated with unphysically polarized gluons.
 
Fig.\ \ref{fig:22}(a) illustrates  the leading regions in a gauge theory
for diagrams that contribute to the structure function tensor
$W^{\mu\nu}(q,p)$. There is a single jet $J$, in the direction
of the incoming particle, and a single hard subdiagram
$H^{\mu\nu}$, containing the hard scattering. Divergences
associated with final state interactions cancel because of
unitarity in the sum over different final state cuts of the same
Feynman graph. Thus we have not included regions in which soft
gluons from the jet $J$ interact with the outgoing particles in the
hard part $H$, even though such regions can give leading
contributions to individual cut graphs.
 
In Feynman gauge, the hard subdiagram is connected to the single
jet by more than one collinear line.  This makes the transition
to the convolution form of Eq.\ (\ref{eq:2}) more complex than in the
scalar case.
 
In physical gauges, the reduced diagram corresponding to an
arbitrary leading region has the same form as for the scalar
theory, Fig.\ \ref{fig:9}.  This simplification is the reason that most
of the original arguments for factorization were given in
physical gauges \cite{EarlyProofs}, where essentially the same
procedure can be used as in Sect.~\ref{subtrsec}. However, it
is important to show how the proof may be carried out in the
covariant gauges, for two reasons. First, as mentioned above, there are
difficulties associated with the unphysical singularities encountered in
physical gauges, which have not been fully understood
yet \cite{difficulties}. Although these are presumably of a
technical nature, and not associated with the content of
factorization, it is surely desirable not to be completely
dependent on this presumption.  Second, physical gauges, because
of their noncovariance, are ill-suited to proofs of
factorization in the crucial case of hadron-hadron scattering.
So, in the interests of generality, we shall discuss deeply
inelastic scattering in the Feynman gauge.  These issues were
not treated in Ref.\ \cite{EarlyProofs}.
 
Let us consider a typical cut Feynman diagram $G^{(C)}$, where
$C$ labels the cut, in the neighborhood of a leading region
$L$.  $L$ is specified completely once we specify how the graph
$G$ is to be decomposed into the subgraphs $J$ and $H$. We shall
write the contribution from region $L$ to $G^{(C)}$ as
$G^{(L,C)}$.
 
Referring to Fig.\ \ref{fig:22}(a), we see that our problem is to organize
the set of longitudinally polarized lines which attach the jets
to the hard parts.  Suppose a set of $n$ gluons of momentum
$l_i^{\alpha_i}$ attaches to the hard part $H$ to
the left of the cut, along with a physically polarized parton of
momentum $k^\mu-\sum_i l_i^\mu$.  Similarly, suppose a set of
$n'$  longitudinally polarized gluons ${l'}_j^{\beta_j}$
attaches to $H$ on the right of the cut, along with a
physically polarized parton of momentum $k^\mu-\sum_j
{l'}_j^\mu$.  Each momentum $l^\mu_i$ is parallel to the
external momentum $p^\mu$ and flows into the hard part.  Each $
{l'}^\mu_j$ is also parallel to $p^\mu$, but flows out.  We sum
over all cuts of the original graph $G$ consistent with this
leading region, with fixed $n$ and $n'$.   We can now represent
the sum over these allowed cuts $C$ of $G^{(L,C)}$ as
\begin{align}
    {\cal G}^{(L)}(\mu) 
{}={}&
    \sum_C\ G^{(L,C)}
\nonumber\\
{}={}&
  \int \frac{\d^4 k }{(2\pi )^4 }\
        \prod_i \int \frac{ \d^4l_i }{(2\pi )^4 }
        \prod_j \int \frac{ \d^4 {l'}_j }{ (2\pi )^4 }
\nonumber\\
& \times
    \sum_{C_H}
    H^{(C_H)}\!\!\left( q^\nu; 
                        k^\mu-\Sigma l_i^\mu,
                        \{l_i^{\alpha_i}\};
                        k^\mu-\Sigma {l'}_j^\mu,
                        \{{l'}_j^{\beta_j}\}
                 \right)^{\{\mu_i,\nu_j\}}_{\eta,\eta'} 
\nonumber\\
&
     \times \sum_{C_J}
     J^{(C_J)}(p^\nu;k^\mu-\Sigma l_i^\mu, \{l_i^{\alpha_i}\};
     k^\mu-\Sigma {l'}_j^\mu,\{{l'}_j^{\beta_j}\})
     _{\{\mu_i,\nu_j\}}^{\eta,\eta'}, 
\numlabel{eq:144}
\end{align}
where $\mu_i$ and $\nu_j$ are polarization indices for the $l_i$
and ${l'}_j$, respectively, and $\eta$ and $\eta'$ are the
polarization indices associated with the physical partons
attaching to the hard part on either side of the cut.   Of
necessity, the sum includes only those cuts which preserve $n$
and $n'$, and we note that it breaks up into independent sums
over the cuts of the hard part and of the jet.
 
The integrals in (\ref{eq:144}) are restricted to the neighborhood of
the region $L$.  We implicitly introduce a variable $\mu$, to
set the scale of $L$.  The integration region in (\ref{eq:144}) is set
by requiring, for instance, that lines within $H$ have
transverse momenta of order at least $\mu$, while those in $J$
have transverse momenta of $\mu$ or less.  $\mu$ will later be
identified with the renormalization scale for the parton
distribution.
 
Because all lines in $H$ are, by construction, far off the mass
shell, we replace the momenta of all its external particles by
lightlike momenta in the corresponding jet direction. Then, if
we keep only leading polarization components, the extra
collinear gluons  which attach the hard part to the jet are
exactly longitudinally polarized.  Corrections are suppressed by
a power of $q^2$.  To formalize this approximation, we introduce
the vectors
\begin{equation}
v^\mu = g^{\mu } _{+}, \ u^\mu = g^{\mu}_{-} ,
\end{equation}
and define
\begin{equation}
u \cdot l_i^\alpha=\lambda_i , \
u \cdot k^\alpha=k , \
u \cdot {l'}_i^\alpha=\lambda'_i.
\end{equation}
In terms of these variables, the approximation is
\begin{multline}
\sum_{C_H}  H^{(C_H)}(q^\nu;k^\mu-\Sigma l_i^\mu,
   \{l_i^{\alpha_i}\}; k^\mu-\Sigma {l'}_j^\mu,\{
   {l'}_j^{\beta_j}\}) ^{\{\mu_i,\nu_j\}}_{\eta,\eta'}
\\
 \rightarrow
   {\hat H}(q^\nu; (k-\Sigma\lambda_i)v^\mu,
   \{\lambda_i v^{\alpha_i}\};(k-\Sigma\lambda'_j)v^\mu,
   \{\lambda'_j v^{\beta_j}\})
   _{\eta,\eta'}
   \prod_i u^{\mu_i}\ \prod_j u^{\nu_j} , 
\numlabel{eq:147}
\end{multline}
where
\begin{multline}
   {\hat H}(q^\nu; (k-\Sigma\lambda_i)v^\mu,
   \{\lambda_i v^{\alpha_i}\};(k-\Sigma\lambda'_j)v^\mu,
   \{\lambda'_j v^{\beta_j}\})  _{\eta,\eta'}
\\
\begin{aligned}
 \equiv &
   \sum_{C_H}
   H^{(C_H)}(q^\nu;k v^\mu-\Sigma \lambda_i v^\mu,
   \{\lambda_i v^{\alpha_i}\}; k v^\mu-\Sigma \lambda'_j v^\mu,\{
   \lambda'_j v^{\beta_j}\})
   _{\eta,\eta'}
   ^{\{\gamma_{i},\delta_{j}\}}
\\
& \times
   \prod_{i'} v_{\gamma_i}\  \prod_{j'} v_{\delta_j} . 
\end{aligned}
\end{multline}
This replacement is analogous to the operator $P$ introduced for the 
scalar theory in Sect.~\ref{subtrsec}.
 
We will now show that the unphysical polarizations of the extra
gluons can be used to factor them from the hard part.  The hard
part will become a function of only the $total$ longitudinal
momentum flowing between it and the jet, as is appropriate for a
factorized form, while the longitudinally polarized gluons will
couple to an eikonal line, which we associate with the jets.
 
\begin{figure}
   \centering
   \includegraphics{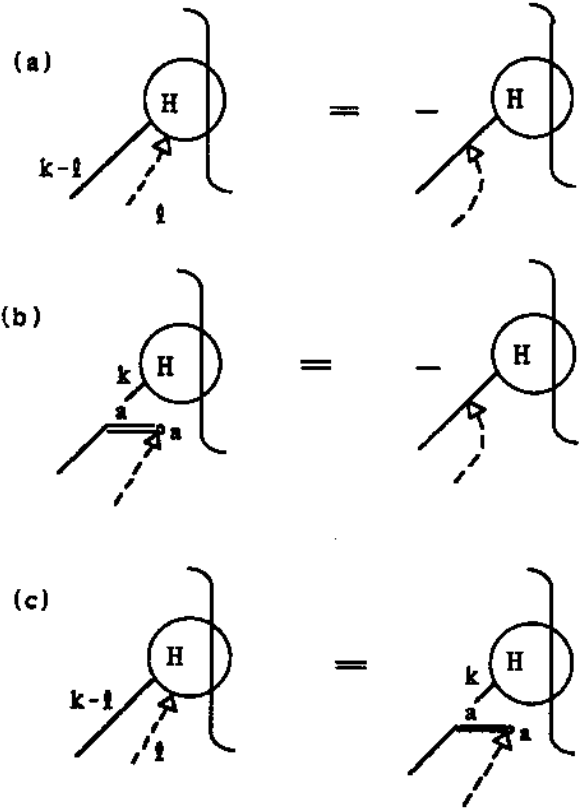}
   \caption{Ward identities for a single gluon. Group sums follow
     repeated indices. (a) Identity for hard part; (b) Eikonal identity;
     (c) Factorization of the gluon.} 
   \label{fig:23}
\end{figure}

Let us show this result first for the diagram on the left-hand
side of Fig.\ \ref{fig:23}(a), with a single longitudinally polarized
gluon of momentum $l^\mu$, which attaches to the hard part along
with the physically polarized parton of momentum $k^\mu-l^\mu$.
(We shall refer to particles by their momentum labels.)   If we
apply the Ward identity, Eq.\ (\ref{eq:131}) to this set of diagrams, we
find the result on the right hand side of Fig.\ \ref{fig:23}(a). The
left-hand side of Fig.\ \ref{fig:23}(a) would vanish, except that the
diagram on the right-hand side, in which the gluon $l^\mu$ is
attached to the physical parton, is not included in $H$ by
construction.  But now consider the identity shown in
Fig.\ \ref{fig:23}(b).  Here we consider a diagram in which the unphysical
line ends in an eikonal line, while $H$ has a single (physically
polarized) external line from $J$, which carries the total jet
momentum $k$.  The right hand side of Figs.\ \ref{fig:23}(a) and
\ref{fig:23}(b) are the same, and we derive the identity of
Fig.\ \ref{fig:23}(c), in which the longitudinally polarized gluon has
been factored onto an eikonal line moving in the opposite
direction from the $A$-jet.
 
To be careful, we should note that in each individual cut
diagram of Fig.\ \ref{fig:23}, the intermediate states are not physical
states, but rather states including on shell
gluons with unphysical polarizations and ghosts.  Once graphs
for a given cut are summed over, however, we may replace the
unphysical states by physical ones \cite{Hooft}. So we may,
without loss of generality, treat the matrix elements as though
they were between physical states.
 
\begin{figure}
   \centering
   \includegraphics{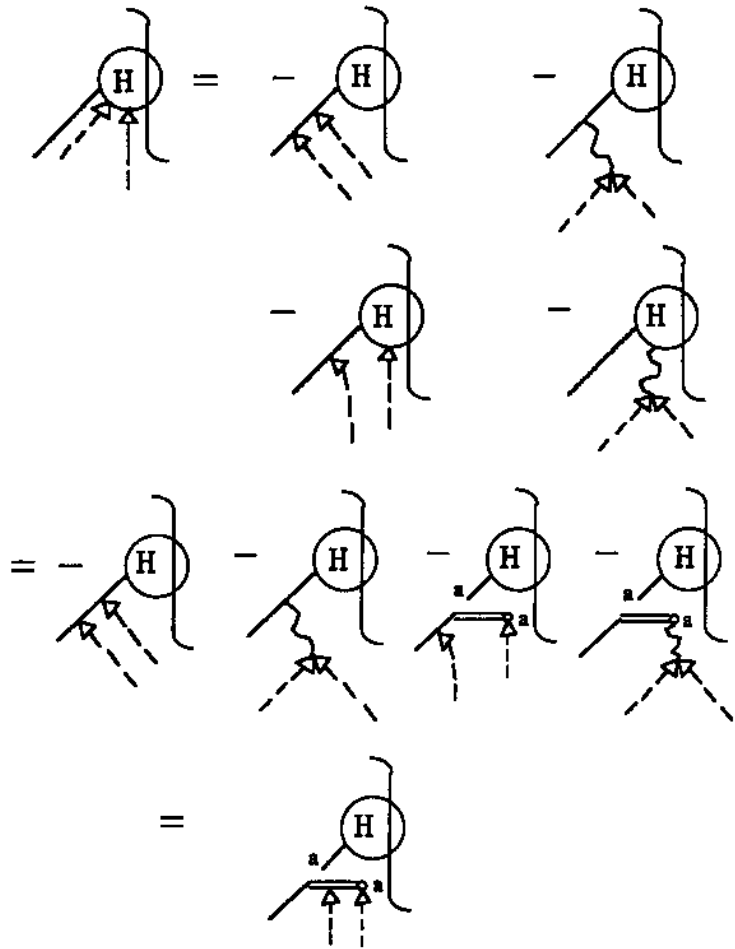}
   \caption{Application of Ward identities to two collinear gluons.}
   \label{fig:24}
\end{figure}
 
\begin{figure}
   \centering
   \includegraphics{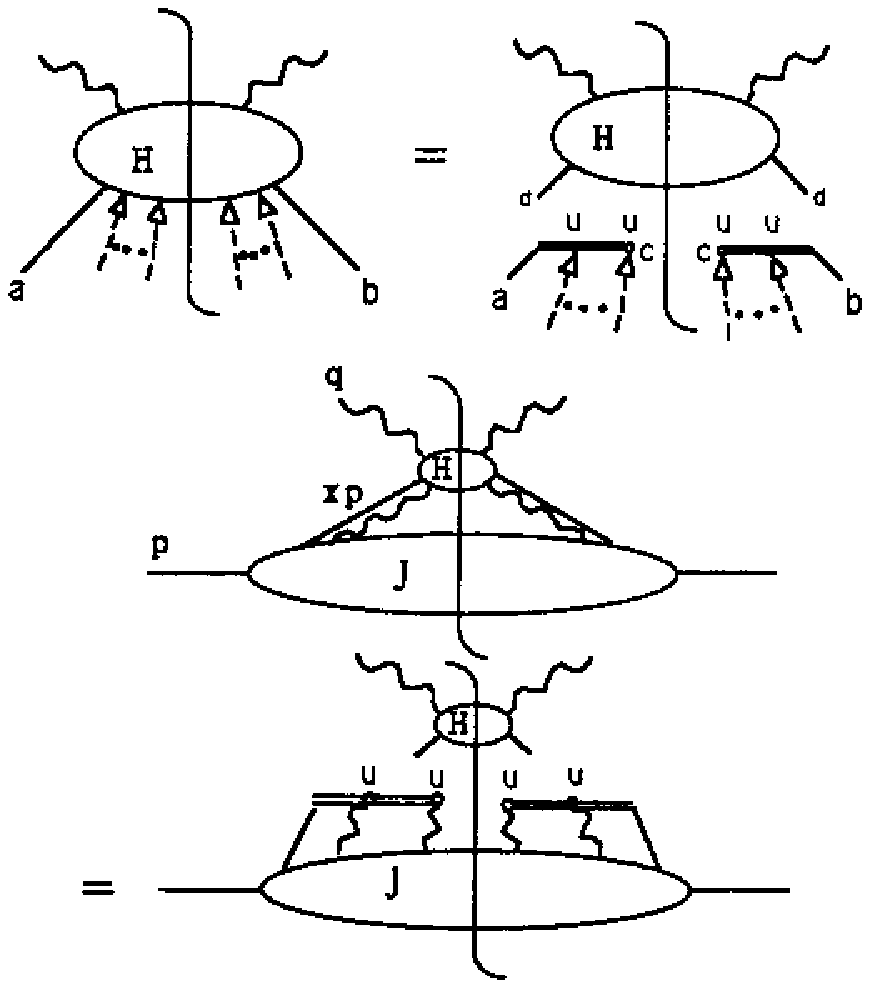}
   \caption{Factorization of collinear gluons.}
   \label{fig:25}
\end{figure}

The extension of this reasoning to two gluons is
straightforward.  We use the identity of Fig.\ \ref{fig:24}, analogous
to Fig.\ \ref{fig:23}. On the right hand side of the first
equality in Fig.\ \ref{fig:24} we have two diagrams in which
only the physical parton attaches to the hard part, and also two
diagrams in which one gluon is still attached to
the hard part. (Diagrams in which the two gluon lines are
interchanged are not indicated explicitly in the figure.) In a
covariant gauge, Lorentz invariance requires that the gluon
entering the hard part in diagram $4$ also be longitudinally
polarized (it has no other vectors on which to depend).  Thus we
can apply the result of Fig.\ \ref{fig:23} for the single gluon entering
the hard part in diagrams $3$ and $4$.  The result is shown in
the second and third equalities in Fig.\ \ref{fig:24}.  This inductive
approach can clearly be extended to arbitrary order, and we
derive Fig.\ \ref{fig:25} for a general leading region.
 
This gives the overall replacement
\begin{eqnarray}
&&
 {\hat H}(q^\nu; (k-\Sigma\lambda_i)v^\mu,
\{\lambda_i v^{\alpha_i}\};(k-\Sigma\lambda'_j)v^\mu,
 \{\lambda'_j v^{\beta_j}\})
_{\eta,\eta'}
\prod_i u^{\mu_i}\ \prod_j u^{\nu_j}
\nonumber\\
&&\hspace*{2cm} \times
J^{(C_J)}(p^\nu;k^\mu-\Sigma l_i^\mu, \{l_i^{\alpha_i}\};
k^\mu-\Sigma {l'}_j^\mu,\{{l'}_j^{\beta_j}\})
_{\{\mu_i,\nu_j\}}^{\eta,\eta'}\nonumber\\
&& \rightarrow
 {\tilde H} (q^\nu,k v^\mu)_{\eta,\eta'}
 \  {\cal E}(u,\{\lambda_i\})^{\{\mu_i\}}
 {{\cal E}^*}(u,\{\lambda'_j\})^{\{\nu_j\}}
\nonumber\\
&&\hspace*{2cm} \times
J^{(C_J)}(p^\nu;k^\mu-\Sigma l_i^\mu, \{l_i^{\alpha_i}\};
k^\mu-\Sigma {l'}_j^\mu,\{{l'}_j^{\beta_j}\})
_{\{\mu_i,\nu_j\}}^{\eta,\eta'} ,
\numlabel{eq:149}
\end{eqnarray}
where ${\cal E}(u,\{\lambda_i\})^{\{\mu_i\}}$ is a lightlike eikonal
line in the $u^\mu$ direction, coupled to $n$ gluons $l_i$, and
similarly ${\cal E}^* (u,\{ \lambda'_j\})^{\{\nu_j\}}$ is an eikonal
in the same direction coupled to the $n'$ gluons ${l'_j}$.  It is
natural to group the eikonal lines with the jet, and to define
(compare Eq.\ (\ref{eq:75}))
\begin{eqnarray}
{\tilde J}(\xi)^{\eta,\eta'}
&=&
 \int \frac{\d^4k }{(2\pi )^4}\
\ \delta (\xi  p \cdot u / k \cdot u - 1)\
\prod_i \int \frac{\d^4 l_i }{(2\pi )^4}\
\prod_j \int \frac{\d^4{l'}_j }{ (2\pi )^4 }
\nonumber\\
&& \times
{\cal E}(u,\{\lambda_i\})^{\{\mu_i\}}
\ {{\cal E}^*(u,\{\lambda'_j\})}^{\{\nu_j\}}
\nonumber\\
&& \times \sum_{C_J}
J^{(C_J)}(p^\nu;k^\mu-\Sigma l_i^\mu, \{l_i^{\alpha_i}\};
k^\mu-\Sigma {l'}_j^\mu,\{{l'}_j^{\beta_j}\})
_{\{\mu_i,\nu_j\}}^{\eta,\eta'} . 
\numlabel{eq:150}
\end{eqnarray}
 
The function $\tilde J$ is linked to the remaining hard
part ${\tilde H}(q^\nu,kv^\mu)$ through only the variable
$\xi$ and the physical polarization indices
$\eta$ and $\eta'$.   Using (\ref{eq:150}) and (\ref{eq:149}) in  (\ref{eq:144}),
we have
\begin{equation}
{\cal G}^{(L)}(\mu) = \int_x^1 \frac{\d \xi }{\xi }\
{\tilde H} (q^\nu,\xi p\cdot u\, v^\mu)_{\eta,\eta'}
{\tilde J}(\xi)^{\eta,\eta'},
\numlabel{eq:151}
\end{equation}
where $x = -q^2/2p\cdot q$ as in Sect.~\ref{introsec}.
Thus in each leading region the cross
section factorizes into an ultraviolet contribution times
a contribution to the distribution of the physical
parton which remains attached to the hard part.  It
is clear that we get every leading region for the parton
distribution in this way.  Note that when we sum over
all leading regions, the perturbative sums for the
hard part ${\tilde H}$ and the factorized jet ${\tilde J}$ are
completely independent.
 
Equation (\ref{eq:151}) contains most of the physics of factorization for deeply
inelastic scattering, but a few more steps are required to obtain
the result (\ref{eq:2}).  First, one argues using Lorentz invariance that
for unpolarized incoming hadrons the hard part are both diagonal
in the spin indices $\eta,\eta'$.  Thus we can sum over the spin
of the partons leaving the jet part and average over the spin of
the parton entering the hard part.  This decouples the two factors
in spin space.  We then sum over all graphs and over the leading
regions $L$ for each graph.  The result is
\begin{equation}
W^{\mu\nu} \sim \sum_a\int_x^1 \frac{\d \xi }{\xi }\
{\cal F}_{a/A}(\xi,\mu)
{\cal H}_a^{\mu\nu}(q^\nu,\xi p\cdot u\, v^\mu,\mu).
\numlabel{eq:152}
\end{equation}
Here a sum over parton types $a$ is indicated, ${\cal F}$ is the
jet part summed over graphs, leading regions, and spins, and
${\cal H}$ is the hard part summed over graphs and leading
regions and averaged over the incoming spins.  Both ${\cal F}$
and ${\cal H}$ depend on the parameter $\mu$ that sets the scale
for the leading regions.
 
We can relate the functions ${\cal F}$ to the \MSbar\
parton distributions $f_{a/A}$ in the following manner.  We
note first that we can carry out exactly the same
factorization procedure for parton distributions defined as
in eqs.~(\ref{eq:43}) and (\ref{eq:44}) as for the deeply inelastic
scattering structure functions above.
Then, in place of (\ref{eq:152}), we find
\begin{equation}
f_{b/A}(\eta,\mu)= \sum_{a}
\int_\eta^1 \frac{\d \xi }{\xi}\,
{\hat g}_{ba} (\eta / \xi) \
{\cal F}_{a/A}(\xi,\mu) ,
\numlabel{eq:153}
\end{equation}
where ${\hat g}_{ba}$ is some new hard part (a matrix in
the space of parton types), while ${\cal F}_{a/A}$ is the
same jet function as in (\ref{eq:152}).  Here we define the scale of
the leading regions to be the same as the renormalization
scale in the parton distribution, and we use the same
notation for both.
 
Using eqs.\ (\ref{eq:152}) and (\ref{eq:153}),
we find the desired result for the structure functions,
\begin{equation}
W ^{\mu \nu}(q^\mu,p^\mu)
\sim \sum_{b}
\int_{x}^1 \frac{\d \eta }{\eta}  \ f_{b/A}(\eta,\mu)\
H_{b}^{\mu\nu}\!\!\left(
q^\mu,\eta p^\mu,\mu,\al(\mu)  \right).
\end{equation}
where the hard part $H^{\mu \nu}$ is defined by the
relation
\begin{equation}
\sum_b \int_x^\xi \frac{\d \eta}{\eta}\
H_{b}^{\mu\nu}\!\!\left(
q^\mu,\eta p^\mu,\mu,\al(\mu)  \right)
\ \hat g_{ba}(\eta / \xi)
={\cal H}_a^{\mu\nu}(q^\nu,\xi p\cdot u\, v^\mu,\mu) .
\numlabel{eq:155}
\end{equation}
 
It should also be possible to demonstrate
this factorization in the more careful manner outlined for
the scalar theory in Sect.~\ref{subtrsec}.   From (\ref{eq:151}), the
leading region $L$ may now be represented by Fig.\ \ref{fig:9},
the canonical form for deeply inelastic
scattering found in the scalar theory and in physical
gauges.  Since the same construction may be carried out for
any leading region, one could define a subtraction
procedure for gauge theories analogous to the one for
scalar theories.  The subtraction operator for a leading
region $L$ then makes the 
replacement (\ref{eq:147}) for the hard part for the region.

\subsection{Single-particle inclusive cross sections and the soft
  approximation} 
 
The leading regions for a single-particle inclusive cross
section, $\e^+ + \e^- \rightarrow A(p) + X$, were
shown in Fig.\ \ref{fig:21}. There is a jet subdiagram $J$ that describes
the jet in which hadron A is observed. The hard subdiagram $H$
contains two short distance interactions (one on each side of
the final state cut) involving highly virtual particles, from
which one or more jets of interacting collinear particles
emerge. Once again, there are extra longitudinally polarized
gluons connecting the jet $J$ to the hard subdiagram $H$.  More
importantly, in contrast with deeply inelastic scattering, there
is a  soft subdiagram $S$ that connects $J$ to $H$. As a result,
the factorization property fails on a graph-by-graph basis.
 
We recall that in any given cut Feynman graph for deeply
inelastic scattering there could be soft partons connecting to
the hard subdiagram, representing soft interactions between
on-shell particles as they enter the final state.  However, we
argued that any leading region containing such soft interactions
gives a cancelling contribution when one  sums over the possible
final state cuts for a given Feynman graph.  Unfortunately,
this rather trivial cancellation mechanism does not work for
single particle inclusive production \cite{CSt}. 
The reason is that we are observing a particle in
the final state rather than summing freely over all
final states. Nevertheless, our aim will be to show that
any leading region with a soft subdiagram connecting
$J$ to $H$ cancels.  The only remaining leading regions are analogous to 
those
already encountered in the case of deeply inelastic scattering, in Sect.\
\ref{gaugeproofsec:DIS.collinear},
so that the proof of factorization sketched there
carries over.
 
We consider the contribution from a leading region $L$ to a
cut Feynman diagram. Each such cut diagram is decomposed into
subdiagrams $J$, $H$, and $S$.  We now sum over all cut graphs
containing the same number of lines connecting the parts $J$,
$H$, and $S$ on each side of the cut and call the result $G$.
Our object is to show that, after summing in addition over where
the lines go relative to the final state cut, $\sum G$ can be
rewritten in a factorized form in the high energy limit.
 
In order to write the kinematic approximations, we pick lightlike
vectors $v^\mu = g^\mu_+$ in the $p^\mu$ direction and $u^\mu =
g^\mu_-$ in the $Q^\mu,p^\mu$ plane, and define the momentum
fraction $\xi$ of the outgoing hadron A by
\begin{equation}
u \cdot k  =  u\cdot p/\xi.
\end{equation}
 
As in the case of deeply inelastic scattering, we use the
longitudinal polarization of the extra collinear gluon lines
which attach the $J$ to the hard part $H$.  We once again
approximate these lines by dominant momentum and polarization
components, so that they appear as longitudinally polarized.
Then, we sum over graphs representing different attachments of
the collinear gluons to $H$ and use Ward identities to remove
them from $H$ and attach them instead to eikonal lines ${\cal
E}$ in the $u^\mu$ direction. They are then grouped with the jet
to form, in this case, a fragmentation function
$d_{A/a}(\xi)$.   We thus derive a form analogous to (\ref{eq:151}),
but with the extra complication of the soft lines,
\begin{eqnarray}
G &\sim & \int \frac{\d \xi}{\xi}
\prod_l  \int \frac{\d^4 q_l }{(2\pi)^4 } 
\prod_j \int \frac{ \d^4{\bar q}_j }{ (2\pi)^4 }
\ \delta^4(\Sigma q_l^\mu + \Sigma {\bar
q}_j^\mu)
\nonumber\\
&& \times
{J}\!\left(
\xi,\{q_l^\nu\}
\right)_{\eta \eta'}^{\{\sigma_l\}}
S(q_l^\alpha,{\bar q}_j^\beta)_{\{\sigma_l,\tau_j\}}
\nonumber\\
&&
\times {H}\!\left(Q^\mu, (u\cdot p/\xi) v^\mu ,\{{\bar
q}_j^\nu\}\right) ^{\eta\eta'\{\tau_j\}} . 
\numlabel{eq:157}
\end{eqnarray}
Here $J$ is analogous to the function $\tilde J$ in
Eq.\ (\ref{eq:150}).  It includes the original jet subgraph together with
eikonal lines attached to the `extra' longitudinally polarized
gluons that formally attached to the hard part.  The indices
$\eta,\eta'$ represent the polarization of the physical parton
that enters the hard part carrying momentum $k^\mu \sim
(u\cdot p/\xi) v^\mu$.  The extra complication compared to deeply
inelastic scattering is the soft subgraph $S$, which couples to
$J$  and $H$ via soft gluons as indicated in
(\ref{eq:157}).
 
It is useful to interpret (\ref{eq:157}) in the language of
Sect.~\ref{gaugeproofsec:DIS.collinear}.  It describes the jet $A$
containing the 
observed hadron $A$ together with the unobserved jets that we have
included in $H$.  All of these jets emerge from a hard
scattering and evolve independently, except for the exchange of
soft partons, which are coupled to the color current of each
jet.  In the frame of jet $A$, for instance, all the charges
within the other jets  are moving at nearly the speed of light.
But then, according to the discussion of
Sect.~\ref{gaugeproofsec:DIS.collinear}, the 
Lorentz transformed field due to these jets should be nearly
gauge equivalent to zero.  Of course, since jet $A$ arises from a
quark or gluon, it is not gauge invariant.  We might expect,
however, that we can  exhibit the gauge nature of this
interaction. To do so, we need a generalization of the eikonal
approximation which we applied in Sect.~\ref{gaugefactsec:IR} to single
parton lines coupled to soft radiation.
 
The relevant generalization of the eikonal
approximation has been termed the ``soft
approximation" \cite{CSS2,CSS4,CSt,CSo}.  For the $A$-jet,it
consists of making the replacement
\begin{equation}
{J}\!\left(
\xi,\{q_l^\nu\}
\right)_{\eta \eta'}^{\{\sigma_l\}}
\rightarrow
{J}\!\left(
\xi,\{\hat q_l^\nu\}
\right)_{\eta \eta'}^{\{\alpha_l\}}
u_{\alpha_1} \cdots u_{\alpha_n}
v^{\sigma_1} \cdots v^{\sigma_n} ,
\numlabel{eq:158}
\end{equation}
in which we define
\begin{equation}
{\hat q}^\alpha = q \cdot v\  u^\mu .
\end{equation}
This approximation replaces each soft gluon
entering the $A$-jet by a fictitious gluon whose
momentum and polarization are both in the
$u$-direction. Before justifying the soft approximation,
let us see what its consequences are.
 
\begin{figure}
   \centering
   \includegraphics{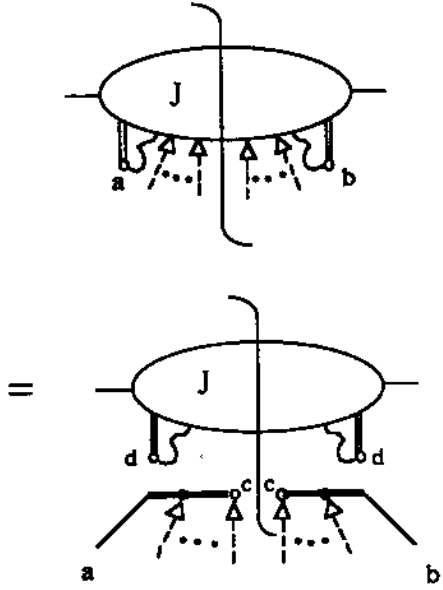}
   \caption{Factorization of soft lines from a jet.}
   \label{fig:26}
\end{figure}

Once we make the soft approximation, each cut jet diagram is a
contribution to a product of matrix elements precisely of the form to
which the Ward identity of Eq.\ (\ref{eq:133}) can be applied, with
the field $\Phi (x)$ now representing the field associated with the
physically polarized parton which couples to the hard part.  As a
result, we have at out disposal a Ward identity, which can be used to
factor soft lines from the jet subdiagram, by an iterative argument
very similar to the one just used to factor longitudinally polarized
collinear gluons from the hard part.  The details of the argument are
slightly more complex because of the extra eikonal line, and we refer
the interested reader to Refs.~\cite{CSS2} and \cite{CSS4} for
details.  Here we simply quote the result, which is illustrated in
Fig.\ \ref{fig:26} and may be expressed as
\begin{multline}
{J}_{ab}\!\left( \xi,\{\hat q_l^\nu\} \right)_{\eta \eta'}^{\{\alpha_l\}}
    u_{\alpha_1} \cdots u_{\alpha_n}
    v^{\sigma_1} \cdots v^{\sigma_n}
\\
=
\frac{1 }{d(R)}\
{J}_{dd}\!\left(\xi\right)_{\eta
\eta'} {\cal E}(v^\mu,{\hat q}^\alpha_L)_{ac}
^{\{\sigma(L)_l\}}
{{\cal E}^*}(v^\mu,{\hat q}^\alpha_R)_{c b}
^{\{\sigma(R)_l\}}
\end{multline}
where ${\cal E}_{ac}(v^\mu,{\hat q}^\alpha_L)$ stands for the
lightlike eikonal line in the $v^\mu$ direction, to which have
been connected those soft gluons to the left of cut $C$, with
momenta ${\hat q}^\alpha_L$ and polarization indices 
$\sigma(L)$. ${\cal E}^*$ is defined similarly in terms of soft
gluons to the right of the cut, with momenta ${\hat q}^\alpha_R$
and polarization indices $\sigma(R)$.  Finally, we have made color
indices $a,b,\cdots$ explicit, and $d(R)$ is the dimension of the
color representation of the physical parton.
 
The complete Green function may now be written in the form
\begin{eqnarray}
   G 
&\sim&
    \int \frac{\d \xi}{\xi}
    \frac{1 }{d(R)}\
    {J}_{dd}\!\left(\xi\right)_{\eta\eta'}
    \prod_l  \int \frac{\d^4 q_l }{(2\pi)^4 }
    \prod_j \int \frac{\d^4{\bar q}_j }{ (2\pi)^4 }\
    \delta^4(\Sigma q_l^\mu + \Sigma {\bar q}_j^\mu)
\nonumber\\
&& \times
{\cal E}(v^\mu,{\hat q}^\alpha_L)_{ac}
^{\{\sigma(L)_l\}}
{{\cal E}^*}(v^\mu,{\hat q}^\alpha_R)_{c b}
^{\{\sigma(R)_l\}}
S(q_l^\alpha,{\bar q}_j^\beta)_{\{\sigma_l,\tau_j\}}
\nonumber\\
&&
\times {H}_{ab}\!\left(Q^\mu,\xi u\cdot q v^\mu
,\{{\bar q}_j^\nu\}\right) ^{\eta\eta'\{\tau_j\}} . 
\end{eqnarray}
Notice that the jet function has now been factored from the rest
of the process. When we now perform a sum over cuts, we can
sum independently over the cuts of $J$ and over the cuts
of the rest of the diagram. In the rest of the diagram, we have
a hard interaction producing the eikonal line ${\cal E}$ and the
jets in $H$.  These are coupled by final state
interactions with the soft gluons in $S$. By the general
reasoning of Sect.~\ref{regionsec:unitarity}, these final state interactions
cancel. Thus any leading region with soft exchanges cancels, and
the factorization reasoning reverts to the arguments which apply
in the scalar theory. We shall skip giving these details, and
will close this subsection with a justification of the
all-important soft approximation, Eq.\ (\ref{eq:158}).
 
The soft approximation consists of an approximation for the
polarizations, and an approximation for soft momenta.  The
former may be justified by detailed power counting
arguments \cite{St78}, but the underlying motivation is simply
that gluon polarizations proportional to $u^\mu$ can couple to
the $A$-jet by contracting into vectors proportional to $p^\mu$,
the momentum of the hadron A. Since $u \cdot p = p^+$ is a
large invariant, it will dominate by a power over invariants
formed from the other internal momentum components present in
$J$.  Note, by the way, that in Feynman gauge gluon
polarizations will have nonvanishing projections onto $u^\mu$
only if the soft subdiagram couples to other parts of the
diagram as well as $J$.
 
The approximation associated with the gluon momentum is more
subtle. Recall that the $A$-jet is in the plus direction.  We
claim that one can neglect the transverse momentum of the gluon
compared to its minus momentum.  As we have seen in
Sect.~\ref{gaugefactsec:IR}, this  is nontrivial.  In fact, regions where
the transverse momentum is nonnegligible are leading by power
counting.
 
Recalling the one-loop discussion of Sect.~\ref{gaugefactsec:IR}, a
typical denominator from the $A$-jet on the left of the cut is of the form of
\begin{equation}
(\ell - q_i)^2 + i\epsilon
\sim \ell^2 - 2\ell^+q_i^- + 2\T\ell \cdot \qt_i
- |\qt_i |^2 +i\epsilon
\numlabel{eq:162}
\end{equation}
with $\ell^\mu$ a typical line momentum in the $A$-jet.
We would like to set $\qt_i $ to zero in all
denominators like (\ref{eq:162}), and all we need for that is
\begin{equation}
|q_i^-|\gg
\frac{|2\T\ell \cdot \qt_i - |\qt_i |^2|
}{ p^+} .
\end{equation}
The relevant question is thus whether the $q^-_i$ momentum
contours are trapped at $q^-_i=0$, at the scale of $(2\T\ell
\cdot \qt_i - |\qt_i |^2)/p^+$. Note that poles of this type can
$only$ come from denominators from the $A$-jet, and not from the
hard part or the soft subdiagram. Now, although every jet line
through which $q^\mu$ flows gives a pole  at a position like
(\ref{eq:162}), close to the origin in the $q_i^-$ plane, all of these
poles are on the same side of the real axis. To see this,
consider how each soft momentum $q_i^-$ flows from the vertex
where it attaches to $J$ to the parton line that attaches $J$ to
$H$.   In general, the $q^-$ pole from any jet line is in the
upper half plane if $q^-$ flows in the opposite sense relative
to the large plus momentum carried by that line.  But $q^\mu$
may always be chosen to flow so that $q^-$ is directed in this
sense for each jet line on which it appears.  This is evident
from Fig.\ \ref{fig:21}.  Soft gluons to the right of the cut may be treated
analogously.  As a result, the $q^-_i$ contours may all be deformed away from
jet poles into a region where $\qt_i $ may be neglected, and the soft
approximation is justified along the deformed contour.  By Cauchy's theorem,
it is also justified in the original integral.  Thus, the factorization
program may be carried out in \epem\ annihilation.
 
\subsection{The Drell-Yan cross section}
 
The thorniest factorization theorems involve two
hadrons in the initial state.  The Drell-Yan cross
section for the process
\begin{equation}
A(p_A) + B(p_B) \rightarrow \ell^+\ell^-(Q^\mu)+X
\end{equation}
is the simplest of these, and has
therefore received essentially all the attention.
$Q^\mu$ will represent the momentum of the lepton pair
$\ell^+\ell^-$.  The step from Drell-Yan to more complex
processes, involving observed hadrons or jets in the final
state is relatively straightforward, as indicated in
Sect.~\ref{regionsec:unitarity}.
 
{\tolerance=5000
 
Factorization for the
Drell-Yan cross section has, at times, been the subject of
con\-tro\-versy \cite{BBL,CSS1,LRS}, although more recent work has,
we believe, established its validity at all
orders \cite{Bo,CSS2,CSS4}. Nevertheless, as we shall observe
below, there is plenty of room for improvement in our
understanding.
 
}
 
The general leading region for the Drell-Yan process is
shown in Fig.\ \ref{fig:22}(b).  After the sum over final
states, all nonforward hadron jets are absorbed into
the hard subdiagram $H$, in the same way as in
deeply inelastic scattering.  In common with the deeply
inelastic scattering and one particle inclusive \epem\
annihilation processes, we can factor collinear gluons from the
hard part.  Once this is done, the sum of cut Feynman diagrams for
the Drell-Yan cross section is very similar to Eq.\ (\ref{eq:157}) for
\epem\ annihilation,
\begin{eqnarray}
G &\sim& \int \frac{\d \xi_A}{\xi_A}  
        \int \frac{\d \xi_B}{\xi_B}
\prod_l  \int \frac{\d^4 q_l }{ (2\pi)^4 } 
\prod_j \int \frac{ \d^4{\bar q}_j }{ (2\pi)^4 }
    \ \delta^4(\Sigma q_l^\mu + \Sigma {\bar
q}_j^\mu)
\nonumber\\
&& \times
{J_A}\!\left(
\xi_A,\{q_l^\nu\}
\right)_{\eta \eta'}^{\{\sigma_l\}}
{J_B}\!\left(
\xi_B,\{q_l^\nu\}
\right)_{\eta \eta'}^{\{\tau_l\}}
S(q_l^\alpha,{\bar q}_j^\beta)_{\{\sigma_l,\tau_j\}}
\nonumber\\
&&
\times {H}\!\left(Q^\mu, \xi_A (u\cdot p_A) v^\mu ,
\xi_B (v\cdot p_B) u^\mu\right)
^{\eta\eta'} , 
\numlabel{eq:165}
\end{eqnarray}
where the lightlike vectors $v^\mu = g^\mu_+,u^\mu=g^\mu_-$ have
been chosen in the $p_A^\mu,p_B^\mu$ directions, respectively
and the parton momentum fractions are defined by $\xi_A = k_A
\cdot u / p_A \cdot u$ and $\xi_B = k_B \cdot v / p_B\cdot v$.
Here ${J_A}(\xi_A,\{q_l^\nu\})$ and  ${J_B}(\xi_B,\{q_l^\nu\})$
are similar to the parton distribution function $\tilde J$
defined for deeply inelastic scattering in Eq.\ (\ref{eq:150}), except
that soft gluons are still attached to them.
Connections between the parton distribution and
a soft subdiagram were absent in the deeply inelastic
scattering cross section, because, after the sum over
cuts, there was only one jet, which cannot by itself
produce large invariants in numerator factors.  In
(\ref{eq:165}) we have a soft subdiagram as in \epem\ annihilation,
but now interacting with the jets associated with the two
incoming hadrons.
 
Our basic problem is the same as in \epem\
annihilation, to show that contributions from any leading region
with a nontrivial soft subdiagram cancel in the sum over final
states and gauge invariant sets of diagrams. Then the remaining
leading regions of Eq.\ (\ref{eq:165}) are just of a form similar to Eq.\ (\ref{eq:144}),
and the arguments for factorization may be given as above for
deeply inelastic scattering, eqs.\ (\ref{eq:152}) to (\ref{eq:155}).
Naturally, we would like to proceed by analogy to \epem\
annihilation. Thus, we would like to apply the soft
approximation to the jets, and factor the soft gluons from
them.  The jets would then contribute to parton distributions as
in Eq.\ (\ref{eq:153}), and, once the remaining soft
contributions cancel, we would derive the desired factorized
form, Eq.\ (\ref{eq:11}).
 
\begin{figure}
   \centering
   \includegraphics{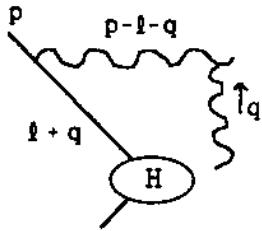}
   \caption{Example illustrating obstacles to the soft approximation.}
   \label{fig:27}
\end{figure}

The main obstacle to this program is shown in Fig.\ \ref{fig:27}, which
illustrates a typical low-order example. It shows a single soft
gluon, $q^\mu$, attached to the A-jet.  The soft momentum flows
through two lines in the $A$-jet, an ``active" jet line $\ell+q$
that carries positive plus-momentum into the hard part, and a
``spectator" line $p-\ell-q$ that carries positive plus momentum
into the final state.  We saw in \epem\ annihilation that the
criterion for the applicability of the soft approximation for
$q^\mu$ is given by $|q^- \ell^+| \gg |\T q \cdot (2{\T \ell} + {\T q})|$,
where $\ell^\mu$ is any line in the jet along which $q^\mu$ may flow.
But this condition may be satisfied along the entire $q^-$ contour
only if the contour is not pinched by poles on opposite sides of
the real line.  In \epem\ annihilation they are not, but in
Drell-Yan they are.  This is illustrated by our example, since
the poles in the $q^-$ plane due to the jet propagators are
approximately at 
\begin{eqnarray}
  q^- &=& \frac{(p-\ell)^2 + 2 (\T p -\T\ell) \cdot \qt
   - |\qt|^2 + i\epsilon }{2(p-\ell)^+ } ,
\nonumber\\
   q^- &=& \frac{-\ell^2 + 2 \T\ell \cdot \qt
   +|\qt|^2 - i\epsilon }{2\ell^+ } ,
\end{eqnarray}
which are on opposite sides of the contour, both at a
distance of order $2\T\ell\cdot \T q/p^+$ from the origin.
Thus, in the Feynman integral associated with Fig.\ \ref{fig:27}, the $q^-$
contour is forced to go through a region in which the soft
approximation fails, and we are unable to apply immediately the
reasoning introduced for \epem\ annihilation.
 
The resolution of this problem is rather technical, and may be found
in Ref.\ \cite{CSS4}. It may be understood most simply as a result of
the Lorentz contraction of the colliding gluon fields, as in
Sect.~\ref{gaugefactsec:classical}.  In addition, we can give an
intuitive picture here, based on semi-classical considerations in the
center of mass frame.  We consider the A-jet to be passing through a
soft color field produced by the B-jet.  Consider first the very
softest part of the color field, with a spatial extent $\sim (1\ {\rm
fm })\ [ p_A^+/ (1\ \GeV)]$.  The self-interactions of the partons in
the A-jet are time dilated, but this field extends so far in space
that it interacts with the partons on the same time scale as that of
their self-interactions.  However, on this distance scale, the soft
gluons cannot resolve the hadron jet into individual partons.  Thus
the jet appears as a color singlet until the time of the hard
interaction, at which time it acquires color because one parton is
annihilated.  The result is that the only interactions of the very
soft color field occur long after the hard scattering event, and such
interactions cancel because of unitarity. Consider now the part of the
color field of smaller spatial extent, say 1~fm.  The point is that
the interactions of this color field with the spectator partons in the
A-jet don't really matter.  The reason is that the self-interactions
of the partons in the A-jet are time dilated, so that the spectator
partons do not interact with the active parton on the 1~fm time scale
in which they interact with the color field.  Since the spectator
partons are not observed, unitarity implies that their interactions
with the color field will not affect the cross section. As a technical
trick, we could as well replace all of the spectator partons with an
equivalent color charge located at $\T x = 0$, right on top of the
active quark.  Then the color field sees a net color singlet in the
initial state and a net colored charge in the final state.  Again, the
only interactions of the color field occur long after the hard
scattering event, and such interactions cancel because of unitarity.
 
Finally, note that the arguments given above are asymmetric
between the two incoming jets.  This is natural, because it is
only necessary that one of the two incoming particles move at
the speed of light for our arguments to apply.  Indeed,
factorization should hold in the (hypothetical) scattering of a
truly lightlike particle with a massive particle at any center
of mass energy.  We should note that explicit two-loop
calculations which show that infrared divergences cancel at
leading power (although not at higher twist), have been carried
out for the most part with one massive and one massless
(eikonal) line \cite{DFT,DGRS}.

%
%

\section{Outlook and Conclusion}
\label{conclsec}

%
%
%
%

In the foregoing, we have described the systematics of
factorization for hard inclusive cross sections in QCD,
and have discussed in some detail the nature of
factorization proofs, first for $(\phi^3)_6$, and then
for gauge theories.  Along the way, we outlined a
systematic approach to perturbative processes at high
energy, based on the classification of leading regions. 

As we have indicated above, the proof of factorization
theorems in gauge theories is by no means a closed
subject.  Factorization proofs for inclusive processes is
the first item of a whole list of subjects in which
progress has been made, but for which important work
remains to be done.  In the following, we briefly discuss
a few other significant 
topics which relate closely to the methods 
discussed in this chapter.  Of great importance are extensions of the
theorems to more general situations. 

\subsection{ Factorization Proofs}

Factorization proofs in nonabelian gauge theories have
reached a certain level of sophistication in Refs.~\cite{Bo},
\cite{CSS2} and \cite{CSS4}. Comparison with the discussion for
$(\phi^3)_6$, however, shows that there is as yet in the literature no 
complete and
systematic subtraction procedure in QCD of the type explained in
Sect.~\ref{subtrsec}, even in the case of deeply inelastic
scattering.  A subtraction algorithm would eliminate
any lingering uncertainty associated with overlaps between
leading regions.  Perhaps even more
importantly, such a procedure should make it
possible to develop bounds on corrections to leading power
factorization theorems, and to prove
factorization theorems for nonleading power
corrections, so-called ``higher twist".  A model for this
program is presumably to be found in the BPHZ formalism for
deeply inelastic scattering cross sections in scalar and
abelian gauge theories developed by Zimmermann \cite{Z},
suitably modified to treat the extra infrared problems
and gauge structure of
 QCD (see Sect.~\ref{subtrsec}).

It should also be noted that the Monte-Carlo
event generators \cite{MC} that are so widespread in analyzing data
depend on generalizations of the factorization theorems; these
generalizations have not yet gone significantly beyond the level of
leading logarithms.

In addition, we should mention that 
additional factorization theorems,
of different but related forms,
are central to the analysis of the
elastic scattering of hadrons,
which decrease as powers of the energy \cite{elastic}.

\subsection{Factorization at Higher Twist }

It has been proposed \cite{Politzer} that
generalized factorization theorems hold beyond the leading
twist for a wide variety of cross sections. Most work on
this possibility has been carried out for deeply
inelastic scattering, where the systematics are best
understood as a generalization of the operator
product expansion \cite{SoldateJaffe,EFP,Qiu}.  In particular,
it has been shown that multiparton distributions may
be defined in a natural way to parameterize soft physics
at higher twist \cite{EFP,Jaffe}.  

In hadron-hadron scattering, factorization at higher twist
is complicated by the infrared structure of perturbative
QCD.  We have seen in Sect.\ \ref{gaugeproofsec} that leading twist
factorization requires the cancellation of infrared
divergences.  It has been shown by explicit calculation
\cite{DFT,DGRS}, however, that infrared divergences do
$not$ cancel beyond a single loop in hadron-hadron
scattering for QCD at higher twist.   This is a
sharp contrast between the nonabelian and abelian
theories.  At two loops, noncancelling divergences occur
at the level $m^4/s^2$ in the Drell-Yan process.  Refering to Sect.\
\ref{gaugefactsec:classical}, this is precisely the level suggested by the 
classical relativistic kinematics of gauge fields.  How one should
interpret this lack of cancellation is not quite clear to us.  The actual
situation, including nonperturbative effects, may be
better or worse than suggested by perturbative
calculations \cite{Nachtmann}.  The fact that perturbation
theory respects factorization at $m^2/s$, however, makes
it possible that
factorization theorems may hold at this level,
even for hadron-hadron scattering \cite{KMS}.

\subsection{  Factorization at the Boundaries of Phase Space}

A rich class of perturbative predictions involve the
summation of corrections near boundaries of phase space
in different processes.  Near some of these boundaries, notably small
$\T Q$ and small $x$, cross sections increase
greatly.

Along these lines, perhaps the
most attention has been given to the Drell Yan cross
section at measured transverse momentum $\d\sigma / \d Q^2
\d^2 Q_\perp$ \cite{DDT,DYlow,CSS5}, with $\T Q \ll Q$ and
the related two-particle inclusive
 cross section for $e^+ + e^- \rightarrow A+B+X$ at measured
transverse momentum \cite{CSo,BBE,KoTr}.  The complete
leading-twist analysis of these cross sections begins
with factorized forms of the type of Eq.\ (\ref{eq:165}),
in which soft partons have been factored from jets,
but not yet cancelled.  At the boundary of phase space the
cancellation of soft gluons outlined in Sect.\ \ref{gaugeproofsec} still
occurs, but is incomplete. All infrared divergences still
cancel at leading power, but finite remainders 
depend on the small parameter in the problem, for
instance the transverse momentum in the cross sections
cited above.  By developing generalizations
of the renormalization group equation for each of the
functions in the factorized form Eq.\ (\ref{eq:165})
\cite{CSo}, it is possible to resum systematically higher order
corrections to these quantities. 

This general approach can be applied in a number of
other physically important situations.  For instance,
the $\tau =Q^2/s \rightarrow 1$ limit in the inclusive
Drell-Yan cross section is related to the normalization of
the Drell-Yan cross section, the ``K-factor"
\cite{Kenyon,Parisi,vanNeer,CataniTrent}.  It is
possible to sum corrections which are singular at $\tau = 1$ 
\cite{St87,AMS}.  An interesting feature of the result is
that it is sensitive to high orders in perturbation theory
\cite{Hooft1} through the running coupling.  Because of this,
it gives a measure of the sensitivity to higher-twist
effects
of perturbative
predictions based on factorization \cite{AMS}.  This
sensitivity is found to be nonnegligible in some, but not
all, regions of physical interest.

Another regime, which is of crucial importance for experiments at the
Tevatron and SSC, is the $x \rightarrow 0$ limit in hadron-hadron
scattering.  The cross sections get into the range of tens of
millibarns, which is enormous compared to typical cross sections at
larger $x$.  So far, much work has concentrated on the behavior of
parton distributions at small $x$
\cite{smallx,Gribov,MuellerQiu,Collins}, assuming the validity of the
standard factorization formulas (\ref{eq:2}), (\ref{eq:3}) and
(\ref{eq:11}).  From a more general point of view, factorization has
been shown to hold explicitly in leading logarithms in $x$
\cite{Milana}.  We would like to suggest, however, that factorization
theorems need a more extensive examination in this region.

In conclusion, we emphasize that essentially
every calculation in perturbative QCD is based on one
factorization theorem or another.
In view of this, 
progress toward developing perturbative QCD as a
quantitative system requires further understanding of the
systematics of factorization.

\section{Acknowledgements}

This work was supported in part by the Department of Energy,
and by the National Science Foundation.


\end{document}